\documentstyle{article}
\newtheorem{proposition}{Proposition}
\newtheorem{corollary}{Corollary}
\newtheorem{remark}{Remark}
\newtheorem{lemma}{Lemma}
\newtheorem{theorem}{Theorem}
\newtheorem{definition}{Definition}

\font\Eul = eufm7 at 12pt

\newcommand{\f}{\underline{f}}

\newcommand{\h}{\underline{h}}
\newcommand{\x}{\underline{x}}
\newcommand{\y}{\underline{y}}

\catcode `\@=11
\@addtoreset{equation}{section}
\def\theequation{\arabic{section}.\arabic{equation}}
\catcode `\@=12
\catcode `\@=11
\@addtoreset{proposition}{section}
\def\theproposition{\arabic{section}.\arabic{proposition}}
\catcode `\@=12
\catcode `\@=11
\@addtoreset{theorem}{section}
\def\thetheorem{\arabic{section}.\arabic{theorem}}
\catcode `\@=12
\catcode `\@=11

\@addtoreset{remark}{section}
\def\theremark{\arabic{section}.\arabic{remark}}
\catcode `\@=12
\catcode `\@=11
\@addtoreset{definition}{section}
\def\thedefinition{\arabic{section}.\arabic{definition}}
\catcode `\@=12
\catcode `\@=11
\@addtoreset{lemma}{section}
\def\thelemma{\arabic{section}.\arabic{lemma}}
\catcode `\@=12
\catcode `\@=11
\@addtoreset{corollary}{section}

\catcode `\@=12

\input epsf
\input mssymb
\date{}
\title{\bf Two-point Functions and Quantum Fields \\ in de Sitter Universe}
\author {Jacques Bros  and Ugo Moschella \\ 
Service de Physique Th\'eorique, C.E. Saclay,
91191 Gif-sur-Yvette, France }
\begin{document}
\maketitle
\begin{abstract}
 We present a theory of general two-point functions 
and of generalized free fields  in $d$-dimensional
de Sitter space-time which closely parallels the corresponding minkowskian
theory.  
The usual spectral condition is now replaced by a certain {\em geodesic spectral
condition}, equivalent to a precise thermal characterization of the
corresponding ``vacuum''states. Our method is based on the geometry of the
complex de Sitter space-time  
and on the introduction of a class of holomorphic functions on this
manifold, called {\em perikernels}, which reproduce { mutatis mutandis}
the structural properties of the two-point correlation functions of the
minkowskian quantum field theory.
The theory contains as basic elementary case the linear massive field models
in their   ``preferred'' representation. The latter are described by the introduction of {\em de Sitter plane waves} in their tube domains which lead to a new integral representation of the two-point functions and 
to   a Fourier-Laplace type transformation on the hyperboloid.
The Hilbert space structure of these theories is then analysed by using this transformation.
In particular we show the Reeh-Schlieder property.
For general two-point functions, a substitute to the Wick rotation is defined both in complex space-time and in the complex mass variable, and substantial results concerning the derivation of K\"allen-Lehmann type representation are obtained.
\end{abstract}
\newpage

\section{Introduction and description of the results}

The de Sitter metric  is an early solution of  
the cosmological Einstein's equations which has 
the same degree of symmetry as the  flat Minkowski solution and    it can 
be seen as a one-parameter deformation of the 
latter which  involves a fundamental length $R$. 
The corresponding space-time  may be represented by a $d$-dimensional 
one-sheeted hyperboloid 
\begin{equation}
X_{d}(R)=X_{d}=
\{x\in {\Bbb R}^{d+1}: {x^{(0)}}^{2}-{x^{(1)}}^{2}-\ldots-{x^{(d)}}^{2}=-R^{2}\}
\label{hyp} 
\end{equation}
embedded in a Minkowski ambient space ${\Bbb R}^{d+1}$, and the
invariance group is $SO_0(1,d)$, the Lorentz group of the ambient space. 
This group plays the same role as the Poincar\'e 
group on  ${\Bbb R}^{d}$. 

Models of Quantum Field Theory (QFT) on the de Sitter space-time have been studied very often
in the past, 
since the symmetry properties of this
universe can be used as a guideline which greatly helps in the otherwise
difficult task of quantizing fields in a gravitational background;
moreover, the radius $R$ may also be regarded as providing a (de Sitter covariant) 
{\em infrared cutoff} for minkowskian 
QFT's, whose removal  regenerates automatically  Poincar\'e covariance.

On the physical side, there has been  a great revival of interest for these 
models at the beginning of the eighties, 
when the de Sitter metric  appeared in the inflationary
cosmological scenario. According to the latter, the universe undergoes a
de Sitter phase in the very early epochs of its life \cite{[L]}.
A possible explanation of phenomena 
having their origin in the very early universe 
then relies on an interplay between space-time curvature and thermodynamics
and a prominent role is played by the mechanisms of symmetry breaking
and restoration in a de Sitter QFT.

In view of the physical importance of these models, and also of the
great theoretical significance of de Sitter space-time, an enormous amount of
work has been done concerning quantum field theory on this space-time and many
mathematical aspects of the latter have been extensively developed (see    \cite{[BD],[A],[KW]}
and references therein.) 
Most of this work has been devoted to the construction of quantum fields satisfying a Klein-Gordon type equation, since even this simplest case already sets non-trivial problems when the background space-time is curved.

In fact, QFT on a curved space-time is plagued by the phenomenon of non-uniqueness of the vacuum state.
In the free field case such ambiguities immediately appear since, whenever 
a second quantization 
procedure is applied, it is based on mode expansions related to
a given choice of local coordinates (which generally cover only a part
of the whole space). 
At a deeper level of investigation
the considerable relativization of the concept of energy operator, 
reduced to be (on curved space-time) the infinitesimal generator of some 
one-parameter group of isometries interpreted as the proper time-evolution for the chosen 
system of coordinates of a certain observer, makes it impossible to 
characterize unambiguously the vacuum state of QFT as a fundamental 
state for the energy in the usual sense; indeed, in general  
{\em no true spectral condition can be  satisfied by 
a  QFT on a curved space-time and this is related to the 
proliferation of quantizations}.

In the case of de Sitter space-time, where such ambiguities are present as in the general case, it has however been discovered that the {\em Hadamard condition} selects a unique vacuum state for quantum fields satisfying the de Sitter covariant Klein-Gordon equation (see \cite{[A],[KW]} for two distinct senses in which this is true).
The Hadamard condition postulates that the short distance behavior of the two-point function 
of the field should be the same for  Klein-Gordon fields on curved space-time as for the corresponding minkowskian free field. The fact that the Hadamard condition selects a unique vacuum state for linear fields has actually been established for  a wide class of 
space-times with bifurcate Killing horizons, of which de Sitter space-time is an 
example (see \cite{[KW]} and references therein); in the de Sitter case the preferred vacuum state thus selected  coincides with the so called ``Euclidean'' \cite{[GH]} or ``Bunch-Davies''  \cite{[BuD]} vacuum state, and it singles out one vacuum in the two-parameter family of quantizations constructed in \cite{[A]}.

However, the  too special character of the Hadamard condition if one wishes to deal with {\em general interacting fields} leads one to search another explanation of the existence of preferred de Sitter vacuum states in the global structure of this space-time manifold and, as we shall see, of the corresponding complex quadric $X^{(c)}_d$ (described by Eq. (\ref{hyp}), with $x$ varying in ${\Bbb C}^{d+1}$). In this connection, the feature of the Hadamard condition which will remain of general value for de Sitter QFT is the requirement that the two-point function is the boundary value of an analytic function ``from the good side'' (i.e. the so-called $i\epsilon$-rule.)

The program which we have in mind concerning QFT in de Sitter space-time deals with the possibility of developping a general approach based on a set of fundamental principles, which should be  completely similar to the Wightman approach for minkowskian fields. This is by no means obvious, since in the latter many important concepts are based on the Fourier representation of space-time, namely the energy-momentum space.
Therefore one major objective to be reached would be to dispose of a global de Sitter-Fourier calculus which would allow one to implement such important notions 
as wave propagation, ``particle states'', etc.. 
However in a preliminary step, one can start from the idea that support properties in Fourier space are closely connected (via Paley-Wiener theorems) to analyticity properties in the original space. In fact, let us recall that the basic momentum space properties of Minkowskian QFT (in particular the spectral condition) can be equivalently expressed as specific analyticity properties of the $n$-point correlation functions 
${\cal W}_n(x_1,\ldots,x_n)$
in complex Minkowski space; moreover it is known that (in view of the ``reconstruction theorem'' \cite{[SW]}) any general sequence of distributions  
$\{{\cal W}_n(x_1,\ldots,x_n)\}$ satisfying the relevant analyticity and positivity properties determine a unique Wightman field acting in the associated ``GNS-vacuum'' representation.

A major success of this analytic framework has been the derivation of the celebrated K\"allen-Lehmann representation of general two-point functions   
which expresses the latter as ponderate superposition of ``elementary'' two-point functions satisfying a Klein-Gordon equation with arbitrary mass $m \geq 0$.

The following important facts, connected with this representation,
have inspired the first steps of our approach to general QFT in the de Sitter universe.
\begin{description}
\item{i)}
{\em The global holomorphic structure of the two-point function } (both in complex space-time variables and in complex energy-momentum variables) is {\em of maximal type} with respect to the principles of locality, covariance and positivity of the energy spectrum; we mean that  the (physical sheet) holomorphy domain of the two-point function, represented by a cut-plane in the relevant Lorentz-invariant variable, could not be enlarged without imposing very special conditions (such as local commutativity at time-like separation or lacunae in the energy spectrum).
One can thus say that at the level of the two-point function, these general principles are completely encoded in the analyticity properties exhibited by the K\"allen-Lehmann formula, which holds for all interacting local fields in the vacuum representation.

\item{ii)} Being given any general two-point function ${\cal W}_2$ satisfying the K\"allen-Lehmann representation there exists a unique {\em generalized free field} which obeys the general principles and admits ${\cal W}_2$ as its two-point function (it is obtained via the GNS construction by imposing that for this field all the truncated $n$-point functions ${\cal W}^T_n$, with $n>2$, and ${\cal W}_1$ vanish).

\item{iii)}  As a by-product of ii), the free fields can be introduced by the previous construction starting from the ``elementary'' two-point functions
which satisfy a Klein-Gordon equation. This way of introducing the free fields presents the advantage of being independent of any particular quantization procedure; it moreover exhibits directly the various properties of the field from those of its two-point function.
\end{description}

In this paper we will show that all the previous facts have their exact counterparts in the de Sitter space-time thanks to the introduction of relevant analyticity domains on the complex quadric   $X^{(c)}_d$ and to the (related) existence of a global Fourier calculus on de Sitter space-time whose primary steps are presented here.

The starting point of our axiomatic approach to de Sitter QFT was precisely the discovery of the existence of two-point kernels on  $X^{(c)}_d$ enjoying maximal analyticity properties (similar to those of minkowskian two-point functions mentioned in i)) and equipped with causal discontinuities of retarded and advanced type on the real de Sitter space $X_d$. Such holomorphic kernels had been introduced in the context
of the  theory of complex angular momentum in minkowskian QFT  \cite{[BV-2],[BV-1]}, but the relevance of this global analytic framework for describing general two-point functions on the de Sitter universe was first advocated by one the authors in \cite{[Br]}, where (as an application of the results of \cite{[BV-2]}) an integral representation of K\"allen-Lehmann type was proved under a certain growth condition related to the dimension of the corresponding de Sitter space-time.

This paper presents several new achievements (partly described in \cite{[BGM]}),  which open the way to a genuine axiomatic approach to de Sitter QFT, and can be shortly described as follows.
The results of \cite{[Br]} concerning the global structure of general two-point functions are  completed not only in themselves (for example, the positivity of the weight of the K\"allen-Lehmann representation is proven here) but by the study of various field-theoretical aspects which they imply.
In particular,  having in view   a possible general approach to de Sitter QFT, we treat the physical interpretation of the maximal analyticity property of two-point functions from the viewpoint of the ``energy content'' of the underlying QFT's (as a substitute to the spectral condition).
Of course,  without any information on the $n$-point functions for $n>2$, one can not infer a physical  property of general interacting fields from the structure of the two-point function alone.
This is why it was reasonable to construct and study at first the {\em generalized free fields}, the latter being uniquely determined by the two-point functions according to the GNS procedure, as in the minkowskian case (see ii) above). Moreover, such a construction presents the major interest of covering the free field case as a by-product, since the maximal analytic framework applies to all the elementary two-point functions which satisfy the de Sitter-Klein-Gordon equation (as in iii)). Our general result can then be stated  as follows: in the maximal analytic framework which we consider, the two-point functions determine generalized free fields in a {\em preferred representation} which is characterized by a {\em well-defined KMS condition}
formulated in appropriate regions of de Sitter universe associated with geodesic observers. The well-known occurrence of Hawking's thermal effects on  de Sitter
space-time \cite{[GH],[KW]} is given here a strikingly simple geometrical interpretation; the KMS condition is actually the ``natural substitute'' to the spectral condition when maximal analyticity   on the temporal geodesics remains satisfied, the linear geodesics of the flat case being here replaced by hyperbolae.

The maximal analytic framework also sheds light on the euclidean properties of the (generalized) free fields considered; in fact all the two-point functions
here considered define Schwinger functions by restriction from $X^{(c)}_d$ to the sphere $S_d$ of ``imaginary times'' $(x^{(0)} = i y^{(0)}$ in Eq. (\ref{hyp})) which plays the same role as the euclidean space in complex Minkowski space. In the free field case, this allows us to identify our model with the pioneering euclidean formulation of the preferred vacua given in \cite{[GH]} (and of course also coincides with the preferred vacua selected by the Hadamard condition). It is perhaps worthwhile to stress  that properties of analytic continuation are at the basis of every treatment of de Sitter field theories based on the functional integral on the euclidean sphere; this includes  the constructive approach to QFT on de Sitter space-time (in this connection see \cite{[FHN]}) or the application that de Sitter theories  may find in minkowskian constructive  field theory, since the de Sitter radius $R$ provides a natural infrared cutoff (the euclidean space is compact!). Without the appropriate analyticity properties all results inferred by euclidean methods would not be relevant for the real de Sitter universe. 

Of course in the general interacting case one cannot expect maximal analyticity properties to hold for $n$-point functions ($n>2$) and again we are left with the task of finding a suitable general approach to de Sitter QFT; to be relevant this approach   should assure the possibility of going to the euclidean sphere, solving there the field equations, coming back to real de Sitter world  (and possibly taking the limit $R \to \infty$ if the de Sitter radius is used as a cut-off).
At the end of this   paper we will present a set of axioms which is based on the previous ideas and which represents one possible solution to this problem. In a forthcoming paper \cite{[BM]} we will show that the euclidean sphere (minus the coinciding points) {\em does belong} to the extended domain of analyticity of the $n$-point functions satisfying these axioms.

A  major contribution of this paper is the introduction of global Fourier variables playing the role of mass and momentum variables, and of relevant Laplace and Fourier-type transformations; in the latter the exponentials will be replaced by appropriate eigenfunctions of the Laplace-Beltrami operator which can be identified with
 {\em de Sitter plane waves}. 
The introduction of de Sitter plane waves in relevant tube domains of $X^{(c)}_d$, first presented by one of the author in \cite{[Mo]}, yields  a new integral representation  of the free field two-point function (and thereby a previously unknown integral representation of Legendre-Gegenbauer functions in their complex domain): this representation,  is the analogue of the standard plane-wave expansion of $D^-(x;m)$ in Minkowski space; it also allows one to control in a very suggestive way the limit of de Sitter QFT to Minkowski QFT when the de Sitter radius tends to infinity. In fact it is at basis of our study of the free field in its various aspects. 
The family of analytic plane waves we will introduce also suggests the
definition of a suitable Fourier type transformation, which maps test functions on $X_d$ to homogeneous functions on its asymptotic cone; similar transforms had also been introduced in the context of harmonic analysis on $X_d$   in the framework of the $SO(1,d)$ representation theory \cite{[Fa2],[M]}, but with our definitions we can derive new mathematical results more adapted to our analytic framework based on the complex quadric $X_d^{(c)}$. 

Besides it will be seen in a further work \cite{[BM2]} that by applying  our Fourier transformation  to the special case  of $SO_0(1,d)$-invariant
retarded propagators  one recovers  the  {\it Laplace transform} of the latter, which coincides with the so-called spherical Laplace transform of invariant Volterra kernels introduced in \cite{[Fa],[FV]}. 
This Laplace transform, which depends on a single variable interpreted as a complex mass, is the genuine analogue of the minkowskian two-point Green function in the invariant mass variable: its property of being holomorphic in a half-plane and  the  characterization of the euclidean Schwinger function by the values of this Laplace transform at discrete imaginary masses (which is a generalization of the so-called imaginary time formalism (ITF)  of thermal QFT)  \cite{[Br],[BV-2]}  are  recalled here in view of their importance. 

In  \cite{[BM2]}, all these notions will be developed so as to produce a genuine Fourier calculus on de Sitter space, which could hopefully be applicable to the $n$-point functions of a general field theory, to investigations on the notion of particle and to the computation of Feynman amplitudes.

Here we make use of the Fourier transform to describe the Hilbert space structure of the (generalized) free field theories and the associated representation of the de Sitter group. In particular 
we  prove the validity of the Reeh-Schlieder property, which is 
another important aspect  of the analyticity framework which we   wish to stress. The latter will be established here for de Sitter (generalized) free fields by arguments which are very close in their spirit to the proof of the original theorem in Minkowski space and differ from the recent proof of \cite{[Ver]}, applicable to free fields on the class of (not necessarily analytic) ultrastatic spacetimes. The principle of uniqueness of analytic continuation (applied under the refined form of the ``edge-of-the-wedge'' theorem in a complex manifold) together with the use of the (analytic) Fourier transformation we have introduced  appear to be the main reasons for the validity of this property, and we believe that this would also hold for a large class of manifolds equipped with a holomorphic structure, provided the two-point functions considered are boundary values of holomorphic functions from relevant tuboid domains (see Appendix A). As a matter of fact, properties of this nature (extended to the $n$-point functions) could represent a reasonable ``microlocal''substitute to the spectral condition for a suitable class of manifolds. one can hope that there exists a  formulation of such an axiomatic approach   in the framework of the analytic wave-front set theory \cite{[BI2]} (where a local substitute of the analytic Fourier transformation would be available)
in the spirit of the one proposed in \cite{[Radz]} (for general space-time manifolds) in the framework of the ${\cal C}^\infty$ wave-front set theory \cite{[Hor]}.
  
\vskip5pt
Our paper is organized as follows:

Section 2 is devoted conjointly to the presentation of general two-point functions on de Sitter space-time enjoying maximal analyticity properties and to the construction and characterization of models of generalized  free fields associated with such two-point functions. In \S 2.1 and 2.2 we fix the notations and introduce the Fock representation for fields satisfying 
only the requirements of locality and de Sitter covariance;  
in \S 2.3  we introduce  the notion of ``normal  analyticity'' for general two-point functions 
and show that this is sufficient to recover the maximal 
analyticity (perikernel property)  of \cite{[Br],[BV-2]};  
for the associated generalized free fields this  is equivalent to postulating the {\em geodesic spectral condition}
or the {\em geometric KMS  condition} of \S 2.4, supplemented by an appropriate ``antipodal condition''  when  the field correlation functions are considered as given on the whole space-time $X_d$.

Sections 3 and 4 continue the study of Section 2 under two different respects
and can therefore be read independently.

Section 3 is a development of the general study of two-point functions,
which is based on various results of 
\cite{[Br],[BV-2],[BV-1],[FV]}.
In \S 3.1 we introduce a Feynman-type algebra and a substitute to 
Wick-rotation in the space-time variables . The Laplace transform of the retarded propagators 
is described in \S 3.2.
We devote \S 3.3 to describe the connection between   this Laplace transform at imaginary masses and the Legendre expansion of the Schwinger function. 
This substitute to the Wick rotation property in momentum space 
(called in  \cite{[BV-2]} the ``Froissart-Gribov property'') 
 is a generalized form (invariant under the de Sitter group) of 
the connection between RTF and ITF in the thermal representations
of QFT in Minkowski space \cite{[BB]}. 
In \S3.4 we present an integral representation of general two-point functions satisfying appropriate growth properties at infinity in $X_d$; for a specially interesting class of two-point functions, it takes the form of a genuine 
 K\"allen-Lehmann representation.   

Section 4 is devoted to  a detailed study of the  
de Sitter-Klein-Gordon fields, which are exhibited as a class of basic
elements of the general framework of Section 2.
Analogues of the plane-wave solutions of the Klein-Gordon equation are 
introduced in  \S 4.1 following \cite{[M]} for the  de Sitter-Klein-Gordon equation, 
and a corresponding analysis of the linear quantum field is  given 
in \S  4.2. The use of the tubular domains of the complex hyperboloid 
is crucial for obtaining 
the relevant ``plane-wave representation'' of the two-point function, 
identified to a  Gegenbauer function of the first kind.
In mathematical terms, we introduce in \S 4.3
a new version of the Fourier  transformation 
on the hyperboloid  (a purely mathematical study of this 
transformation will be developed in \cite{[BM2]}).
This transformation allows us to completely characterize in \S 4.4
the Hilbert space of ``one-particle'' states and the corresponding 
irreducible unitary representations of the de Sitter group.
We will also discuss in a simple and direct way the contraction of the 
unitary irreducible representations of the de Sitter group (giving rise 
to  {\em reducible}  representations of the Poincar\'e group \cite{[MN]}). 
 
In Section 5 we complete our discussion of general two-point functions. 
In \S 5.1 we 
prove  the positivity property of the  
K\"allen-Lehmann representation obtained in \S 3.4 for 
the relevant class of  de Sitter two-point functions   
by using the Fourier type transform    introduced in \S 4.3.
We then discuss in  \S 5.2 the flat limit of de Sitter GFF: we  show that 
the plane-wave representation introduced in \S 4.2 tends by contraction to the corresponding 
Fourier representation of the minkowskian Wightman function; the same property is derived for de Sitter general two-point functions having a K\"allen-Lehmann decomposition.   It follows from this analysis that  group-theoretical contraction  is not a suitable procedure to draw consequences on the structure of the limiting quantum field theory; in particular no problem of negative energy arises in the latter.
\S 5.3 is devoted to show the Reeh-Schlieder property for de Sitter GFF's. 
We conclude Section 5 by some remarks on the QFT's obtained by taking the restriction to $X_d$ of a Minkowski QFT in the ambient space.

In our final  Section 6 we outline a possible 
Wightman-type axiomatic approach to the theory 
of interacting fields on the de Sitter space-time, which is formulated in terms of global analyticity properties of the $n$-point functions of the fields. 
 Consequences of this axiomatic approach will be discussed in a 
forthcoming paper \cite{[BM]}. 

Throughout our work, the Wightman functions of the fields are described as distributional boundary values of analytic functions from relevant domains of the complex de Sitter manifold, which we call ``tuboids'' \cite{[BI]}; a general survey of these mathematical notions is given in our Appendix.

\section{Two-point functions and generalized free fields in the maximal analyticity framework}

\subsection{Notations}

It is useful to describe the main properties of the de Sitter space-time 
$X_{d}$ (see Eq. (\ref{hyp})) in terms of the minkowskian structure of
 the ambient  space ${{\Bbb R}}^{d+1}$, whose scalar product is denoted  by 
\begin{equation}
x\cdot y =  {x^{(0)}}{y^{(0)}}-{x^{(1)}}{y^{(1)}}-\ldots-{x^{(d)}}y^{(d)}.
\label{product}
\end{equation}
with as usual $x^2 = x\cdot x.$

$X_{d}$  is then equipped with a {\em causal} ordering relation (induced by
that of ${{\Bbb R}}^{d+1}$): let 
\begin{equation}
\overline{V^{+}} = \left\{x\in {{\Bbb R}}^{d+1}:
{x^{(0)}}\geq\sqrt{{x^{(1)}}^{2}+\ldots +{x^{(d)}}^{2}}\right\}
; \end{equation}
then, for 
$
x,y \in X_{d},\;\;x\geq y\; \leftrightarrow \; x-y \in \overline{V^{+}}.
$
The future (resp. past) cone of a given point or ``event'' $x$ in $X_{d}$
will be denoted by $\Gamma^+{(x)}$ (resp.$\Gamma^-{(x)}$)
$
\Gamma^\pm{(x)}= \{y \in X_{d}: y \geq x \;\;\;( y \leq x)\}.
$
The ``light-cone''$ {\partial \Gamma}(x)$,x namely the boundary set of   
$\Gamma^+{(x)}\cup\Gamma^-{(x)}$  is the union of all linear generatrices of 
$X_{d}$ containing the point $x$ (equivalently: 
${\partial \Gamma}(x)=\{y \in X_{d}: (x-y)^2=0\})$.
Two events $x$ and $y$ of $X_{d}$ are in ``acausal relation'', or 
``space-like separated'' if $y \not\in \Gamma^+{(x)}\cup\Gamma^-{(x)}$, i.e. if 
$x\cdot y > -R^2$.

The (pseudo)-distance in the de Sitter space-time is implicitely defined by 
\begin{equation}
\cosh  \frac{d(x,y)}{R} = -\frac{x\cdot y}{R^{2}}
\label{distance}
\end{equation}
 for $x$ and $y$ time-like separated; it is defined by 
\begin{equation}
\cos \frac{d(x,y)}{R} = -\frac{x\cdot y}{R^{2}}
\label{distance1}
\end{equation}
 for $x$ and $y$ space-like separated and such that $|x\cdot y| < R^2$;
it is not defined  for $x$ and $y$ space-like separated and such that
$x\cdot y > R^2$. Correspondingly, for any pair $(x,y)$ with $x\geq y$, 
there exists a unique arc of hyperbola in $X_{d}$ in the two-plane 
determined by  $0,x$ and $y$, which is a time-like geodesic with end-points 
$x$ and $y$, and $d(x,y)$ (defined by Eq. (\ref{distance})) is the proper 
time which elapses from $y$ to $x$ for an observer sitting on this geodesic.
When $x$ and $y$ are space-like separated, there exists a unique (connected) 
geodesic with end-points $x$ and $y$ if and only if  $|x\cdot y| < R^2$;
in this case, this geodesic is an arc of ellipse   in the two-plane 
determined by  $0,x$ and $y$, and $d(x,y)$ (defined by Eq.
 (\ref{distance1})) is the spatial distance between $x$ and $y$.
In both cases, the distance is defined as an ``angle'' (respectively hyperbolic or circular) multiplied by $R$.     

The relativity group of the de Sitter space-time is the pseudo-orthogonal group
$SO_{0}(1,d)$ leaving invariant the scalar product (\ref{product}),
and each of the sheets of the  cone  $C = C^{+}\cup C^{-}$:
\begin{equation}
C^{\pm}= \{x\in {{\Bbb R}}^{d+1} :{x^{(0)}}^{2} - {x^{(1)}}^{2} - \ldots -
{x^{(d)}}^{2} = 0
,\;{\rm sgn}{x^{(0)}}=\pm \}\label{cone} 
\end{equation}
This group is  denoted by $G_d$.
The   $G_d$-invariant volume form on $X_{d}$ will be denoted by 
$d\sigma(x)$, and normalized as follows (in Leray's
notations \cite{[Le]}): 
\begin{equation}
d\sigma(x)=\left.\frac{dx^{(0)}\wedge dx^{(1)}\wedge \ldots \wedge dx^{(d)}}
{d(x^2 + R^2)}\right|_{X_{d}}
\label{leray}
\end{equation}
Since the group $G_d$ acts in a transitive way on $X_{d}$, it will be
convenient to distinguish a {\em base point} $x_{0}$ in $X_{d}$, which will play the 
role of the origin in Minkowski space-time; we choose the point 
$x_{0}=(0,\ldots,0,R)$, and consider the tangent space $\Pi_d$ to  $X_{d}$,
namely the hyperplane  $\Pi_d = \{ x \in {{\Bbb R}}^{d+1}: x^{(d)}=R\}$ as the
$d$-dimensional Minkowski space-time (with pseudo-metric 
${dx^{(0)}}^{2} - d{x^{(1)}}^{2} - \ldots d{x^{(d-1)}}^{2}$) onto which the 
 de Sitter  space-time can be contracted in the zero-curvature limit.

Let then $L_d$ be the subgroup of $G_d$ which leaves the point $x_0$ invariant:
$L_d$ is isomorphic to the Lorentz group $SO_0(1,d-1)$ of $\Pi_d$ and its orbits
 are the $(d-1)$-dimensional hyperboloids $\omega_{\alpha}$ with equations 
${x^{(0)}}^{2} - {x^{(1)}}^{2} - \ldots -
{x^{(d-1)}}^{2} = R^2(\alpha^2 - 1)$ in the corresponding hyperplanes
 $\Pi_d(\alpha) = \{ x \in {{\Bbb R}}^{d+1}: x^{(d)}=R\alpha\}$.
For $\alpha >-1$, each hyperboloid $\omega_{\alpha}$ represents a set of points 
$x$ in $X_{d}$ which are equidistant from the base point $x_0$, namely 
$\omega_{\alpha}= \{x \in X_{d} : d(x,x_0) = d_{\alpha}\}$ where (in view of Eqs. (\ref{distance}) and (\ref{distance1})): for $\alpha\geq 1$,  
 $\cosh d_{\alpha}/R = x^{(d)}/R =\alpha$; for $|\alpha|<1$,  
$\cos d_{\alpha}/R = x^{(d)}/R =\alpha$.

\subsection{Local scalar fields in de Sitter space-time; generalized free fields}
The fields $\phi(x)$ which we wish to consider on $X_d$ are expected to be \cite{[SW],[J]} 
operator-valued distributions on $X_d$ acting in a Hilbert space $\cal H$ 
and satisfying the following property of {\em local commutativity} 
(on a suitable dense domain): $[\phi(x),\phi(y)]=0$, if $ x$ and $y$ are 
spacelike separated in the sense of \S 2.1. Moreover, there should exist
a continuous 
unitary representation $U(g)$ of the de Sitter group $G_d$ acting on $\cal H$ 
under which the field $\phi$ is transformed covariantly, i.e. 
$
U(g)\phi(x)U(g)^{-1} = \phi(gx)
$
(we limit ourselves here to scalar bosonic fields) and a distinguished vector 
$\Omega \in {\cal H}$, cyclic for the polynomial algebra of field operators and
 invariant under the representation $U(g)$ of $G_d$, which we call ``the 
vacuum''. For a more elaborate formulation of this theoretical framework, one is led to consider the Borchers-Uhlmann \cite{[B]} algebra $\cal B$ of terminating sequences of 
test-functions ${\bf f} = (f_0,f_1(x_1),\ldots,f_n(x_1,\ldots,x_n),\ldots)$ on $X_d$,
where the $f_n$ are ${\cal C}^\infty$ test-functions with compact support on the cartesian product of
$n$ copies of $X_d$. This $\star$-algebra $\cal B$ (in which the involution is 
$ {\bf f}\to {\bf{\bar f}} =(\bar{f_0},\bar{f_1}(x_1),\ldots,\bar{f_n}(x_n,\ldots x_1),\ldots))$
is equipped in a natural way with a representation $\alpha_g$ of the de Sitter group $G_d$, given by 
$\alpha_g({\bf f})=
{\bf f}_g=(f_0,f_1(g^{-1}x_1),\ldots,f_n(g^{-1}x_1,\ldots,g^{-1}x_n),
\ldots)$, and there exists a ``locality'' ideal ${\cal I}_{loc}$ in ${\cal B}$, 
which is defined as in the minkowskian case (but in terms of the de Sitter 
space-like separation). Each scalar local field theory is then uniquely defined by a positive linear functional $\omega$ on $\cal B$ satisfying the following two properties:

 i)  $G_d$-{\em invariance}: $\omega({\bf f}) = \omega({\bf f}_g)$, $\forall g \in G_d$

 ii) {\em locality}: $\omega({\bf f})=0$, for every $f \in {\cal I}_{loc}$.
While the functional $\omega$ is defined in terms of the quantum field $\phi(x)$
by the sequence of its ``Wightman functions'' $\{ {\cal W}_n(x_1,\ldots,x_n) = 
\langle\Omega, \phi(x_1)\ldots\phi(x_n) \Omega\rangle, \linebreak[0] n\in {\Bbb N}\}$,
the field theory is conversely reobtained from the functional $\omega$ via 
the  G.N.S.-type  construction of a triplet $({\cal H}, \Phi, \Omega)$
satisfying all the required (covariance and locality) properties; in the latter,
the representation ${\bf f} \to \Phi({\bf f})$ 
(of ${\cal B})$ contains the field $\phi(x)$ through the relation $ \phi({ f_1})= \int \phi(x)f_1(x) d\sigma(x)
= \Phi\left((0,f_1,0,\ldots)\right)$. For every open set ${\cal O}$ of $X_d$
the corresponding polynomial algebra ${\cal P}({\cal O})$ of the field $\phi$ 
is then defined as the subalgebra of $\Phi({\cal B})$ whose elements 
$\Phi(f_0,f_1,\ldots,f_n,\ldots)$ are such that for all $n\geq 1$ 
supp$f_n(x_1,\ldots, x_n)\subset {\cal O}^{\times n}$.
The set ${\rm D} = {\cal P}(X_d)\Omega$ is a dense subset of ${\cal H}$.

  In this framework a {\em generalized free field} on $X_d$ can then be defined 
by a functional $\omega = \{ {\cal W}_n; n\in {\Bbb N}\}$, such that $W_1=0$ and all
the truncated functions ${\cal W}^{\rm tr}_{n}$ vanish for $n>2$. The theory is then 
entirely encoded in the two-point function ${\cal W}={\cal W}_2(x_1,x_2)$ which 
should be a 
distribution on $X_d\times X_d$ satisfying the following conditions:
\vskip5pt
a) {\em Positivity}: $\forall f \in {\cal D}(X_{d})$
\begin{equation}
\int_{X_d\times X_d} {\cal W}(x,y)\overline{f}(x)f(y)d\sigma(x)d\sigma(y)\geq 0;
\label{positivity}
\end{equation}
\vskip5pt
b) {\em locality}: ${\cal W}(x,y)={\cal W}(y,x)$ {\em for every space-like separated pair} $(x,y)$;
\vskip5pt
c) {\em covariance}: ${\cal W}(gx,gy) ={\cal W}(x,y)$,  $\forall g \in G_d$.
\vskip5pt
In this case, the G.N.S. triplet associated with the functional $\omega$ can be 
constructed explicitly: it is the Fock representation of a field
$\phi(x)$ satisfying the following commutation relations:
\begin{equation}
\left[\phi(x),\phi(y)\right] = C(x,y) \times {\bf 1} \label{CR}
\end{equation}
where the commutator function is 
$
C(x,y)={\cal W}(x,y)-{\cal W}(y,x). \label{CR1}
$
The Hilbert space ${\cal H}$ of the representation can be described as the
Hilbertian sum ${\cal H}_0 \oplus [ \oplus_n S({\cal H}_1)^{\otimes n}]$ 
(with S
denoting the symmetrization operation), in which ${\cal H}_0 =\{\lambda \Omega,
\lambda \in {{\Bbb C}}\}$ and ${\cal H}_1$ is defined as follows: a regular
element $\dot{h}_1$ of ${\cal H}_1$ is a class of functions $h_1(x)$ in 
${\cal D}(X_d)$ modulo the functions $g$ such that 
$\int_{X_d\times X_d} {\cal W}(x,y)\overline{ g}(x)\linebreak[0]g(y)d\sigma(x)d\sigma(y)=0$;
for such element $h_1$ the corresponding norm is defined by the formula
\begin{equation}
\|\dot{h}_1\|^{2}=\langle \dot{h}_1,\dot{h}_1\rangle=
\int_{X_d\times X_d} 
{\cal W}(x,y)\overline{h_1}(x)h_1(y)d\sigma(x)d\sigma(y)\geq 0;
\label{hjk}
\end{equation}
The full Hilbert space ${\cal H}_1$ is then obtained by completion 
(for this norm) of the space of regular elements.
An analogous description follows for the regular elements $\dot{h}_n$ of 
the various subspaces ${\cal H}_n = S({\cal H}_1^{\otimes n})$.

Each field operator $\phi(f)$ can then be decomposed into ``creation'' and 
``annihilation'' parts $\phi(f)=\phi^{+}(f)+\phi^{-}(f)$, the action
 of the latter on the dense subset of ``regular elements'' of the form 
$\dot{h} = 
(\dot{h}_0, \dot{h}_1,\ldots \dot{h}_n,\ldots, \dot{h}_N, 0,0,\ldots)$
being specified  by
\begin{equation}
\left(\phi^{+}(f)\dot{h}\right)_n(x_1,\ldots,x_n)=\frac{1}{\sqrt{n}}
\sum_{k=1}^{n}f(x_k)\dot{h}_{n-1}(x_1,\ldots, \hat{x}_{k},\ldots,x_n)\label{crea}
\end{equation}
$\left(\phi^{-}(f)\dot{h}\right)_n(x_1,\ldots,x_n)=$
\begin{equation}
={\sqrt{n+1}}\int_{X_d\times X_d} f(x){\cal W}(x,y)
\dot{h}_{n+1}(y,x_1,\ldots,x_n)
d\sigma(x)d\sigma(y) \label{azzo}
\end{equation}
The latter formulae imply the commutation relations (\ref{CR}) and therefore (in
view of condition b)
on $\cal W$) the field $\phi(x)$ satisfies
the property of local commutativity. Finally, condition c) implies the 
$G_d$-covariance of the field $\phi(x)$, the representation $U(g)$ of $G_d$ in 
$\cal H$ (namely the one trivially induced by $\alpha_g$) being unitary in view of 
the same condition c) and of formula (\ref{hjk}).

\subsection{The property of maximal analyticity of two-point functions; perikernels on $X^{(c)}_d$}

The previous construction of the  Fock representation of generalized free fields 
has not necessitated (a priori) any postulate playing the role of the positivity of the energy spectrum; it is in fact independent of the properties of the underlying space-time and differs from the usual presentation  of GFF in 
Minkowski spacetime (see e.g. \cite{[Wig],[J]}) in which the positivity 
of the energy spectrum is built-in via the use of momentum-space and the 
choice of a spectral measure $\rho(m)$.

In the case of de Sitter space-time, where no momentum-space interpretation 
is given a priori, we shall keep from the minkowskian case the idea that 
 the {\em properties of analytic 
continuation} of the theory (i.e. of the two-point function) in the {\em
complexified space-time} are directly related to the energy content
of the model considered;
 we shall therefore take as a postulate certain 
analyticity properties of ${\cal W}(x,y)$ on the complexified de Sitter space-time 
$X^{(c)}_d$ which are completely similar to those implied by the usual spectral 
condition in the complex Minkowski space-time ${{\Bbb C}}^{d+1}$.
The interpretation of these properties in terms of the energy content of our 
models of GFF on de Sitter universe will then be given in the next \S 2.4; moreover we  think that, as in Minkowski space, such global analyticity of the two-point functions, should   also hold for {\em general} interacting fields on de Sitter space-time.

Let us first introduce some geometrical notions concerning the 
complex hyperboloid 
\begin{equation}
X^{(c)}_{d}(R)=X^{(c)}_{d}=
\{z = x+iy \in {{\Bbb C}}^{d+1}: {z^{(0)}}^{2}-{z^{(1)}}^{2}-\ldots-{z^{(d)}}^{2}=-R^{2}\},
\label{hypec} 
\end{equation}
 equivalently described as the set 
\begin{equation}
X^{(c)}_{d}=
\{(x,y)\in {\Bbb R}^{d+1} \times 
{\Bbb R}^{d+1}: {x}^{2}-{y}^{2}=-R^{2}, x\cdot y = 0\}.
\label{hypec1} 
\end{equation}
We define the following open subsets of $X^{(c)}_{d}$: 
\begin{equation}
{\cal T}^{+} = \makebox{\rm T}^+\cap X^{(c)}_{d},\;\;\;\;\;
{\cal T}^{-} = \makebox{\rm T}^-\cap X^{(c)}_{d},
\label{tubi1}
\end{equation}
 where {\rm T}$^{\pm} = 
 {\Bbb R}^{d+1} + i {V^{\pm}}$ are the so-called forward and backward 
 tubes\footnote{These are the (minimal) analyticity domains 
 of the Fourier-Laplace 
transforms of tempered distributions $\tilde{f}(p)$ with support contained in 
$\overline{V^+}$ or in $\overline{V^-}$, obtained in connection with the 
spectral condition in Wightman QFT in ${{\Bbb R}}^{d+1}$ \cite{[SW],[J]}.}
in ${{\Bbb C}}^{d+1}$. 
${\cal T}^{+}$ (resp. ${\cal T}^{-}$) can be characterized as the set
 of points $z=x+iy \in X^{(c)}_{d}$ such that $-R^{2}< x^{2}<0$ or $x=0$ 
(equivalently
$0<y^{2}\leq R^{2}$) and $y^{(0)} > 0$ (resp. $y^{(0)} < 0)$.
In the  same way as $\overline{\makebox{\rm T}^+}\cup
\overline{{\rm T}^-}$ contains the ``euclidean subspace''
${\cal E}_{d+1}=\{ z=(iy^{(0)},x^{(1)},\ldots, x^{(d)}): (y^{(0)},
x^{(1)},\ldots,  x^{(d)}) \in {{\Bbb R}}^{d+1} \}$ of the complex Minkowski 
space-time ${{\Bbb C}}^{d+1}$, one easily checks that 
$\overline{{\cal T}^+}\cup
\overline{{\cal T}^-}$ contains the  sphere
${ S}_{d}=\{z=({iy^{(0)}},{x^{(1)}},\ldots{x^{(d)}}):\;\; 
{y^{(0)}}^{2}+{x^{(1)}}^{2}+\ldots+{x^{(d)}}^{2}=R^{2}\}$. 
We will also call  ${\cal T}^{+}$, ${\cal T}^{-}$ and ${ S}_{d}$ the 
``forward'' and ``backward tubes'' and the ``euclidean sphere'' of 
$X^{(c)}_{d}$.

In order to complete the parallel with the tubes ${\rm T}^+$ and ${\rm T}^-$ of Minkowski space, it is important to show that ${\cal T}^+$ and ${\cal T}^-$ are 
domains of $X^{(c)}_{d}$ from which one can take the boundary value on $X_{d}$
of analytic functions (in the sense of distributions) in a well-defined way; the relevant geometrical notion which generalizes those of tube and local tube is the notion of {\em tuboid} in a complex manifold
(see definition A.3, the corresponding notion of boundary value being fully characterized in Theorem A.2.).

\begin{proposition}
The ``tubes'' ${\cal T}^{+}$ and  ${\cal T}^{-}$ are tuboids above $X_{d}$ in 
$X^{(c)}_{d}$, whose profiles are respectively 
$ \Lambda^+ = \bigcup_{x\in X_{d}} (x,\Lambda^+_x)$ and
$ \Lambda^- = \bigcup_{x\in X_{d}} (x,\Lambda^-_x)$, where:
\begin{equation}
\Lambda^+_x = - \Lambda^-_x = T_x X_{d} \cap V^+,\label{profiles}
\end{equation}
$T_x X_{d}$ denoting the tangent space to $ X_{d}$ at the point x,  identified here to a hyperplane of the vector space 
$T_x{{\Bbb R}}^{d+1}\approx {{\Bbb R}}^{d+1}$. 
Similarly the set
\begin{equation}
{\cal T}_{12} = \{(z_1,z_2): z_1\in {\cal T}^{-},  z_2\in {\cal T}^{+}\}
\label{tub12}
\end{equation}
is a tuboid above  $X_{d} \times X_{d}$ in 
$X^{(c)}_{d} \times X^{(c)}_{d}$, whose profile is 
$  \Lambda^-\times\Lambda^+$.
\label{prop1}
\end{proposition} 
 {\bf Proof.}
For each point $\x \in X_{d} $  we introduce the  subset 
$\Lambda_{\x,R}=\{\y \in T_{\x}X_d ; \;\;    \y^2<R^2\}$ 
of the tangent space to 
$X_d$ at the point $\x$: 
$T_{\x}X_d=\{\y\in R^{d+1};\;\;\x\cdot\y=0\}$ ($T_{\x}X_d$ is a $d$-dimensional minkowskian subspace of ${\Bbb R}^{d+1}$). 
Correspondingly, we put $\Lambda_R= \bigcup_{\x\in X_{d} }\left(\x,\Lambda_{\x,R}\right)$; $\Lambda_R$ is an open subset of the tangent bundle $ TX_{d} $ to  $X_d$. We then introduce the following mapping $\delta$ from $\Lambda_R$ into $X^{(c)}_{d}$:
\begin{equation}
z = \delta((\x,\y)) = \frac{\sqrt{R^2-\y^2}}{R} \x +i\y.\label{etre}
\label{delta}
\end{equation}
In view of Eq. (\ref{hypec1}) it is clear that $\delta$ is a one-to-one 
${\cal C}^{\infty}$-mapping from $\Lambda_R$ onto the following subset $Z_R$ of 
$X^{(c)}_{d}$ (considered as a $2d$-dimensional real manifold):
\begin{equation}
Z_R = X^{(c)}_{d}\setminus Y_R, \;\;\makebox{with}\;\;
Y_R =\{z \in X^{(c)}_d;\; z=iy,  (y^2=R^2)\}
\end{equation}
one  moreover checks that at each point $(\x,\y)$ of $\Lambda_R$, the corresponding tangent mapping $T\delta$ is of maximal rank, so that $\delta$ is in fact a diffeomorphism from $\Lambda_R$ onto $Z_R$.

If one now introduces the following open and connected subset $\Lambda^+_R= \{
(\x,\y) \in \Lambda_R;\; \y\in V^+\}$ of $\Lambda_R$, one easily deduces from Eq. (\ref{delta}) that $\delta(\Lambda^+_R) = {\cal T}^{+}\setminus Y_R$; since $\delta$ is a diffeomorphism, $ {\cal T}^{+}\setminus Y_R$ is an open connected subset of $X^{(c)}_d$, which also entails the connectedness of ${\cal T}^+$;
the tubes  ${\cal T}^+$ and  ${\cal T}^-$ are therefore domains of  $X^{(c)}_d$. In order to show that  ${\cal T}^+$ and  ${\cal T}^-$ are 
tuboids above $X_d$ with respective profiles $\Lambda^+$ and $\Lambda^-$ (defined by Eq. (\ref{profiles})), one just has  to notice that the {\em global} diffeomorphism $\delta$ provides (in the neighbourhood of all  points $\x \in
X_d$) {\em admissible local} diffeomorphisms in the sense of definition A.1 
(conditions a) and b) of the latter are satisfied in view of Eq. (\ref{etre})).
Properties a) and b) of definition A.3 are then satisfied by ${\cal T}^{+}$ and 
${\cal T}^{-}$ as a by-product of the equalities $\delta(\Lambda^{\pm}_{R})=
{\cal T}^\pm\setminus Y_R$ and $\delta(\Lambda_{R}\setminus\Lambda^{\pm}_{R})=
Z_R \setminus {\cal T}^\pm$ (resulting from the previous analysis).
The  statement concerning the set ${\cal T}_{12}$ is  a direct 
 application of the previous facts and  of proposition A.2. 
   \\  

We can now formulate the following postulate of ``normal analyticity'' for general two-point functions  on de Sitter space-time, which supplements the postulates a), b), c) stated in \S 2.2:
\vskip5pt
d) {\em postulate of normal analyticity: \\
the two-point function
${\cal W}(x_1,x_2) =\langle \Omega, \phi(x_1) \phi(x_2) \Omega \rangle$ is 
the boundary value (in the sense of distributions according to Theorem A.2) of a function 
${\rm W}(z_1,z_2)$ which is analytic in the  domain ${\cal T}_{12}
$ of 
$X^{(c)}_{d}\times X^{(c)}_{d}$.}
\vskip5pt
From this postulate and from the locality and covariance conditions satisfied 
by ${\cal W}$, we deduce the following properties of ${\rm W}$:
\begin{proposition} (property of maximal analyticity)
\noindent 1) ${\rm W}(z_1,z_2)$ can be analytically continued in the ``cut-domain''
$\Delta = X^{(c)}_{d}\times X^{(c)}_{d} \setminus \Sigma^{(c)}$ where 
``the cut'' $ \Sigma^{(c)}$ is the set $\{(z_1,z_2): z_1\in 
X^{(c)}_{d}\times X^{(c)}_{d}: (z_1-z_2)^2=\rho,\forall \rho\geq 0\}$;\\
2)  ${\rm W}(z_1,z_2)$ satisfies in $\Delta$ the complex covariance condition:
\begin{equation}
 {\rm W}(gz_1,gz_2)= {\rm W}(z_1,z_2) \label{covcom}
\end{equation}
for all $g\in G^{(c)}_d$, the complexified  of the group 
$ G_d$;\\
3) the ``permuted Wightman function'' ${\cal W}(x_2,x_1) = 
\langle \Omega, \phi(x_2) \phi(x_1) \Omega \rangle$ is the boundary value of 
${\rm W}(z_1,z_2)$ from the domain 
${\cal T}_{21} = \{(z_1,z_2): z_1\in {\cal T}^{+}, \; z_2\in {\cal T}^{-}\}$.
\label{maxan}
\end{proposition}
 {\bf Proof.} In view of Eq. (\ref{tubi1}), 
the tubes ${\cal T}^{+}$ and   ${\cal T}^{-}$ are (like 
$ \makebox{\rm T}^+$ and
$ \makebox{\rm T}^-$) invariant under   
$G_d$; the covariance postulate c) then implies the corresponding covariance property
for the analytic function W in its domain ${\cal T}_{12}$, namely
\begin{equation}
\forall g \in G_d,\;\;\;{\rm W}(gz_1,gz_2) = {\rm W}(z_1,z_2). \label{covzero}
\end{equation}
In fact, the boundary value on $X_d\times X_d$ (from ${\cal T}_{12}$) of every derivative 
$D{\rm W}$ of ${\rm W}$ with respect to any one-parameter subgroup of $G_d$ is the corresponding 
distribution $D{\cal W}$, which is zero, in view of postulate c); this entails
(by a refinement of the analytic continuation principle given in Corollary A.1) that $D{\rm W}=0$, and therefore proves Eq. (\ref{covzero}).
The extension of Eq. (\ref{covzero}) to  the complexified   group 
$g\in G^{(c)}_d$
then follows by analytic continuation 
in the group variables.
This implies that the function ${\rm W}$ can be analytically continued
in the domain ${\cal T}^{ext}_{12} = \{(z_1,z_2) \in X^{(c)}_d\times
X^{(c)}_d;\;\;
z_1=gz'_1,\;
z_2=gz'_2;\;
(z'_1,z'_2)\in {\cal T}_{12}, \;g \in G^{(c)}_d\}$; the fact that this set
coincides with $\Delta$ is a by-product of the study of the extended 
tube in two vector variables 
${\rm T}^{ext}_{12}$  
 (in ${{\Bbb C}}^{d+1}$) (see e.g.
\cite{[KaW]}, where it is shown that the section of 
${\rm T}^{ext}_{12}$  
by $z^2_1=z^2_2=-R^2$, namely 
${\cal T}^{ext}_{12}$, 
is described by the cut-plane ${{\Bbb C}}\setminus
\overline{{\Bbb R}}^+$ in the third Lorentz-invariant variable $z^2_3$,
where $z_3 = z_1-z_2$). The relation (\ref{covcom}) then obviously holds in
$\Delta$ as a consequence of (\ref{covzero}), with $g\in G^{(c)}_d,
(z_1,z_2)\in 
 {\cal T}^{ext}_{12}$. In view of d), it is clear that    the {\em permuted
Wightman function}
${\cal W}'(x_1,x_2)
={\cal W}(x_2,x_1)$ is the boundary value of a function 
${\rm W}'(z_1,z_2)$ 
which is analytic in the tuboid 
${\cal T}_{21}$; then the same argument as above applied to ${\rm W'}$
shows that ${\rm W'}$ can also be analytically continued in the domain
$\Delta$ in which it satisfies the covariance condition 
${\rm W}'(gz_1, gz_2)=
{\rm W}'(z_1, z_2),\;\;\;(\forall g \in G^{(c)}_d)$.
Now the domain $\Delta$ of
$ X^{(c)}_{d}\times
X^{(c)}_d$ 
contains a real open set $\cal R$ (of 
$ X_d\times
X_d$) in which ${\rm W = W'}$. This is the set of space-like separated
points ${\cal R} = \{ (x_1,x_2) \in 
X_d\times
X_d; (x_1-x_2)^{2}< 0\}$, where the previous equality holds as a
consequence of the locality condition b). It follows that the analytic
functions ${\rm W}$ and ${\rm W'}$ coincide, which implies property 3)
and ends the proof of the proposition.    \\  

Properties 1) and 2) of proposition \ref{maxan} characterize ${\rm W}$ as being an 
{\em invariant perikernel} on $X^{(c)}_d$ in the sense of [BV-2] (with domain $\Delta = D_0$). By using the transitivity of the group 
$G^{(c)}_d$ on  $X^{(c)}_d$,  one can identify any such perikernel 
${\rm W}(z_1,z_2)$ with an $ L^{(c)}_d$-invariant function $\underline{\rm{W}}(z)= {\rm W}(z,x_0)$, $x_0$ being the base point $(0,\ldots,0,R)$ introduced in \S 2.1, and  $L^{(c)}_d$ the 
subgroup of  $G^{(c)}_d$ leaving invariant $x_0$; $\underline{\rm{W}}$ is 
analytic in the domain 
\begin{equation}
\underline{\Delta} = \left\{ 
z\in X^{(c)}_d, \; z^{(d)}=-\frac{z\cdot x_0}{R} \in 
{{\Bbb C}}\setminus [R,\infty[\right\}.
\end{equation}
In view of its $ L^{(c)}_d$-invariance property, 
$\underline{\rm{W}}(z)$ can also be identified with a function 
${\rm w}( \alpha)$ analytic in the cut-plane ${{\Bbb C}}\setminus [1,\infty[$
by putting ${\rm w}(\frac{z_d}{R})= \underline{\rm W}(z)$ (with $z_d = R \alpha$). $\underline{\rm W}$ and ${\rm w}$ will be called reduced forms of ${\rm W}$.
According to theorem A.2, 
the postulate d) of normal analyticity implies that the analytic function ${\rm W}(z_1,z_2)$ is locally of moderate growth, i.e. is bounded  by an inverse power of the distance from its real boundary $X_d\times X_d$. It then follows that the analytic functions 
$\underline{\rm W}(z)$ and ${\rm w}(\alpha)$ are themselves locally of moderate  growth near their (respective) real boundary sets $X_d$ and ${{\Bbb R}}$. 
In view of theorem A.2, 
the function $\underline{\rm W}(z)$ thus admits two distribution-like boundary values $\underline{\cal W}$ and 
$\underline{\cal W}'$ on $X_d$ from the respective tubes 
 ${\cal T}^{-} $ and  ${\cal T}^{+}$ such that formally 
$\underline{\cal W}(x)$ =${\cal W}(x,x_0)$ and 
$\underline{\cal W}'(x)$ =${\cal W}(x_0,x)$. The corresponding jump of $\underline{\rm W}$, namely the distribution
$\underline{\cal C} = \underline{\cal W} 
-\underline{\cal W}'$ (such that  $\underline{\cal C}(x)
= {\cal C}(x,x_0)$), has its support contained in 
$\Gamma^{+}(x_0)\cup\Gamma^{-}(x_0) = \{ x \in X_d;\;\;x^{(d)} \geq R\}$, 
see fig.1); in the open set ${\cal R}_{x_0}= \{x\in X_d;\;\;x^{(d)} < R\}$ the distributions $\underline{\cal W}$ and $\underline{\cal W}'$ coincide and are identical with the restriction of the analytic function  $\underline{\rm W}$ to 
${\cal R}_{x_0}$.

\bigskip

\begin{figure}
\epsfysize=8.cm{\centerline{\epsfbox{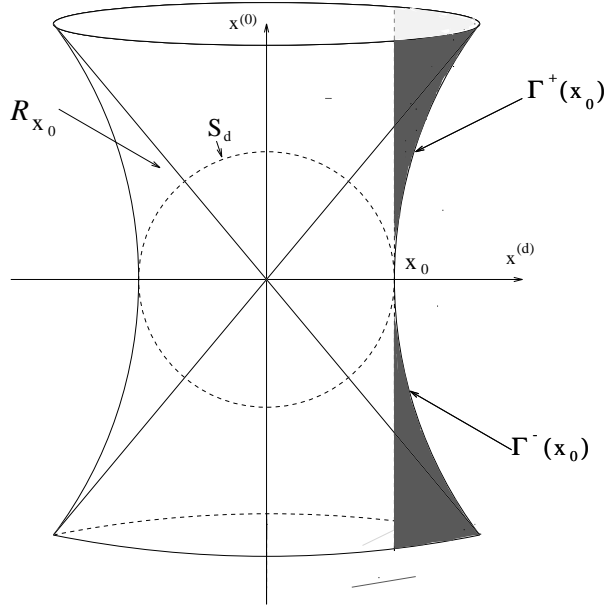}}}
\caption{Causal domains associated with the point $x_0$}
\label{fig0}
\end{figure}

The distributions  $\underline{\cal W}$, $\underline{\cal W}'$ and 
$\underline{\cal C}$ are invariant under the subgroup $L_d$ of $G_d$, but their identification with the corresponding boundary values ${\rm w}_+$, ${\rm w}_-$
(resp. the jump c=${\rm w}_+$-${\rm w}_-$) of the analytic function of moderate growth ${\rm w}(z^{(d)}/R)$ on (resp. across) ${{\Bbb R}}$ must be done carefully, namely separately on the open half-line $]-\infty,1[$, 
considered as the projection of ${\cal R}_{x_0}$, and on the open half-line 
$]1,\infty[$, considered as the projection of the interior either of $\Gamma^{+}(x_0)$ or  of $\Gamma^{-}(x_0)$.

The previous analysis closely follows the corresponding presentation of two-point functions in Minkowski space-time (from the viewpoint of analytic functions and of the 
affiliated distributions); here, the transition to the $L_d$-invariant variable $\alpha=\cos \frac{d_\alpha}{R}=  \frac{z^{(d)}}{R}$ (see \S 2.1) 
is the exact analogue of the passage from the vector variable ${\rm z}i
= ({\rm z}^{(0)}, \vec{z})$ to the Lorentz-invariant variable 
${\rm z}^2= ({{\rm z}^{(0)}}^{2}- {\vec{z}}^{2} )$ in the minkowskian case. 
This analysis will also be similarly completed
by the following results for the Green functions and for the 
euclidean propagators.
\vskip5pt
{\em The Green functions}
\vskip5pt
The retarded propagator ${\cal R}(x_1,x_2)$ (resp. its reduced form 
$\underline{\cal R}(x)={\cal R}(x,x_0)$. is introduced by 
splitting the support
of ${\cal C}(x_1,x_2)$ (resp. $\underline{\cal C}(x)$) corresponding to the formal definitions 
\begin{equation}
{\cal R}(x_1,x_2)= i\theta (x^{(0)}_1-x^{(0)}_2) {\cal C}(x_1,x_2),
\end{equation}
\begin{equation}
\underline{\cal R}(x)= i\theta (x^{(0)}) \underline{\cal C}(x).
\end{equation}
$\underline{\cal R}(x)$ has its support contained in the ``future cone'' $\Gamma^{+}(x_0)$ of $X_d$; its rigorous definition as an $L_d$-invariant distribution is obtained in the local coordinates $(x^{(0)},\ldots,x^{(d-1)})$
along the same line as in the minkowskian case (see e.g. \cite{[EG]}); 
in fact the singular character of $\underline{\cal C}$ at the point $x=x_0$ 
necessitates a two-step procedure for splitting its support, namely:\\
i) $\underline{\cal R}$ is defined on a subspace of test-functions vanishing up to a certain derivation order  at $x=x_0$;\\
ii) $\underline{\cal R}$ is then continued (up to a distribution with support 
at $x=x_0$) to the whole space ${\cal D}(X_d)$ by virtue of the Hahn-Banach theorem.
The other Green functions and their reduced forms are defined in terms of 
${\cal R}$ and  $\underline{\cal R}$ by the usual formulae, namely:\\
-advanced propagator: ${\cal A}= {\cal R}-i{\cal C}$ (resp.
 $\underline{{\cal A}}$= $\underline{\cal R}$-$i\underline{\cal C}$);\\
-chronological propagator: 
$\tau = -i{\cal R}+{\cal W}'= -i{\cal A}+{\cal W}$ (resp.
 $\underline{\tau}$ = $-i\underline{\cal R}$+$\underline{\cal W}'$);\\
-antichronological propagator: $\overline{\tau} = {\cal W}+i{\cal R}= 
{\cal W}'+i{\cal A}$ (resp.
 $\underline{\overline{\tau}}$ = $\underline{\cal W}$+$i\underline{\cal R}$).\\
As for the distributions ${\rm w}_+,\;{\rm w}_-$ and c considered on the whole real line (in $\alpha$), they are respectively the one-variable representatives of $L_d$-invariant distributions $\underline {\overline{\tau}}$, $\underline{\tau}$ and $\underline {\overline{\tau}}$- $\underline{\tau}$=
$i(\underline{\cal A} + \underline {\cal R}$), 
the latter being formally equal to 
$\epsilon(x^{(0)}){\cal C}(x)$; however this calls for the following

\begin{remark}{\em       although apparently absent from the definition of 
${\rm w}_+,\;{\rm w}_-$ and c (directly obtained from boundary values of 
the analytic function w($\alpha$)), the  problem splitting the support 
(necessitating a two-step procedure of the type described above) reappears in the reconstruction of 
$\underline{\tau}$ and $\underline{\overline{\tau}}$ 
as distributions on $X_d$ from their one-variable representatives 
 ${\rm w}_-,\;{\rm w}_+$. 
}\end{remark}  
\vskip5pt 
\begin{remark}{\em     
As in the minkowskian space-time, 
the two-point function 
of a hermitian field satisfies the identity
${\cal W}'(x_1,x_2) =\overline{{\cal W}(x_1,x_2)}$.\\ 
It follows that the function ${\rm W}(z_1,z_2)$ is real analytic, 
i.e. for $z_1,z_2 \in \Delta$ $\overline{{\rm W}(z_1,z_2)} ={{\rm W}(\overline{z_1},\overline{z_2})}$, and that the Green functions ${\cal R}(x_1,x_2)$ and ${\cal A}(x_1,x_2)$ are real valued.
}\end{remark}  
{\em The Euclidean propagators or Schwinger functions}
\vskip5pt
It is easily seen that the intersection of the domain $\Delta$ with the 
``euclidean  subset''  ${\cal E} = S_d \times S_d$ of $X^{(c)}_d \times 
X^{(c)}_d $ is the set $\dot{\cal E} = {\cal E} \setminus \delta$, where 
$\delta$ is the set of ``coinciding points'', namely 
${\delta} = \{(z_1,z_2)\in X^{(c)}_d \times 
X^{(c)}_d, \; z_1 =z_2\}$: in fact, if $z_i = (iy^{(0)}_i,\vec{x}_i), i=1,2$, one has $(z_1-z_2)^2=-(y^{(0)}_1-y^{(0)}_2)^2 - (\vec{x}_1-\vec{x}_2)^2 \leq 0$,
which implies that $(z_1,z_2)\in \Sigma^{(c)}$ iff $z_1=z_2$. It thus follows from Proposition 2.2 that the ``euclidean propagator'' ${\rm S}={\rm W}|_{S_d \times S_d}$ is defined as an analytic function in $\dot{\cal E}$; moreover, 
since ${\rm W}(z_1,z_2)$ is of moderate growth with respect to the distance of
$(z_1,z_2)$ from $\Sigma^{(c)}$ (due to the distribution character of the boundary values of W), S is itself of moderate growth on $\dot{\cal E}$ with respect to the distance of $(z_1,z_2)$ from 
$\delta_{\cal E} = {\cal E}\setminus\dot{\cal E} = \delta \cap {\cal E}$.
As in the minkowskian case, a standard argument based on the construction of a suitable primitive of S, shows that S can be extended as a distribution to the whole euclidean space ${\cal E}$.
S can be seen as a distribution kernel on $S_d$,
with singular support on the diagonal, which is moreover 
invariant under the subgroup $SO(d)$ of $G^{(c)}_d$,
in view of the $G^{(c)}_d$-invariance of W. It follows that S admits the 
reduced forms $\underline{\rm S}(z) = {\rm S}(z,x_0)$ and ${\rm s}(\cos\theta)$
defined by
$\underline{\rm S}(z)$, 
$={\rm s}(-\frac{z\cdot x_0}{R^2})$
for $z\in S_d$; 
$\underline{\rm S}$ is in fact a distribution on $S_d$ with singular 
support at $x_0$, invariant under the stabilizer of  $z_0$, and is also obtained by taking 
the restriction of $\underline{\rm W}$ to $S_d$; similarly one can say that 
${\rm s}|_{[-1,1[} = {\rm w}_\pm |_{[-1,1[}$.
 
\subsection{The thermal interpretation and the geodesic spectral condition for generalized free fields}

In this subsection we shall study the GFF's whose two-point functions satisfy the general postulates a), b), c), d) (this is what is meant when we will say that these GFF satisfy a), b), c), d)). As seen in \S 2.2 a), b), c) imply that they are scalar local fields acting in a Hilbert space.

In order to investigate the enrgy content of such fields, which is encoded in d), we shall adopt the viewpoint of an observer moving on the geodesic 
of the point $x_0$ contained in the $(x^{(0)},x^{(d)})$-plane, namely the branch of
hyperbola 
\begin{equation}
h(x_0)= \{x=x(t);\;\;\;x^{(0)}=R\sinh\frac{t}{R},\;
x^{(1)}=\ldots x^{(d-1)}=0,\;x^{(d)}=R\cosh\frac{t}{R}\}
\label{geod}\end{equation}
where $  t\in {{\Bbb R}}$.  The set of all events of  $X_R$ which
can be connected  with the observer by the reception and the emission of   light-signals is the region: 
\begin{equation}
{\cal U}_{h(x_0)} = \{x \in X_d: \;\;\;x^{(d)}>|x^{(0)}|\};
\end{equation}
this region admits two boundary parts $H^+_{h(x_0)}$ and $H^-_{h(x_0)}$,
respectively called the ``future'' and ``past horizons'' of the observer with geodesic
${h(x_0)}$, namely 
\begin{equation}
H^\pm_{h(x_0)} = \{x \in X_d: \;\;\;x^{(0)}=\pm x^{(d)},\;\;x^{(d)}>0\}.
\end{equation}
Since the parameter $t$ of the representation (\ref{geod}) is interpreted as the proper
time of the observer sitting on the geodesic ${h(x_0)}$ (see \S (2.1)), 
one can call ``time-translation group relative to ${h(x_0)}$'' 
the one-parameter subgroup $T_{h(x_0)}$ of $G_d$ 
(isomorphic to the one-parameter Lorentz subgroup of $SO_0(1,d)$) 
whose transformations are hyperbolic rotations  parallel to the 
$(x^{(0)},x^{(d)})$-plane. The action of $T_{h(x_0)}$ on the domain
${\cal U}_{h(x_0)}$ is defined below. Let $x=x(\tau,\vec{\x})$ be any point 
in ${\cal U}_{h(x_0)} $ represented as follows:
\begin{equation}
x(\tau,\vec{\underline{x}})  
=
\left[
\begin{array}l
x^{(0)} =  \sqrt{R^2 - \vec{\underline{x}}^2 }\sinh \frac{\tau}{R}\\
( x^{(1)},\ldots,x^{(d-1)}) = \vec{\underline{x}}\\
x^{(d)}=\sqrt{R^2 - \vec{\underline{x}}^2 } \cosh \frac{\tau}{R} 
\end{array}
\right. , \;\;\;\;\;\tau\in {\Bbb R},\;\;{\vec{ \x}}^2<R^2.
\label{parahor}
\end{equation}
For $t\in {{\Bbb R}}$, one puts 
\begin{equation}
T_{h(x_0)}(t) [x(\tau,\vec{\underline{x}})]  
=x(t+\tau,\vec{\underline{x}})\equiv x^{t},
\label{old}
\end{equation}
(with $x^0 = x$ in this notation). $T_{h(x_0)}$ thus defines a group of 
isometric automorphisms of 
${\cal U}_{h(x_0)}$ whose orbits are all branches of hyperbolae of 
${\cal U}_{h(x_0)}$ in two-dimensional plane sections parallel to the 
$(x^{(0)},x^{(d)})$-plane (see \cite{[KW]} for a general 
discussion of this kind of structure). 
Note however, that in this family of orbits of $T_{h(x_0)}$,
${h(x_0)}$ is the only one which is a  geodesic of $X_d$, so that the time-translation
interpretation of the group $T_{h(x_0)}$ is, strictly speaking, only relevant for observers
sitting on ${h(x_0)}$ (or in a neighbourhood of ${h(x_0)}$ which is ``small 
with respect to the radius $R$ of $X_d$).
One can also consider the action of the  group $T_{h(x_0)}$ on the horizons  
$H^\pm_{h(x_0)}$ which are foliated by linear orbits of this
 group (these orbits 
are the light rays on $X_d$ which are parallels to the asymptots of ${h(x_0)}$).
We can now establish the following property of the generalized free fields on 
$X_d$:
\begin{proposition} (geometrical KMS condition)\\
For every pair of bounded regions ${\cal O}_1$, ${\cal O}_2$ of 
${\cal U}_{h(x_0)}$, all the (two-point) correlation functions between 
elements of the corresponding
polynomial algebras  ${\cal P}({\cal O}_1)$, ${\cal P}({\cal O}_2)$ of a 
generalized free field on $X_d$ satisfying the previous postulates a), b), c), d) 
enjoy a KMS condition with respect to the time-translation group $T_{h(x_0)}$ whose 
temperature parameter is equal to $1/2\pi R$.
\label{kms}
\end{proposition}

 {\bf Proof.} 
We first consider the correlation functions with respect to             $T_{h(x_0)}(t) $ between one-component
elements $\Phi(0,f^{(i)},0,\ldots)=\phi(f^{(i)})$ of ${\cal P}({\cal O}_i)$, 
$i=1,2$, namely
\begin{equation}\langle\Omega, \phi(f^{(1)})\phi(\alpha^t_{h(x_0)}f^{(2)})\Omega\rangle= {\cal W}_{12}(t),
\;\;\langle\Omega, \phi(\alpha^t_{h(x_0)}f^{(2)})\phi(f^{(1)})\Omega\rangle= 
{{\cal W}'}_{12}(t)
\label{mamma}
\end{equation}
where we have put $[\alpha^t_{h(x_0)}f]{(x)}=f(x^{-t})$.
In view of Eq. (\ref{old}), we can then write
\begin{equation}
{\cal W}_{12}(t)=\int {\cal W}(x_1,x^{t}_{2})f^{(1)}(x_1)f^{(2)}
(x_2)d\sigma(x_1)d\sigma(x_2),
\end{equation} 
which defines ${\cal W}_{12}$ as a ${\cal C}^{\infty}$ function of $t$.
Let us  now consider, for every point $(x_1,x_2(\tau,\vec{\x}))$ 
$(\tau\in{{\Bbb R}})$ in ${\cal O}_1\times{\cal O}_2$
the corresponding (complex) curve $h_{x_1,x_2 }$ of $X_d\times X^{(c)}_{d}$ defined as follows:
\begin{equation}
h_{x_1,x_2 }=\{(x_1,z_2 )\,;\, z_2= z(t+\tau,\vec{\x})\,,\, \;t\in{{\Bbb C}}\}
\end{equation}
where $z(t+\tau,\vec{x})\equiv z^t$ denotes the complexified form of the parametric representation (\ref{parahor}).
The curve  $h_{x_1,x_2 }$ is a complex hyperbola whose representation enjoys the following properties (easily checked from Eq. (\ref{parahor})):\\
i) ($2\pi R$)-periodicity along the imaginary axis of the $t$-plane (since  
 $z(t  ,\vec{x})= z(t+2i \pi  R,\vec{x}))$\\
ii) for each $t$ in the strip $\Sigma_+=\{t:\;0<\Im t <\pi R\}$ (resp.
 $\Sigma_-=\{t:\;\pi R<\Im t <2\pi R\}$) the corresponding point  
$z(t+\tau,\vec{x})$ belongs to the domain ${\cal T}^{+}$ (resp. ${\cal T}^{-}$) of
$X^{(c)}_{d}$;\\
iii) for each $t$ such that $\Im t =\pi R$  the point $z_2^t$ 
is real and such that $-z^t_2\in {\cal U}_{h(x_0)}$. For all such points one then checks that $(x_1-z_2)^2<0$;\\
iv) if $(x_1-x_2)^2<0$ then there exists  a  finite interval $I_{x_1,x_2}$ containing the origin in the real $t$-axis on which 
 $(x_1-x^t_2)^2$ remains negative.

From these geometrical properties and from the analyticity properties of 
${\rm W}(z_1,z_2)$, one then deduces that the restriction of W to the complex curve $h_{x_1,x_2 }$ defines a $2 i \pi R $-periodic analytic function of $t$, namely 
${\rm W}_{x_1,x_2 }(t)= {\rm W}(x_1,z^t_2)$
whose domain is the periodic cut-plane
\begin{equation}
{{\Bbb C}}^{cut}_{x_1,x_2 }=\{t\in {{\Bbb C}};\;\Im t \not= 2n\pi R,\; n\in 
{\Bbb Z}\}
\cup \{t;\; t-2 i n\pi R\in I_{x_1,x_2}, \; n\in {\Bbb Z}\}.\label{ccut}
\end{equation}
It must be noted that the {\em extended domain} $\Delta$ of ${\rm W}$ is 
used here, since the sets $X_d\times {\cal T}^\pm$ and ${\cal R}$ 
(whose traces in 
$h_{x_1,x_2 }$ yield respectively $\Sigma_+$, $\Sigma_-$ and 
${{\Bbb R}}+i{\pi R}$) belong to the {\em boundaries} of the primitive 
``tubes of normal analyticity'' ${\cal T}_{12}$ and ${\cal T}_{21}$.
However these sets are contained in  the extended domain $\Delta$.
Moreover it is easily checked that the boundary values 
${\cal W}^{\pm}_{x_1,x_2 }(t)= \lim_{\epsilon\to 0^{+}}{\rm W}_{x_1,x_2 }
(t\pm i\epsilon)$
of ${\rm W}_{x_1,x_2 }(t)$ on ${{\Bbb R}}$ 
(defined as distributions in the variable $t$, for each fixed values of 
${x_1,x_2 }$) represent the restriction to $h_{x_1,x_2} \cap (X_d\times X_d)$
 of the corresponding two-point Wightman distributions, namely
\begin{equation}
{\cal W}^{+}_{x_1,x_2 }(t)={\cal W}(x_1,{x^t_2 })\,,\;\;\;\;\;\;
{\cal W}^{-}_{x_1,x_2 }(t)={\cal W}({x^t_2 },x_1);
\end{equation}
therefore the $2i\pi R$ periodicity of the function ${\rm W}_{x_1,x_2 }(t)$ induced by the geometry of the complex hyperbola $h_{x_1,x_2 }$ implies in turn the following K.M.S.-type condition:
\begin{equation}
{\cal W}(x(t+\tau,\vec{\x}),{x_1 })=\lim_{\epsilon\to 0^{+}}{\rm W}(x_1,z(t
+\tau +2i\pi R - i\epsilon,\vec{\x})).
\label{kmcond}\end{equation}
Since all these properties of the function ${\rm W}_{x_1,x_2 }(t)$ hold for any choice of the point $(x_1,x_2)$  in ${\cal O}_1\times{\cal O}_2$, one easily 
obtains the corresponding result for the integral 
\begin{equation}
{\rm W}_{12}(t)=\int {\rm W}_{x_1,x_{2}}(t)f^{(1)}(x_1)f^{(2)}(x_2)d\sigma(x_1)d\sigma(x_2).
\end{equation} 
The latter is analytic and periodic in the cut plane
\begin{equation}
{{\Bbb C}}^{cut}({\cal O}_1,{\cal O}_2)= 
\bigcap_{x_1 \in {\cal O}_1 ,x_2\in {\cal O}_2 } {{\Bbb C}}^{cut}_{x_1,x_2}
\end{equation}
and its boundary values 
${\cal W}_{12}(t)$ and ${\cal W}'_{12}(t)$  satisfy the corresponding KMS relations:
\begin{equation}
{\cal W}_{12}(t)=\lim_{\epsilon\to 0^{+}}{\rm W}_{12}(t+i\epsilon),\;\;\;\;\;\;
{\cal W}'_{12}(t)=\lim_{\epsilon\to 0^{+}}{\rm W}_{12}(t+2i\pi R -i\epsilon).
\label{gms}\end{equation}
One notices that the cut-plane ${{\Bbb C}}^{cut}({\cal O}_1,{\cal O}_2 )$  is connected if 
${\cal O}_1$ and ${\cal O}_2 $ are space-like separated (since then 
$\cap_{{x}_{1}\in{\cal O}_{1},{x}_{2}\in{\cal O}_{2}}I_{x_1,x_2}$ contains the 
point $t=0$).
In the general case one considers correlation functions between arbitrary elements 
$\Phi(f^{(i)}_{0}, f^{(i)}_{1}, \ldots, f^{(i)}_{n}, \ldots)$ of
${\cal P}({\cal O}_{i})$ (i=1,2); this can easily be seen to reduce to the previous one-component case; in fact any general two-point function 
$ \langle\Omega, \Phi({\bf f}^{(1)})\Phi(\alpha^t_{h(x_0)}{\bf f}^{(2)})\Omega\rangle$
can always be decomposed (via the standard contraction formalism of Fock space)
as a polynomial of elementary two-point functions ${\cal W}(x_{i,1},x^t_{j,2})$ 
(integrated with the appropriate test functions of the variables $x_{i,1},x_{j,2}$).
Since the previously described analytic structure (including $2i\pi R$-periodicity  and KMS condition of the boundary values) in the $t$-plane is stable 
under multiplication and integration over the real parameters 
${{x}_{i,1}\in{\cal O}_{1},{x}_{j,2}\in{\cal O}_{2}}$, 
one concludes that there exists an analytic function ${\rm W}_{{\bf f}^{(1)}{\bf f}^{(2)}}(t)$, analytic and periodic in  ${{\Bbb C}}^{cut}({\cal O}_1,{\cal O}_2)$ and such that 
\begin{equation}
 \langle\Omega, \Phi({\bf f}^{(1)})\Phi(\alpha^t_{h(x_0)}{\bf f}^{(2)})\Omega\rangle = \lim_{\epsilon \to 0^+} 
{\rm W}_{{\bf f}^{(1)}{\bf f}^{(2)}}(t+i\epsilon)
\label{kmsvec1}
\end{equation}
\begin{equation}
 \langle\Omega, \Phi(\alpha^t_{h(x_0)}{\bf f}^{(2)})\Phi({\bf f}^{(1)})\Omega\rangle = \lim_{\epsilon \to 0^+} 
{\rm W}_{{\bf f}^{(1)}{\bf f}^{(2)}}(t + 2i\pi  R -i\epsilon).
\label{kmsvec2}
\end{equation}
   \\

It must be noted that in the course of the proof of this KMS property, the following additional property, relating by analytic continuation the region 
${\cal  U}_{h(x_0)}$ with the antipodal region 
${\cal  U}_{h(-x_0)} = \{x\in X_d, -x\in   {\cal  U}_{h(x_0)}\}=$
$\{x = (x^{(0)},\vec{\x},x^{(d)}) \in X_d, \check{x} = (-x^{(0)},\vec{\x},-x^{(d)}) \in   {\cal  U}_{h(x_0)}\}$
 has been obtained as a by-product (see the property iii) of $h_{x_1,x_2}$ in this proof).
\begin{proposition}
(antipodal condition)\\
Under the condition that all the test functions have their support contained in
${\cal  U}_{h(x_0)}$, the following identities hold 
\[{\rm W}_{12}(i\pi R)=\int {\rm W}_{x_1,x_{2}}(i\pi R)f^{(1)}(x_1)f^{(2)}(x_2)d\sigma(x_1)d\sigma(x_2)=\]
\begin{equation}
=\int {\cal W}({x_1,x_{2}})f^{(1)}(x_1)f^{(2)}(\check{x}_2)d\sigma(x_1)d\sigma(x_2)
\label{antipod}
\end{equation} 
and more generally $\forall t \in {\Bbb R}$  
\begin{equation}
{\rm W}_{{\bf f}^{(1)}{\bf f}^{(2)}}(t+i\pi R) =
\langle\Omega, \Phi({\bf f}^{(1)})\Phi(\alpha^t_{h(x_0)}{\bf \check{f}^{(2)}})\Omega\rangle 
\label{antipod2}
\end{equation} 
where the sequence $ {\bf \check{f}^{(2)}}=({f}^{(2)}_0, ({f}^{(2)}_1(\check{x}_1),
\ldots,{f}^{(2)}_n(\check{x}_1,\ldots,\check{x}_n),\ldots)$ is localized in the antipodal region ${\cal  U}_{h(-x_0)}$.
\label{prantipod}\end{proposition}

We shall now introduce an ``energy operator'' 
${\cal E}_{h(x_0)}$ associated with the geodesic 
${h(x_0)}$ by considering in ${\cal H}$ the continuous unitary representation 
$\{ U^t_{h(x_0)};\linebreak[0] t\in {{\Bbb R}}\}$ of the time-translation group 
$T_{h(x_0)}$ and its spectral resolution 
\begin{equation}
U^t_{h(x_0)} = \int_{-\infty}^\infty e^{i\omega t} dE_{h(x_0)}(\omega);
\label{unitario}\end{equation}
This defines (on a certain dense domain of ${\cal H}$) the unbounded operator
\begin{equation}
{\cal E}_{h(x_0)}=\int_{-\infty}^\infty \omega dE_{h(x_0)}(\omega).
\end{equation}
For any pair of one-component vector states 
$\Psi^{(1)}= \phi(\bar{f}^{(1)})\Omega$,
$\Psi^{(2)}= \phi({f}^{(2)})\Omega$, the corresponding
correlation function given in  (\ref{mamma})  can be written as follows:
\begin{equation}
{\cal W}_{12}(t) = 
\langle \phi(\bar{f}^{(1)})\Omega, U^t_{h(x_0)}
\phi({f}^{(2)})\Omega\rangle, 
\label{ruota}
\end{equation}
which shows that ${\cal W}_{12}(t)$ is a continuous and bounded function.
In view of Eq. (\ref{unitario}) it can be expressed as the Fourier transform of the bounded measure 
\begin{equation}
\langle \phi(\bar{f}^{(1)})\Omega,dE_{h(x_0)}(\omega) \phi({f}^{(2)})\Omega
\rangle = \tilde{\cal W}_{12}(\omega).\label{tutor}
\end{equation}
We also obtain  similarly that ${\cal W}'_{12}(t)$ is the Fourier
transform of 
\begin{equation}
\langle \phi(\bar{f}^{(2)})\Omega,dE_{h(x_0)}(-\omega) \phi({f}^{(1)})\Omega
\rangle = \tilde{\cal W}'_{12}(\omega).\label{dicendo}
\end{equation}
Eqs. (\ref{tutor}) and (\ref{dicendo}) are valid for arbitrary test-functions 
${f}^{(1)}$ and ${f}^{(2)}$ in ${\cal D}(X_d)$. Now, if ${f}^{(1)}$ and ${f}^{(2)}$
belong to  ${\cal D}({\cal U}_{h(x_0)})$, the functions 
${\cal W}_{12}(t)$ and ${\cal W}'_{12}(t)$ satisfy the KMS relations (\ref{gms}),
and their Fourier transforms satisfy (as bounded measures) the following relation,
which is equivalent to (\ref{gms}):
\begin{equation}
\tilde{\cal W}'_{12}(\omega)= e^{-2\pi R \omega} \tilde{\cal W}_{12}(\omega).
\label{byby}\end{equation}
Since the KMS relations hold similarly for the correlation functions 
between arbitrary vector
states  
$\Psi^{(1)}= \Phi(\bar{\bf f}^{(1)})\Omega$,
$\Psi^{(2)}= \Phi({\bf f}^{(2)})\Omega$
of D (see the end of the proof of  proposition 2.3.) we could reproduce completely 
the previous argument and computations in the general case and therefore obtain
the following

\begin{proposition}
(geodesic spectral condition) \\
For every pair of  states 
$\Psi^{(1)}= \Phi(\bar{\bf f}^{(1)})\Omega$, 
$\Psi^{(2)}= \Phi({\bf f}^{(2)})\Omega$
in ${\cal P}({\cal  U}_{h(x_0)})\Omega $, the corresponding matrix elements of the spectral measure
$dE_{h(x_0)}(\omega)$ satisfy the following relation 
\begin{equation}
 \langle \Phi(\bar{\bf f}^{(2)})\Omega,dE_{h(x_0)}(-\omega) \phi({\bf f}^{(1)})\Omega
\rangle=e^{-2\pi R \omega}
 \langle \Phi(\bar{\bf f}^{(1)})\Omega,dE_{h(x_0)}(\omega) \phi({\bf f}^{(2)})\Omega
\rangle\label{tyty}
\end{equation}
\label{pr4}\end{proposition}

\begin{remark}{\em     
This geodesic spectral condition gives a precise content to the statement that 
in the region ${\cal U}_{h(x_0)}$ corresponding to  an observer living 
on the geodesic ${h(x_0)}$, the  energy measurements (relative to this observer)
give exponentially damped expectation values in the range of negative energies. 
In the limit of flat space-time  the l.h.s. of Eq. (\ref{tyty}) would be  
equal to zero for $\omega >0$, which corresponds to recovering the usual 
spectral condition of ``positivity of the energy''.
}\end{remark}  
\begin{remark}{\em     
 A  statement similar to proposition \ref{pr4} making use of the same unitary group
$\{{  U}^t_{h(x_0)}\}$ would hold for vector states 
$\Psi^{(1)} $
$\Psi^{(2)} $ in ${\cal P}({\cal  U}^t_{h(-x_0)})\Omega$, i.e. states localized in the antipodal  region of ${\cal  U}^t_{h(x_0)}$.
However, as far as the corresponding correlation functions are concerned,
the KMS analyticity strip would be replaced by $\{t -2\pi R < \Im t < 0\}$
and Eqs. (\ref{gms}) by the following ones:
\begin{equation}
{\cal W}_{12}(t)=\lim_{\epsilon\to 0^{+}}{\rm W}(t-i\epsilon),\;\;\;\;\;\;
{\cal W}'_{12}(t)=\lim_{\epsilon\to 0^{+}}{\rm W}(t-2i\pi R +i\epsilon).
\label{gmsy}
\end{equation}
whose equivalent form for the Fourier transforms is 
\begin{equation}
\tilde{\cal W}'_{12}(\omega)= e^{2\pi R \omega} \tilde{\cal W}_{12}(\omega).
\end{equation}
In writing a geodesic spectral condition of the form (\ref{tyty}) for such states in 
${\cal P}({\cal  U}_{h(-x_0)})\Omega$ the factor $e^{-2\pi R \omega}$ (at the r.h.s.)
should therefore be replaced by $e^{2\pi R \omega}$,  which would correspond
to an exponential
damping of $\langle dE_{h(x_0)}(\omega)\rangle$ in the range of positive 
$\omega$'s. It must however be remarked that 
for an observer sitting on the geodesic $h(-x_0)$ (antipodal to $h(x_0)$) 
the natural time and energy variables should be respectively equal to $-t$ and 
$-\omega$, since the corresponding time-translation group $T_{h(-x_0)}$ (obtained 
from $T_{h(x_0)}$ e.g. by a conjugation of the form 
$T_{h(-x_0)}= rT_{h(x_0)}r^{-1}$, $r$ being a rotation of angle 
$\pi$ in a plane orthogonal
to the $x^{(0)}$-axis) is such that 
$T_{h(-x_0)}(t)=T_{h(x_0)}(-t)$. 
}\end{remark}  
We can also reexpress the antipodal condition of proposition \ref{prantipod} under the following equivalent form, which asserts that the spectral measure 
$ dE_{h(x_0)}$ has exponentially damped matrix elements between states localized in the mutually antipodal regions ${\cal  U}_{h(x_0)}$ and 
${\cal  U}_{h(-x_0)}$:
\begin{proposition}
(antipodal spectral condition).\\
The following relation holds
for every pair of vector states 
$\Psi^{(1)}= \Phi(\bar{\bf f}^{(1)})\Omega$, 
$\Psi^{(2)}= \Phi({\bf f}^{(2)})\Omega$ $\in$  
${\cal P}({\cal  U}_{h(x_0)})\Omega$ \begin{equation}
 \langle \Phi(\bar{\bf f}^{(1)})\Omega,dE_{h(x_0)}(\omega) \Phi({\bf \check{f}}^{(2)})\Omega
\rangle=e^{-\pi R \omega}
 \langle \Phi(\bar{\bf f}^{(1)})\Omega,dE_{h(x_0)}(\omega) \Phi({\bf f}^{(2)})\Omega
\rangle, \label{tyty7}
\end{equation}
$\Phi({\bf \check{f}}^{(2)})\Omega$ being in 
${\cal P}({\cal  U}_{h(-x_0)})\Omega$.
\label{prantipod2}
\end{proposition}
 {\bf Proof.} The proof is obtained by writing that the Fourier transform of 
${\rm W}_{{\bf f}^{(1)}{\bf f}^{(2)}}(t+i\pi R) $ (given by Eq. (\ref{antipod2})) is equal to the Fourier transform of 
$\lim_{\epsilon \to 0}{\rm W}_{{\bf f}^{(1)}{\bf f}^{(2)}}(t+i\epsilon) $  (given by Eq. (\ref{kmsvec1})) multiplied by 
$\exp(-\pi R \omega)$.\hfill    \\  
\begin{proposition}
A sufficient condition for a generalized free field satisfying the postulates a),b),c) to enjoy the full analytic structure entailed by the normal analyticity postulate d) is that it satisfy either \\
e) the geometrical KMS condition together with the antipodal condition,\\
or\\
e') the geodesic spectral condition together 
with the antipodal spectral condition.
\label{prpostulates}
\end{proposition}
 {\bf Proof.} It is sufficient to show that either Eqs. (\ref{gms}) or Eq. (\ref{byby}) expressing equivalently the KMS condition or the geodesic spectral condition for arbitrary one-component vector states 
$\phi(\bar{f}^{(1)})\Omega$ and $\phi({f}^{(2)})\Omega$ in 
${\cal P}({\cal U}_{h(x_0)})\Omega$ imply the analyticity properties of ${\rm W}(z_1,z_2)$ described in \S 2.3.

According to the analysis given in the proof of proposition \ref{kms} Eqs. (\ref{gms}) exactly express the fact that on the open set 
${\cal U}_{h(x_0)}\times{\cal U}_{h(x_0)}$ the distributions 
 ${\cal W}(x_1,x_2)$ and  ${\cal W}'(x_1,x_2)$= ${\cal W}(x_2,x_1)$ are 
 the boundary values of an analytic function of {\em one} complex variable (depending on 
$2d-1$ real variables) 
${\rm W}(x_1,z^t_2)$  defined in the union of the domains of the complex hyperbolae $h_{x_1,x_2}$ (with  $x_1, x_2$ varying in ${\cal U}_{h(x_0)}$) represented by $
{{\Bbb C}}^{cut}_{x_1,x_2 }$ (see Eq. (\ref{ccut})). It is then clear that this information is sufficient (and even redundant) for constructing an invariant perikernel ${\rm W}(z_1,z_2)$ through its reduced form w($\alpha$) (see \S 2.3).
In fact by considering the special hyperbola  $h_{x_0,x_0}$ (i.e. 
$x_1=x_0$, $z_2(t)=(R\sinh\frac{t}{R},\vec{0}, R\cosh\frac{t}{R})$ we obtain the following identification
\begin{equation}
{\rm w}\left(\cosh\frac{t}{R}\right)= {\rm W}(x_1,z^t_2)|_{x_1=x_2=x_0}
\label{a255}
\end{equation}
which defines the analytic function  w($\alpha$) in the cut-plane ${{\Bbb C}}\setminus [1,\infty[.$ For arbitrary points  $x_1, x_2$  in ${\cal U}_{h(x_0)}$
represented by the parametrization (\ref{parahor}) one would obtain more generally 
\begin{equation}
{\rm w}\left(\frac{\sqrt{R^2-\vec{\x}^2_1}}{R}\frac{\sqrt{R^2-\vec{\x}^2_2}}{R} \cosh\frac{\tau_2-\tau_1+t}{R}+ \frac{\vec{\x}_1\cdot\vec{\x}_2}{R^2}\right)= {\rm W}(x_1,z^t_2), 
\label{a256}
\end{equation}
the consistency of this set of relations being ensured by the $G_d$-covariance 
of the boundary values of ${\rm W}(x_1,z^t_2)$ (for $\Im t\to 0^{+}$), namely 
${\cal W}(x_1,x^t_2)$=${\rm w}_{\epsilon}\left(-\frac{{x_1}\cdot{x^t_2}}{R^2}\right)$ with the specification $\epsilon=\makebox{sgn} (x^{(0)}_2 -x^{(0)}_1)$. The reduced form w($\alpha$) thus obtained defines a unique invariant perikernel
 ${\rm W}(z_1,z_2)$=${\rm w}\left(-\frac{{z_1}\cdot{z_2}}{R^2}\right)$, whose boundary values from the normal analyticity domains coincide with the given distributions  ${\cal W}(x_1,x_2)$ and  ${\cal W}'(x_1,x_2)$ on 
${\cal U}_{h(x_0)}\times{\cal U}_{h(x_0)}$ (by construction, in view of the proof of proposition \ref{kms}). The fact that the coincidence relations extend to the whole space $X_d\times X_d$ necessitates the use of the antipodal condition (or of the equivalent antipodal spectral condition) since the latter allows one to identify the (coinciding) boundary values of  ${\rm W}(z_1,z_2)$
on ${\cal U}_{h(x_0)}\times{\cal U}_{h(-x_0)}$ 
with the given distribution  ${\cal W}(x_1,x_2)$ = ${\cal W}'(x_1,x_2)$ in that region. In fact one  notes that $(x_1-x_2)^2<0$
if $x_1 \in {\cal U}_{h(x_0)}$ $x_2 \in {\cal U}_{h(-x_0)}$ while 
$(x_1-x_2)^2>-2R^2$
if $x_1 \in {\cal U}_{h(x_0)}$ $x_2 \in {\cal U}_{h(x_0)}$.
Therefore,  the right identification of the boundary values of W on 
${\cal U}_{h(x_0)}\times{\cal U}_{h(x_0)}$ {\em and }
${\cal U}_{h(x_0)}\times{\cal U}_{h(-x_0)}$ determines by $G_d$-covariance its right identification on the whole space $X_d\times X_d$, since 
 these two subsets fix the right boundary values of the reduced form w($\alpha$) of W on the whole real $\alpha$-axis (from both half-planes $\Im \alpha >0$ and 
$\Im \alpha <0$ and with the coincidence region $\{\alpha <1\}$), 
i.e. on the image in $\alpha=-\frac{{x_1}\cdot{x_2}}{R^2}$ 
of the whole product  $X_d\times X_d$.  
   \\  
\begin{remark}{\em      The fact that the perikernel ${\rm W}(z_1,z_2)$
is entirely determined by the knowledge of ${\cal W}(x_1,x_2)$ and ${\cal W}'(x_1,x_2)$ in the region  ${\cal U}_{h(x_0)}\times{\cal U}_{h(x_0)}$ (the antipodal condition appearing as a self-consistency condition) suggests the following alternative presentation of proposition \ref{prpostulates}.}\end{remark}  
\begin{proposition}
Let us consider a generalized free field theory whose Wightman distributions are given {\em only} on the classically accessible  region  ${\cal U}_{h(x_0)}$ of an observer living on the geodesic ${h(x_0)}$ of $X_d$ and satisfy the postulates a),b),c) (restricted to ${\cal U}_{h(x_0)}$) together with the corresponding geometrical KMS condition (or equivalently the geodesical spectral condition).

Then this theory can be extended in a unique way to a generalized free field theory on the whole de Sitter space-time which enjoys the full analytic structure entailed by the postulate d), and satisfies (as a by-product) the antipodal condition.
\end{proposition}
\begin{remark}{\em     
In this connection we mention  the interesting 
approach to two-dimensional constructive de Sitter
QFT's given in \cite{[FHN]}, where Wightman functions are introduced
(with space-time cutoffs)
only in the region  ${\cal U}_{h(x_0)}$. The previous statements clarify the situation concerning the extension of those theories satisfying suitable properties to the whole de Sitter space-time. 
In particular, in the  treatment of Klein-Gordon  fields   
of \cite{[FHN]} the use of space cut-offs which breaks the de Sitter 
invariance 
leaves somewhat implicit the $G^d$-invariant limiting theory.
Our results show that these limiting theories fit within the whole analytic
structure described in this section, and can therefore be identified
with the Klein-Gordon fields of \S 4.2.
}\end{remark}  

\begin{remark}{\em     
All the features that have been discussed in this section are also naturally interpreted  in terms of the existence of an antiunitary involution $J$ 
relating the algebras ${\cal P}({\cal U}_{h(x_0)})$ and  ${\cal P}({\cal U}_{h(-x_0)})$ and the validity of the corresponding Bisognano-Wichmann duality theorem for the Von Neumann algebras ${\cal A}({\cal U}_{h(x_0)})$ and 
 ${\cal A}({\cal U}_{-h(x_0)})$ \cite{[Ara],[BW]}.
We will not further dwell on these facts here. 
A complete discussion of this class of results will be given in a 
following paper \cite{[BM]}
in which we will deal with the general properties of (free or interacting) de Sitter quantum field theories.
}\end{remark}  
\section{
Feynman-type algebra, double analytic structure, and complex 
mass representation of the de Sitter propagators}

\subsection{ De Sitterian propagators: Feynman-type algebra and a
substitute to the ``Wick-rotation''}

The previously described analytic structure of two-point functions in 
complex de
Sitter space-time (see \S  2.3) is the exact analogue of the
standard analytic structure of general two-point functions in complex
Minkowski space-time implied by the Wightman axioms. As in the latter
case, the complex-space time function W$(z_1,z_2)$ presents two
important aspects in agreement with the common wisdom of QFT, namely the
euclidean propagator S (here a kernel on the sphere $S_d$) and the
retarded propagator ${\cal R}$ 
(called in \cite{[BV-1]} a ``Volterra'' or
``causal kernel'' on
$X_d$). Any general two-point function in the
complexified de Sitter space-time is thus characterized in a suggestive
way as a triplet (W,S,${\cal R}$).  
The function ${\rm W}(z_1,z_2)$ (or ``perikernel'') is analytic in the cut domain $\Delta = X^{(c)}_d \times  X^{(c)}_d \setminus \Sigma^{(c)}$ (see Proposition 2.2); ${\rm S}(z_1,z_2)$ is the restriction of W to $S_d \times S_d$
extended to the set of  coinciding points $(z_1=z_2)$; ${\cal R}(x_1,x_2)$ is a 
kernel on $X_d$ with support  $\{(x_1, x_2)\in X_d\times X_d; \;x_1\geq
x_2\}$ obtained by taking the discontinuity of W across the real trace 
$\Sigma = \{(x_1, x_2)\in X_d\times X_d; \;(x_1-x_2)^2 \geq 0\} $ of 
$\Sigma^{(c)}$ and restricting the latter to the future component 
$(x^{(0)}_1\geq x^{(0)}_2)$.

The algebraic properties of the class of triplets ({\rm W},{\rm S},${\cal R}$) which have
been derived in particular in \cite{[BV-1]} (see also \cite{[Br]}) can
then immediately be interpreted in the usual language of Feynman
diagrams and provide in fact a substitute to the standard Wick-rotation
procedure when one deals with de Sitterian propagators. We shall now
describe these results. 
\begin{figure}
\epsfxsize=12.cm{\centerline{\epsfbox{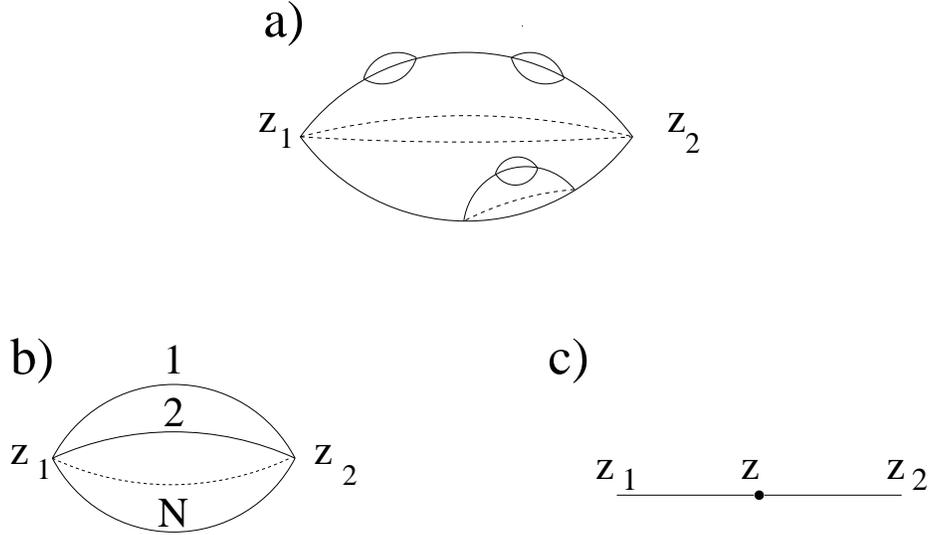}}}
\caption{Self-energy diagrams.}
\label{fig1}
\end{figure}

We shall consider a class of self-energy diagrams 
$\Gamma$
whose vertices are labelled by de Sitter space-time
variables $z_i$ and whose lines $(i,j)$ represent general two-point
functions 
${\rm W}^{\alpha}(z_i,z_j)$
(namely also the associated triplets 
(${\rm W}^{\alpha}$,
${\rm S}^{\alpha}$,
${\cal R}^{\alpha}$));
this class is the one 
generated by the two operations described below (bubble insertion and
vertex convolution, 
see Fig.\ref{fig1}a). 
Our results, obtained by an iterated application of the properties
listed below, can be stated as follows:
with every diagram $\Gamma$ of this class,
it is possible to associate 
a  two-point function 
${\rm W}_{\Gamma}$
namely a triplet (${\rm W}_{\Gamma}$,
${\rm S}_{\Gamma}$,
${\cal R}_{\Gamma}$) 
whose euclidean propagator 
${\rm S}_{\Gamma}$,
is obtained
in terms of the various euclidean propagators 
${\rm S}^{\alpha}$ involved in the diagram by the same integration rule
as in the minkowskian space (the sphere $S_d$ playing now the role of
the euclidean space).\\
i) {\em Bubble insertion}: with any diagram 
$\Gamma_N$ (represented on Fig.\ref{fig1}b) is associated the two-point function
${\rm W}_{\Gamma_N}(z_1,z_2)=\prod_{i=1}^{N} {\rm W}^{\alpha_i}(z_1,z_2)$
whose analyticity domain is $\Delta$ and whose corresponding triplet  belong to the
considered class.\\
ii) {\em Vertex convolution}: this operation, represented by the
elementary diagram of Fig \ref{fig1}c), corresponds to the
$*^{(c)}$-composition product of two perikernels introduced in
\cite{[BV-1]}. 
Let 
(${\rm W}^{\alpha_i}$
${\rm S}^{\alpha_i}$,
${\cal R}^{\alpha_i}$),
$i=1,2$, be the triplets associated with two given
two-point functions 
${\rm W}^{\alpha_1}$,
${\rm W}^{\alpha_2}$; there exists a two-point function W, denoted by 
${\rm W}= 
{\rm W}^{\alpha_1}*^{(c)}
{\rm W}^{\alpha_2}$,
whose associated triplet 
(${\rm W}$
${\rm S}$,
${\cal R}$),
enjoys the following properties:
\begin{enumerate}
\item ${\rm S} = {\rm S}^{\alpha_1}*
{\rm S}^{\alpha_2}$
is the convolution of euclidean propagators, defined as the standard
composition of kernels on the sphere $S_d$.
\begin{equation}
{\rm S}(z_1,z_2) = \int_{S_d}
{\rm S}^{\alpha_1}(z_1,z)
{\rm S}^{\alpha_2}(z,z_2)d\sigma(z)
\label{cp1}
\end{equation}
\item ${\cal R} = {\cal R}^{\alpha_1}\Diamond  
{\cal R}^{\alpha_2}$
is the convolution of retarded propagators, introduced in
\cite{[FV],[BV-2]} as the composition of Volterra kernels on  
$X_d$ by the following formula:  
\begin{equation}
{\cal R}(x_1,x_2) = \int_{\Diamond(x_1,x_2)}
{\cal R}^{\alpha_1}(x_1,x)
{\cal R}^{\alpha_2}(x,x_2)
\label{cp2}
\end{equation}
where $
{\Diamond(x_1,x_2)} = \{x\in X_d;\, x_1\geq x \geq x_2\}$
  is the ``double cone'' with vertices $x_1,x_2$ and 
$d\sigma(x)$ is the volume form on $X_d$ defined by Eq. (\ref{leray}).\\

\item  For $(z_1,z_2)$ varying in (a dense subdomain of) $\Delta$, there
exists a compact integration cycle
$\gamma(z_1,z_2)$,
obtained from $S_d$ by continuous
distortion, such that 
\begin{equation}
{\rm W}(z_1,z_2) = \int_{\gamma(z_1,z_2)}
{\rm W}^{\alpha_1}(z_1,z)
{\rm W}^{\alpha_2}(z,z_2)d\sigma(z).
\label{cp3}
\end{equation}
\end{enumerate}
In Eq. (\ref{cp1}) and (\ref{cp3})
$d\sigma(z)$
denotes the complexified form of 
$d\sigma(x)$, i.e. a $G^{(c)}$-invariant $d$-form on $X^{(c)}_d$.

The following comment can be added concerning this result. The
geometrical argument of contour distortion (given in its complete form
in 
\cite{[BV-1]} and sketched  in \cite{[Br]}) is of a
different nature from the standard Wick-rotation argument of the flat
case; the fact that only  {\em compact} cycles
are used in the present case (namely in  Eqs.
(\ref{cp1})-(\ref{cp3})) indicates that the chronological propagators
cannot be involved in the argument, the Wick rotation being now
unapplicable between the non-compact real space-time $X_d$ and the compact
euclidean sphere $S_d$.
On the contrary, the connection between the euclidean and the retarded
convolutions $*$ and $\Diamond$ relies on a  more general distortion 
procedure,
which remains also valid in the flat minkowskian limit (up to the
non-compactness of the euclidean integration in that case). 
\begin{remark}{\em     
The analyticity domains $\Delta_\mu$ of the perikernels considered in 
\cite{[BV-1]} involved the cut 
$\Sigma^{(c)}_{\mu}= \{(z_1,z_2);\;(z_1-z_2)^2\geq 2R^2(\cosh\mu - 1)
\}$ with $\mu>0$, thus giving a weaker version of the previous results
since $\Delta_0=\Delta\subset\Delta_\mu $.
The case $\mu=0$ which is relevant for us necessitates some
distribution-theoretical refinement due to the singularities at
coinciding points but can be treated along the lines indicated in \cite{[Br]};
these technical aspects do not spoil the purely geometrical essence of the
argument.
}\end{remark}

\subsection{The Laplace transform of the retarded propagators}

The causality and de Sitter covariance properties of the retarded
propagators ${\cal R}(x_1,x_2)$ (see \S  2.3)   allow the
introduction of a Laplace-type transform of the latter enjoying 	
analyticity 
properties with respect to the relevant conjugate variable $\nu$, proportional to a complex mass.
We shall obtain in this way an   analogue of the analyticity domain
of the Minkowskian propagators in the mass variable; however, as in that case it is necessary to assume that the propagators 
satisfy  suitable growth conditions  at infinity on $X_d$ (i.e. on their support $\{(x_1,x_2)\in X_d\times X_d;\;x_1\geq x_2\})$, which will 
be specified below. 

The retarded propagators are invariant Volterra 
kernels to which the methods of \cite{[BV-2],[Fa],[FV]} 
can be applied. According to the latter, the 
Laplace transform ${\rm G}(\nu)$ of ${\cal R}(x_1,x_2)$  can  be
defined by the following formula\footnote{ where ${\rm G}(\nu) = \tilde{F}(-\frac{d-1}{2} - i\nu)$, 
$\tilde{F}(\lambda)$ being the Laplace transform of $f(v) = r(\cosh v) $ introduced in \cite{[BV-2]}.} :
\begin{equation}
{\rm G}(\nu) = \omega_{d} 
\int_{0}^{\infty}r(\cosh v){Q}^{(d+1)}_{-\frac{d-1}{2}-i\nu}(\cosh v) (\sinh v)^{d-1} dv \label{34}
\end{equation}
where
$\cosh v = -\frac{x_1\cdot x_2}{R^2}$ and 
$\omega_d =2\pi^{\frac{d}{2}}/\Gamma(d/2)$;
 $r\left(-\frac{x_1\cdot x_2}{R^2}\right) = {\cal R}(x_1,x_2)$
 has its support contained in $[1,\infty[$.
 $\;{Q}^{(d+1)}_{-\frac{d-1}{2}-i\nu}$ denotes a generalized Legendre function of the second kind \cite{[BV-2]},   (proportional to 
a Gegenbauer function of the second kind \cite{[ER]} admitting  the
following  representation: 
\vskip5pt
a) for $d$ even:
\[
{Q}^{(d+1)}_{-\frac{d-1}{2}-i\nu}(\cosh v )  = 
\]
\begin{equation}
\frac{2(-1)^{\frac{d}{2}}
i\nu \omega_{d+1}}{(2\pi)^{\frac{d}{2}+1}{h_{d+1}(-\frac{d-1}{2}-i\nu)}}
\left(\frac{d}{d\cosh v}\right)^{\frac{d}{2}-1}
\int_v^\infty e^{i\nu w}
[2(\cosh w -\cosh v)]^{-\frac{1}{2}}dw
\label{gege2}
\end{equation}
b) for $d$ odd:
\begin{equation}
-\frac{\omega_{d+1}}{(2\pi)^{\frac{d}{2}}h_{d+1}(-\frac{d-1}{2}-i\nu)}
\left(-\frac{d}{d\cosh v}\right)^{\frac{d-1}{2}}
  e^{i\nu v}
\label{gege3}
\end{equation}
In these expressions $h_{d+1}$ denotes the following polynomial:
\begin{equation} 
h_{d+1}(\lambda)= \frac{2\lambda + d - 1}{(d-1)!}\frac{\Gamma(\lambda + d
-1)}{\Gamma(\lambda + 1)}.
\label{gege4}
\end{equation}
 Eq. (\ref{gege2}) (resp.  (\ref{gege3}))  
defines a  function $\nu \to h_{d+1}\left(-\frac{d-1}{2}-i\nu\right){Q}^{(d+1)}_{-\frac{d-1}{2}-i\nu}(\cosh v)$ analytic in the half plane 
$\Im \nu > -\frac{1}{2}$ (resp. the whole plane). 
It can then be shown that, if the behaviour of $r(\cosh v)$ at infinity
is governed by $e^{mv}$ 
 then the product $h_{d+1}\left(-\frac{d-1}{2}-i\nu\right){\rm G}(\nu)$ is analytic in the half-plane
$\{\nu:\,\Im \nu > m + \frac{d-1}{2}\}$ and admits a continuous boundary
value on the line $\{ \Im \nu = m + \frac{d-1}{2}\}$.
\vskip5pt
 {\em Geometrical interpretation}
\vskip5pt
As a matter of fact, the expression (\ref{34}) has been obtained {a posteriori} in \cite{[BV-2]}; at first, the Laplace transformation was introduced  by a more
geometrical method (initiated in \cite{[Fa]}) which exhibits the
function ${\rm G}(\nu)$ as the one-dimensional Laplace transform of  a Radon-type transform $\hat{r}(t)$ of
$\underline{\cal R}(x)={\cal R}(x,x_0)=r(x^{(d)}/R)$. 
In this transformation,  the integration is taken on the following family of plane sections of
$X_d(R)$ called ``horocycles'':
$h_t = \{x\in X_d(R);\, x^{(0)} + x^{(d)} = R e^{\frac{t}{R}}\}$.
 Horocycles are  parametrized in the following way:
\begin{equation}
x=x_R(t,\vec{\rm x}) \to \left\{
\begin{array}{ccl}
x^{(0)} & = & R\sinh \frac{t}{R} + \frac{1}{2R}\vec{\rm x}^{2}  e^{\frac{t}{R}}\\ {\bf x}= (x^{(1)}\ldots,x^{(d-1)}) & = & \vec{\rm x}  e^{\frac{t}{R}}\\
x^{(d)} & = &R \cosh \frac{t}{R} - \frac{1}{2R}\vec{\rm x}^{2}  e^{\frac{t}{R}}\\
\end{array}
\right.
\label{tytyty}\end{equation}

For each $t$ real, $h_t$ is a paraboloid ( parametrized by the vector
$\underline{\vec{x}}$) 
whose intersection with the support  of $\underline{\cal R}(x)$,
namely the ``cone'' $\Gamma^{+}(x_0)$, is either empty (if $t<0$)
or equal to the ball $B_t$ represented by 
$B_t=\{\underline{\vec{x}} \in {{\Bbb R}}^{d-1};\,
|\underline{\vec{x}}|\leq R(1-e^{-\frac{t}{R})}$.
For $t\geq 0$ one then defines:
\begin{equation}
\hat{r}(t) = 2 (-1)^{d-1}e^{-(d-1)\frac{t}{R}}\int_{h_t}\underline{\cal R}(x)
d\sigma(x) = 
\int_{B_t}\underline{\cal R}(x_t(\vec{\rm x}))d \vec{\rm x}
\label{Lapla}. \end{equation}

As proven in \cite{[BV-2]} (Eq. (3.20), Corollary of Proposition
13 and Eq. (4.4)) there holds the following 
\begin{proposition}
i) $ \hat{r}(t) $ can be expressed by the following Abel-type tranformation
\begin{equation}
 \hat{r}(t) = \omega_{d-1} 
e^{-(d-1)\frac{t}{2R}}
\int_0^\frac{t}{R} r(\cosh v)[2(\cosh ({t}/{R}) -\cosh v)]^{\frac{d-3}{2}} \
\sinh v dv. \label{llap1}
\end{equation}
\noindent ii) If $e^{-mv}r(\cosh v) \in
{L^1({{\Bbb R}}^+,dv)}$ with $m>-1$
 then $e^{-mt/R}\hat{r}(t) \in {L^1}$.\vskip5pt

\noindent iii) under the assumption of ii) the transform ${\rm G}(\nu)$ of $r$ is defined in the half-plane $\{\nu; \, \Im \nu > m +\frac{d-1}{2}\}$    
as the following one-dimensional Laplace transform of $\hat{r}$:
\begin{equation}
{\rm G}(\nu) = \frac{1}{R} 
\int_0^{\infty} 
e^{(\frac{d-1}{2} +i \nu)\frac{t}{R}}\hat{r}(t) dt
\label{llap2}
\end{equation}
\end{proposition}
 
The similarity of the previous geometrical definition of ${\rm G}(\nu)$ with that of 
the  Laplace transform of the 
minkowskian retarded propagator as a function of the complex mass variable can be exhibited as follows. If 
${\cal R}_M({\rm x}_1, {\rm x}_2) =  r_M(( {\rm x}_1-  {\rm x}_2)^2)$ denotes the Minkowski 
propagator in dimension $d$ (with $ {\rm x}_i = ( {\rm x}^{(0)}_i,\vec{x}_i)\in {{\Bbb R}}^d$,
$ {\rm x}^2_i = { {\rm x}^{(0)}_i}^2-{\vec{x}_i}^2, i=1,2$), 
its Laplace transform can be 
written as a function of the mass variable as follows:
\begin{equation}
{\rm G}_M(\kappa) 
= \int_{\overline{V}^+}e^{i\kappa  {\rm x}^{(0)}} r_M( {\rm x}^2)d {\rm x}^{(0)}d\vec{x}
\label{lap1}
\end{equation}
or, by introducing the family of space-like hyperplanes 
$h^{(M)}_t=\{ {\rm x}\in {{\Bbb R}}^d,  {\rm x}^{(0)}=t\}$:
\begin{equation}
{\rm G}_M(\kappa) = \int_{0}^{\infty}e^{i\kappa t} \hat{r}_M(t) dt,
\label{lap2}
\end{equation}
where $\hat{r}_M(t)$ denotes the following Radon transform of ${r}_M( {\rm x}^{2})$:
\begin{equation}
\hat{r}_M(t)=\int_{\{ {\rm x}\in  h^{(M)}_t; |\vec{x}|\leq t\}} r_M( {\rm x}^2) d\vec{x} =
\frac{\omega_{d-1}}{2}\int_{0}^{t^2}r_M(v^2)(t^2-v^2)^{\frac{d-3}{2}}dv^2.
\label{rad1}   
\end{equation}
Formulae (\ref{lap2}) and (\ref{rad1}) 
are the respective analogues of  
(\ref{llap2}) and 
(\ref{Lapla}),(\ref{llap1}),  
and they define ${\rm G}_M(\kappa)$ (under the usual temperateness assumption 
on ${\cal R}_M$) as an analytic function in the half plane $\Im \kappa > 0$.
 Moreover, 
${\rm G}_M$ is expressed in terms of $r$ by the following integral transformation, (obtained by inserting the r.h.s. of Eq. (\ref{rad1}) in the integral of Eq. (\ref{lap2}) and inverting the order of the integrations):
\begin{equation}
{\rm G}_M(\kappa) = \omega_{d-1}\int_{0}^{\infty}r_M(v^2)
H_d(\kappa v)v^{d-1}dv.
\label{lap3}
\end{equation}
In this formula, which is the analogous of Eq. (\ref{34}), the kernel 
$H_d(\kappa v)$ is a Hankel-type function 
defined for $\Im \alpha > 0$ by the integral representation
\begin{equation}
H_d(\alpha) = \int_{1}^{\infty}(y^2-1)^{\frac{d-3}{2}}e^{i\alpha y}dy = 
\frac{\Gamma(\frac{d-1}{2})}{\Gamma(\frac{1}{2})}(-i\alpha)^{-\frac{d-2}{2}}
K_{\frac{d-2}{2}}(-i\alpha),
\label{hank}
\end{equation}
$K_{\frac{d-2}{2}}(-i\alpha)$ being the function introduced in \S 6.15 of  \cite{[Wat]}

\subsection{The euclidean propagators in the mass variable. The FG property}

We shall now state the de Sitterian analogue of the following well-known 
property of the minkowskian propagator in momentum-space: the restriction of 
the latter to imaginary values of the masses is the Fourier transform of the 
euclidean propagator $ {\rm S}(z_1,z_2) = {\underline{\rm S}}(z)$, 
with $z=z_1-z_2 = (iy^{(0)}, \vec{x})$, namely
\begin{equation}
{\rm G}_M(i\kappa) = 
\int^{ }_{{ {\Bbb R}}^d} {\rm e}^{-i\kappa \ y^{(0)}}{\underline{\rm S}} \left(y^{(0)},\vec x
\right) d y^{(0)} d\vec x
\end{equation}
Since in the de Sitter case, $ {\underline{\rm S}} 
(y^{(0)},\vec x) $ is defined on
the euclidean sphere $ S_d $,one can only associate with it a {\it discrete} Fourier transform, namely the 
sequence of coefficients $ \left\{ g_n;\ n\in \Bbb N \right\} $ of its
Fourier-Legendre expansion:
\begin{equation}
{\underline{\rm S}} \left(y^{(0)},\vec x \right) = 
\left. {\rm s} \left(\frac{x^{(d)}}{R} \right) 
\right\vert_{ x^{(d)}=R\ \cos\theta}  = \frac{1}{\omega_{ d+1}} \sum^{ }_{ }
g_nh_{d+1}(n)P^{(d+1)}_n( \cos\theta) ,
\label{a318}\end{equation}
given by:
\begin{equation}
g_n = \omega_ d \int^\pi_0 s( \cos\theta)  P^{(d+1)}_n( \cos 
\theta)( \sin  \theta)^{ d-1} d \theta.\label{huhuhu}
\end{equation}
\begin{proposition}  Under the assumptions   of proposition 3.1 the 
Laplace transform $ {\rm G}(\nu) $ of the retarded propagator $ r( \cosh  v) $
and the Fourier-Legendre coefficients $g_n$ of the euclidean propagator $s(\cos\theta)$ are related by the 
following relations \footnote{for $m$ integer the case $n=m$ requires some additional regularity assumption on ${\rm G}$ on the line $\nu = i(m +\frac{d-1}{2})$ (Theorem 3' of \cite{[BV-2]})}:
\begin{equation}
\forall n \in  {\Bbb N}\ ,\ \ \ \ n \geq m\ ,\ \ \ \ 
{\rm G} \left(i \left({d-1 \over 2}+n \right) \right) = g_n
\end{equation}
Moreover $ {\rm G}(\nu) $ 
is the (unique) carlsonian interpolation of the sequence
of values 
(\ref{huhuhu}) in the complex half-plane $ \left\{ \nu ;\ \Im  \ \nu  > m + {d-1
\over 2} \right\} . $
\end{proposition}
The proof of this result given in \cite{[BV-2]} consists in first
showing the property 
for the one-dimensional Fourier case (i.e. $ d=1) $ by a simple
contour-distortion 
argument, and then proving that the general $ d $-dimensional case reduces to
the 
previous one by applying the Radon-Abel transformation $ r \rightarrow 
\hat r $ described in \S 3.2.

\begin{remark}{\em      One can also give a  direct proof based 
on a contour distortion argument in the $ \cos\theta $ 
plane relating the formulae (\ref{34}) and (\ref{huhuhu}) thanks to the properties of the 
functions $ P^{(d+1)}_\lambda $ and $ Q^{(d+1)}_\lambda $; an argument of this 
type was in fact produced (for $d=2)$ in \cite{[Fr]} and \cite{[G]} in the 
context of complex angular momentum analysis of scattering amplitudes in 
S-matrix theory, involving a somewhat similar mathematical situation;
this is why we shall call ``Property FG'' 
the property  stated in Proposition 3.2.
}\end{remark}  

\begin{remark}{\em       As it will be suggested by the detailed study of the de
Sitterian 
linear fields given in Section 4, the most significant assumptions on $ r(
\cosh \ v) $ 
should correspond to the choice $ m = - {d-1 \over 2} $ for the behaviour at
infinity and to 
the inclusion of a distribution with singular support at $ \cosh \ v = 1
$ and maximal 
order $ E \left({d-4 \over 2} \right) $ if $ d \geq  4. $
Therefore, strictly speaking, it is only in the case $ d=2, $ that such
requirements 
allow to apply the statements of Propositions 3.1 and 3.2 in a fully
satisfactory 
way. However, it is possible to extend the proofs of these statements given in
$\cite{[BV-2]}$, with a small amount of technicalities, so as to
include the relevant 
requirements for $ d>2; $ this will be presented elsewhere.
}\end{remark}  
\vskip5pt 
{\em Analogy with the formalism of thermal QFT}
\vskip5pt 
The KMS-type interpretation of the analyticity of the two-point functions in 
complex temporal geodesics (see \S 2.4) is accompanied (as in the usual 
Matsubara formalism: see e.g. \cite{[LV]} and references therein)  by the phenomenon of quantization of the
associated 
imaginary energies (this is the so-called imaginary-time-formalism (ITF)). 
The FG property is actually a phenomenon of similar mathematical nature (in 
fact, {\it the same} phenomenon for $ d=1, $ see e.g.  
\cite{[BB]}). 

However, in contrast with the previous geodesical thermal property, the FG 
property is {\it global} (i.e. coordinate independent) on de Sitter 
space-time; in fact, as it will be seen below, the imaginary mass variable 
which is quantized in the euclidean formalism is actually a de 
Sitter-invariant quantity.

\subsection{  Inversion of the Laplace transformation and K\"allen-Lehmann-type 
representations}

In order to complete the description of the double analytic structure of the 
de Sitter two-point functions (namely, analyticity in space-time variables on 
$ X^{(c)}_d $ and analyticity in the mass variable $ \nu $ for the relevant
Laplace 
transform), we must also give the inversion formulae which allow one to 
represent all the elements of a given triplet $ ({\rm W},{\rm S},{\cal R}) $ in terms of
the 
corresponding Laplace transform $ {\rm G}(\nu ). $
The following statement has been established in $ \cite{[BV-2]}$

\begin{proposition} For triplets of moderate growth $ ({\rm W},{\rm S},
{\cal R}) $
whose 
associated Laplace transform $ {\rm G}(\nu ) $ is analytic\footnote{It is indeed sufficient that the product $h_{d+1}(-\frac{d-1}{2} -i\nu){\rm G}(\nu)$ be analytic
in this half-plane.}
 in the half-plane $ {\rm
Im} \ \nu >m+{d-1 \over 2} $ 
and satisfies appropriate regularity conditions on the boundary line $ {\rm
Im} \ \nu =m+{d-1 \over 2}, $ 
the following inversion formulae hold:
\[\underline{\cal R}(x) \equiv r \left(\frac{x^{(d)}}{R}\right) = \frac{1}{\omega_{d+1}}\int_{-\infty}^{+\infty} {\rm G}\left(\nu +
i\left(m + {d-1 \over 2}\right)\right)\cdot
\]
\begin{equation} 
\cdot h_{d+1} (m -i\nu) [P^{(d+1)}_{m-i\nu}(\cosh v)-2\sin\frac{\pi d}{2}
Q^{d+1}_{m-i\nu}(\cosh v)]
d\nu, 
\label{a320}
\end{equation}
\[
\underline{\rm W}(z) = 
-\frac{1}{2\omega_{d+1}}\int_{-\infty}^{+\infty}\frac{{\rm G}\left(\nu +
i\left(m + {d-1 \over 2}\right)\right) h_{d+1} (m -i\nu) P^{(d+1)}_{m-i\nu}
(\cos(\theta-\pi))} {\sin\pi(m -i\nu)}  
d\nu \]\begin{equation}
+ 
\frac{1}{\omega_{d+1}}\sum_{n=0}^{m-1} 
{g_n} h_{d+1} (n ) P^{(d+1)}_{n}(\cos \theta) = \left. {\rm w}\left( \frac{z^{(d)}}{R}\right)\right|_{z^{(d)}= R\cos\theta}.  
\label{a321}
\end{equation}

\end{proposition}
The functions 
$P^{(d+1)}_{\lambda}$ are generalized Legendre functions of the first
kind \cite{[BV-2]} (proportional to Gegenbauer functions of the first kind \cite{[ER]}) 
given by the following integral representation (valid for $\cos \theta
\in {{\Bbb C}} \setminus ]-\infty, -1]$):
\begin{equation}
P^{(d+1)}_{\lambda}(\cos \theta)
=\frac{2\omega_{d-1}}{\omega_d}(\sin\theta)^{2-d} 
\int_0^\theta\cos[(\lambda +\frac{d-1}{2})\tau]
[2(\cos\tau-\cos\theta)]^{\frac{d-3}{2}}d\tau
\label{legendre1}
\end{equation}
Formulae (\ref{a320}) and (\ref{a321}) apply respectively to the
following ranges: $x\in X_d, \, x>x_0$ and $z\in X^{(c)}_d; (z-z_0)^2
\not= \rho, \, \rho \geq 0$; moreover, for $z\in S_{d}\setminus
{z_0}$ (i.e. $\theta$
real), Eq. (\ref{a321}) yields the Schwinger function 
${\underline{\rm S}(z) }\equiv
{\rm s}(z^{(d)}/R)$, equivalently expressed by the Fourier-Legendre
expansion (\ref{a318}), with $g_n = {\rm G}(i(d-1+2n)/2)$ for $n\geq m$.

These formulae can be considered as {\it complex mass representations} valid 
for all two-point functions $ {\rm W} $ (resp. retarded propagators $ {\cal R}) $ of
de\nobreak\ Sitter 
QFT's satisfying the general analytic structure of perikernels and moderate 
growth conditions (in the sense of proposition 3.1). 
Since the functions 
$P^{(d+1)}_{\lambda}(z_1\cdot z_2/R)$ represent a family of special perikernels \cite{[BV-2]} on $X_d$,  Eq. (\ref{a321}) appears as a decomposition of general two-point functions on a basis of elementary two-point functions on $X_d$ admitting the same type of growth (governed by $\exp mv$).
By now making use of the 
hermiticity of the underlying fields, one can show the following 
result.

\begin{theorem}If the Laplace transform $ G(\nu) $ of the retarded propagator
$ {\cal R} $ of a 
de\nobreak\ Sitter QFT is analytic in the half-plane $ {\ Im} \ \nu  > 0, $
then $ {\cal R} $ and the 
associated complex space-time two-point function $ {\cal W} $ satisfy the
following 
K\"allen-Lehmann-type representations:
\[
{\underline{\cal R}}(x)\equiv r( {\cosh} v) =
\frac{1}{\omega_{d+1}}
\int^{ +\infty}_ 0{\rho}_0(
\nu) 
i h_{d+1}\left(-\frac{d-1}{2} -i\nu\right)\cdot\]
\begin{equation}
\cdot\left[P^{(d+1)}_{-{d-1 \over 2}-i\nu}(  {\cosh v}) - 2\  {\sin} 
{\pi d \over 2} Q^{(d+1)}_{-{d-1 \over 2}-i\nu}(  {\cosh v}) \right] {d} \nu, 
\label{a324a}\end{equation}
\begin{equation}
 \underline{\rm W}(z) = w( {{\cos}}\theta)  = \frac{1}{\omega_{d+1}}
\int^{ \infty}_{0 }{ {\rho}_0(
\nu)i
 h_{d+1}(-\frac{d-1}{2} -i\nu) P^{(d+1)}_{-{d-1 \over 2}-i\nu}( \cos(\theta -\pi)) \over 2\nu {\rm
sin} \ \pi \left({d-1 \over 2} + i\nu \right)} {d} \nu \ ,
\label{a325}\end{equation}
\end{theorem}
 {\bf Proof.} By writing Eq.(\ref{a320} for the case $ m = - {d-1 \over 2} $ and by
taking into account  that 
$ h_{d+1}(-\frac{d-1}{2} -i\nu)
P^{(d+1)}_{-{d-1 \over 2}-i\nu}( {{\cos}}\theta) / \sin(\pi(\frac{d-1}{2} -i\nu) $
is an odd function of $\nu$ (see Eqs. (\ref{gege4}) and (\ref{legendre1})),
one readily  obtains directly Eq. (\ref{a325}), with
\begin{equation}
{\rho}_0( \nu)  = -i\nu[{G(\nu)  - \ G(-\nu) }] 
\label{a326}
\end{equation}
Eq.(\ref{a324a}) is obtained similarly from Eq.(\ref{a320}) or from Eq. (\ref{a325}) by using a discontinuity formula
for $ P_\lambda( \cos(\theta -\pi)) $ 
across the line $ \theta =iv $ (\cite{[BV-2]}, Eqs.(3.104),
(3.105')),
 since $
{\underline{\cal R}} = i \left({\underline{\cal W}}_+-{\underline{\cal W}}_-
\right)$.
 As in the minkowskian case, the hermiticity of the field $ \phi $ implies that
$ r(  {\cosh v})\equiv i\theta \left(x^{(0)}_1-x^{(0)}_2 \right)\times
\left\langle \Omega , \left[\phi \left(x_1 \right),\phi \left(x_2 \right)
\right]\Omega \right\rangle $ 
is a real-valued function; the property of reality of $ {\rho}_0( \nu) $ then
follows  from 
Eqs. (\ref{34})-(\ref{gege4}),   and (\ref{a326}).

Proving the positivity of ${\rho}_0$ from the positive-definiteness of the Wightman functions requires a longer argument which relies on the introduction of a full Fourier-Laplace transformation. This will be given in \S 5.1.
   \\

\section{Quantum theory of free   fields}
\subsection{The de Sitter-Klein-Gordon equation and its plane-wave type
solutions}

It is common  to call free (bosonic, scalar) quantum field on a given riemannian manifold a distributional solution of the Klein-Gordon equation adapted to the given manifold. These theories are also called linear (as opposed to ``free'') because   there is an ``interaction'' between the field and the (unquantized) metric which gives rise to linear field equations.  Given the metric tensor $g$, the corresponding  Laplace-Beltrami  is obtained by the well-known formula
$
\square_{g} = {(-g)}^{-\frac{1}{2}}
 {\partial_{\alpha}}
\left({(-g)}^{\frac{1}{2}}g^{\alpha\beta}
{\partial_{\beta}}
\right)
$. In our case an easy way to deal with this operator is  to consider
the  d'Alembert  operator in the ambient space:
$
\square = \frac{\partial^{2}}{\partial {x^{(0)}}^{2}}
-\frac{\partial^{2}}{\partial {x^{(1)}}^{2}}-\ldots 
-\frac{\partial^{2}}{\partial {x^{(d)}}^{2}}.
 $
The  Laplace-Beltrami  $\square_{d}$ associated with $
X_d$ is seen to be the
trace of the   d'Alembertian  on $X_d$. This is constructed as follows:
let  $f\in {\cal C}^{2}(X_d, {{\Bbb C}})$. We associate with it  a
function
$\hat{f} \in {\cal C}^{2}({\cal U}, {{\Bbb C}})$ defined on the  open set
${\cal U}= \{x\in {{\Bbb R}}^{d+1}:\; x\cdot x<0\}$,  homogeneous of degree zero
and such that $ \hat{f}|_{X_d} = f$; it follows that 
$
\square_{d} f =
\square \hat{f}|_{X_d}.
$
The de Sitter-Klein-Gordon field equation is then written 
\begin{equation}
\square_{d}\phi +{\mu^{2} } \phi=0, \label{kg}
\end{equation}
where $\mu$ is a mass parameter which also includes the coupling to the gravitational field, and we look for an operator-valued distributional \cite{[SW]} solution for this equation satisfying the properties specified in \S 2.2 and \S 2.3.

There is an  important class of solutions   of the associated equation $(\square_{d} +\lambda )\psi=0$, where now $\lambda$ is real or complex, which is going to play the same role of {\em plane wave basis} as the exponentials in the Minkowski case, but with an important difference: in contrast with the minkowskian exponentials, these waves are singular on  three-dimensional 
light-like manifolds and can at first  be  defined only on suitable halves of the hyperboloid. We will need an appropriate $i\epsilon$-prescription (indicated below) to obtain global waves. Here is the relevant definition \cite{[BGM],[Fa2],[M],[Mo]}: let $\xi\in C^+$ and 
consider the function
\begin{equation} \psi^{\xi}_{+}(x,s)=\left(\frac{|x\cdot \xi|}{mR}
\right)^{s}
,\;\;s\in {{\Bbb C}},
\end{equation}
defined for those $x \in X_d$ such that $x\cdot \xi > 0$ (same definition for $\psi^{\xi}_{-}(x,s)$  for those $x \in X_d$ such that $x\cdot \xi < 0$);      $m$ is an auxiliary mass parameter introduced here for dimensional reasons, but it will also have  a minkowskian physical interpretation. The associated homogeneous function is given by
$F_{+}(x,s)=
\left(\frac{x\cdot\xi}{m\sqrt{-x\cdot x}}\right)^{s},
$
and it is defined in a corresponding open subset of $\cal U$; it follows that
$\square_d F_{+}(x,s) =  {m^{-s}}{s(d-1+s)}(-x\cdot x)^{\frac{s}{2}-1}
(x\cdot\xi)^{s}$ and therefore 
\begin{equation}
\square_{d} \psi^{\xi}_{+}(x,s)=\frac{s(d-1+s)}{R^{2}}
\psi^{\xi}_{+}(x,s)
\end{equation}
Physical values of the parameter $s$ are given by 
\begin{equation}
s = -\frac{d-1}{2} + \kappa,\;\;\;\;
{\rm with}\;\;\;\; 
\left\{
\begin{array}{ll}
\kappa =  i\nu, & \nu \in {\Bbb R}\\
\kappa = \nu, & \nu \in {\Bbb R}, \;\;|\nu| \leq \frac{d-1}{2}
\end{array}
\right.
\end{equation}
corresponding in the first case to  $-s(d-1+s)=\left(\frac{d-1}{2}\right)^2
+ \nu^2 = \mu^2 R^2   $  
and in the second case  to  $-s(d-1+s)=\left(\frac{d-1}{2}\right)^2
- \nu^2 = \mu^2 R^2  $.   
This will correspond to the existence of two qualitatively different families 
of Klein-Gordon fields, characterized by  a mass $\mu$ greater or lower than the ``geometrical'' mass $\mu_d = \frac{d-1}{2R}$.
Correspondingly, the plane waves have an oscillatory or power law behaviour. 
Borrowing a terminology used in the context of the $SO_0(1,d)$ representation theory  we may call {\em principal} plane waves (and consequently {\em principal} quantum fields) the plane waves belonging to the first class   and {\em complementary} plane waves those belonging to the second class. We will further comment on this distinction. For the time being the following argument, based on the large $R$ behaviour, makes it clear that principal waves are more closely similar to minkowskian plane waves.
In fact, 
parameterizing $\xi$ by the wave-vector of a (minkowskian) 
particle of mass $m$, i.e. 
$\xi=[{ \rm k}^{0} , \vec{\rm  k}, -{m}]$, with
${\rm k}^{0}={\sqrt{\vec{\rm  k}^{2} + m^{2}}}$,
yields
\begin{equation}
\lim_{R \rightarrow \infty}
\left(\frac{x_R({{\rm x}) } \cdot\xi}{mR}\right)^{-\frac{d-1}{2}+i{mR}}
=\exp(i{\rm k}\cdot {\rm x});\label{parcondicio}
\end{equation}
in this equation, points in the de Sitter universe must be described using the minkowskian  space-time variable ${\rm x} \equiv ({\rm x^{(0)}},\vec{x})$ measured in units of the de Sitter radius $R$;
for instance we can choose the global parametrization
\begin{equation}
x_R({\rm x}) =
\left[
\begin{array}l
x^{(0)} = R \sinh \frac{{\rm x}^{(0)}}{R} \\
x^{(i)}= 
R \cosh \frac{{\rm x}^{(0)}}{R} \cos \frac{{\rm x}^{(1)}}{R}\ldots\cos \frac{{\rm x}^{(i-1)}}{R} \sin \frac{{\rm x}^{(i)}}{R},\;\;i=1,\ldots,d-1\\
x^{(d)}=R \cosh \frac{{\rm x}^{(0)}}{R} \cos \frac{{\rm x}^{(1)}}{R}\cos \frac{{\rm x}^{(2)}}{R}\ldots \cos \frac{{\rm x}^{(d-1)}}{R}
\end{array}\; 
\right],
\label{para1}
\end{equation}
or equivalently the one given by Eq.(\ref{tytyty}).

From the other side, important field  theories, like the massless field conformally coupled to the de Sitter metric belong to the  complementary  class.

\subsection{Plane-wave analysis of the free fields}

Let us for the moment restrict our attention to principal Klein-Gordon fields.
In Section 3, we have seen that the two-point function $ {\rm W }\left(z_1,z_2
\right) $ of any 
de\nobreak\ Sitter GFF satisfying the  double analytic structure
 and suitable growth conditions admits the 
 K\"allen-Lehmann-type representation (\ref{a325}) in terms of the elementary 
perikernels $ P^{(d+1)}_{-{d-1 \over 2} +i\nu} \left({z_1\cdot z_2 \over R^2}
\right) $,  (with $ z_1\cdot z_2=-R^2 {\cos}  \theta , $ 
according to the notations of Theorem 3.1).

We shall now enhance this analogy with the Wightman QFT's in Minkowski 
space-time by proving the following results. For each positive number $ \nu
=mR, $ 
the function $ P_{-{d-1 \over 2}+i\nu} \left({z_1\cdot z_2 \over R^2} \right)
$ can be identified up to a constant factor   
with the two-point function $ {\rm W}_\nu \left(z_1,z_2 \right) $ of a {\it linear}
de\nobreak\ Sitter QFT 
satisfying Eq.(\ref{kg}) with the mass parameter $ \mu^ 2 = {1 \over R^2} \left[
\left({d-1 \over 2} \right)^2+\nu^ 2 \right] = \left({d-1 \over 2R}
\right)^2+m^2; $ 
this constant factor  is then determined unambiguously by imposing
the 
canonical commutation relations (CCR) or the local Hadamard behaviour (see \cite{[H],[KW]} and references therein) 
with the relevant
coefficient of the dominant term  (i.e. the value of the corresponding 
quantity in the minkowskian case).
 Moreover, the function $ {\rm W}_\nu $ will
be 
introduced by an integral representation in terms of principal de\nobreak\ Sitter
plane 
waves which is completely analogous 
to the standard Fourier representation of 
the two-point function of the minkowskian free field.

Writing such a plane-wave representation immediately sets the problem of 
extending the previous plane waves $ \psi^ \xi_{ \pm} $ of \S 4.1 to the whole
(real) 
hyperboloid. 
In particular, we need to specify the correct phase factors to glue together 
$\psi^{\xi}_{+}$ and $\psi^{\xi}_{-}$. 
Following \cite{[BGM],[Mo]} we solve this problem by making use of  the properties of the complex hyperboloid $X^{(c)}_{d}$
which have been described in \S 2.3; 
in particular the tuboids  ${\cal T}^{+}$ and 
${\cal T}^{-}$  play here a crucial role.
Indeed,  when $\xi \in C^+$ is fixed and  $z$ varies in ${\cal T}^{+}$ or in ${\cal T}^{-}$ then the plane waves 
\begin{equation}
\psi^\xi(z,s)=\left(\frac{z\cdot \xi}{mR}
\right)^{s}
,\;\;s\in {{\Bbb C}} \label{waves}
\end{equation}
are globally defined because  
$\Im(z\cdot \xi)$ has a fixed
sign. 
The phase is chosen such that $ bv\ \psi^ \xi\vert_{ x\cdot \xi >0} = \psi^
\xi_ +, $ with $ \psi^ \xi_ +>0 $ for $ s $ real.
These global waves reproduce the analyticity properties of the minkowskian 
plane-waves $ {\rm e}^{i{\rm kz}} $ in the corresponding tubes $ {\rm T}^+ $ and $ {\rm T}^- $
(for $ {\rm k} $ in $ V^+). $

Let us now introduce for convenience the notion of an \lq\lq orbital
basis\rq\rq\ $ \gamma $ of $ C^+ $ 
with respect to a subgroup $ L_e $ of $ {\rm SO}_0(1,d) $ which is the
stabilizer of a unit 
vector $ e $ $ \left(\vert e^2\vert=1 \right) $ in $ {\Bbb R}^{d+1}: $

$\alpha$) 
if $ e \in  V^+, $ $ \gamma $ is the section of $ C^+ $ by a hyperplane of
the form $ e\cdot \xi =a $ $ (a>0), $ 
namely an orbit (of spherical type) of the corresponding subgroup $ L_e\approx
{\rm SO}(d). $

$\beta$) 
if $ e^2=-1, $ $ \gamma $ is the union of the sections of $ C^+ $ by two
hyperplanes of the 
form $ e\cdot \xi =\pm b $ $ (b>0), $ namely the union of two hyperboloid
sheets which are 
orbits of the corresponding subgroup $ L_e\approx {\rm SO}_0(1,d-1). $
 
We shall make a special use of the following bases $ \gamma_ 0 $ and $ \gamma_
d $ of respective 
types  $\alpha$ and $\beta$. 
\begin{eqnarray}
\gamma_ 0&=& \left\{ \xi  \in  C^+;\ \xi^{( 0)}=m \right\} \ , \\ \gamma_ d&=&
\gamma^ +_d \cup  \gamma^ -_d = \left\{ \xi \in C^+;\ \xi^{( d)}=m \right\} 
\cup  \left\{ \xi \in C^+;\ \xi^{( d)}=-m \right\}\label{4.8}
\end{eqnarray}
All the orbital bases $\gamma$ of either type $\alpha)$ or $\beta)$, 
equipped with a suitable orientation, belong to the same homology class $\dot{\gamma}$ in $H_{d-1}(C^{+}; \Phi)$, where the latter denotes a homology group associated with a suitable family $\Phi$ of supports, sufficiently regular at infinity \cite{[Le]}. For example, the compact cycle $\gamma_0$ and the non-compact cycle $\gamma_d$are easily seen to be homologous, as enclosing the region of the cone $C^+$ described as follows: $\{\xi = (\lambda,\lambda u_1,\ldots, \lambda u_d);\;u_1^2+\ldots+u_d^2=1, m<\lambda<m/|u_d|\}$
We can now define in a consistent way on each orbital basis $\gamma$ a measure $d\mu_\gamma(\xi_\gamma)$ which is invariant under the corresponding subgroup
$L_e$ by putting $d\mu_\gamma = 
i_\Xi\omega_{C^+}|_\gamma$;
 here 
$i_\Xi\omega_{C^+}$ denotes the (d-1)-form defined on $C^+$ as the contraction of the vector field $\Xi$ (where $\Xi$ is the restriction to $C^+$ of the vector field whose cartesian components are $\underline{\Xi^{(j)}} = \xi^{(j)}$) and the volume form $\omega_{C^+}
= \left.\frac{d\xi^{(0)}\wedge d\xi^{(1)}\wedge \ldots \wedge d\xi^{(d)}}
{d(\xi^2 )}\right|_{C^+}$. We shall now rely on the following lemma, which can be directly checked.
\begin{lemma}
For every function $f$ on $C^+$ which is homogeneous of degree $1-d$, the $(d-1)$-form $f\,i_\Xi\omega_{C^+}$ is closed.
\end{lemma}

We can then state:

\begin{theorem} \label{maintheorem}
For each  $ m\geq
0, $ there exists a 
de\nobreak\ Sitter GFF satisfying all the properties a), b), c), d) 
described in 
section 2 which is a solution of the linear field equation (\ref{kg}) with $ \mu^
2= \left({d-1 \over 2R} \right)^2+m^2. $

The corresponding two-point function $ {\rm W}_\nu \left(z_1,z_2 \right), $ labelled
by the dimensionless 
parameter $ \nu  = mR, $ is given by the   integral 
representation:
\begin{equation}
{\rm W}_\nu \left(z_1,z_2 \right) = c_{ d,\nu} \int^{ }_ \gamma
\left(z_1\cdot \xi_ \gamma \right)^{-{d-1 \over 2}+i\nu} \left(\xi_ \gamma
\cdot z_2 \right)^{-{d-1 \over 2}-i\nu} d\mu_ \gamma \left(\xi_ \gamma
\right)\ ,\label{wig1}
\end{equation}
in which $ \gamma $ denotes {\it any} orbital basis of $ C^+ $   and  $ c_{ d,\nu} $ is a positive constant.
\end{theorem}
 
 {\bf Proof.} From the analyticity properties of the plane-waves
(\ref{waves}), 
it is clear that the function $ {\rm W}^{(\gamma)}_ \nu \left(z_1,z_2 \right) $
(depending {\it a priori} on $ \gamma $) 
defined by Eq.(\ref{wig1}) is analytic in the tuboid $ {\cal T}_{12} = {\cal
T}^-\times{\cal T}^+ $ of $ X^{(c)}_d \times  X^{(c)}_d $.
Moreover, $ {\rm W}^{(\gamma)}_ \nu$ is of moderate growth when $(z_1,z_2)$ tends to the real from the tuboid ${\cal T}_{12}$, and therefore it admits a 
distributional 
boundary value $ {\cal W}^{(\gamma)}_ \nu \left(x_1,x_2 \right) $ on
$ X_d\times X_d $ (Theorem A.2.).
Now, fixing $(z_1,z_2) \in {\cal T}_{12}$ and putting
$f^{(\nu)}_{z_1z_2}(\xi) = 
c_{ d,\nu} 
\left(z_1\cdot \xi\right)^{-{d-1 \over 2}+i\nu} \left(\xi
\cdot z_2 \right)^{-{d-1 \over 2}-i\nu}$
we can write 

\begin{equation}
{\rm W}_\nu \left(z_1,z_2 \right) = c_{ d,\nu} \int^{ }_ \gamma
f^{(\nu)}_{z_1z_2}(\xi)i_\Xi\omega_{C^+}(\xi)|_\gamma;
\label{wig111}
\end{equation}
since Lemma 4.1 obviously applies to $f= f^{(\nu)}_{z_1z_2}$, it follows that the value of the integral (\ref{wig111}) is independent of the orbital basis $\gamma \in \dot{\gamma}$, the integrability at infinity on non-compact bases of type $\beta)$ being also ensured by the homogeneity properties of $f^{(\nu)}_{z_1z_2}$.
To see this explicitly, consider for example the basis
$\gamma_d$ in Eq. (\ref{4.8}) 
which may be parametrized as follows: 
\begin{equation}
\xi^+(\vec{k}) =\{ {k^0}, \vec{k},  m\},\;\;\;\;
\xi^-(\vec{k}) =\{ {k^0}, -\vec{k}, - m\},\;\;\;\;
k^0=\sqrt{{\vec k}^2 + m^2};
\end{equation} 
 the corresponding invariant measures are written 
$d\mu_{\gamma^+}(\vec{k})=m\,d\vec{k}/{k^0}$ and  
$d\mu_{\gamma^-}(\vec{k})\linebreak[0] =m\,d\vec{k}/{k^0}$. 
In these variables Eq. (\ref{wig1}) is rewritten as the following absolutely convergent integral:
\begin{equation}
{\rm W}^{\gamma_d}_{\nu}(z_1,z_2) = 
c_{d,\nu}\sum_{l=+,-}\int 
\left({z_1\cdot \xi^{l}(\vec{k})}\right)^{-\frac{d-1}{2} + 
i\nu}\left({\xi^{l}(\vec{k})\cdot
z_2}\right)^{-\frac{d-1}{2} - i\nu}\frac{m d{\vec{k}} }{{k^0}}.\label{wig11}
\end{equation}
We conclude that for all $\gamma$, Eq. (\ref{wig1}) defines the same analytic function ${\rm W}_\nu ={\rm W}_\nu^{(\gamma)}$ in ${\cal T}_{12}$.  Let us now check that its boundary value ${\cal W}_\nu$ satisfies  
 the 
positivity condition.
 We consider for each test function $f \in {\cal D}(X_d)$ the integral 
\begin{equation}
{\cal I}(f,\epsilon) =\int_{X_d\times X_d}{\rm W}_{\nu}
(\overline{z(x_1,\epsilon)},z(x_2,\epsilon)) \overline{f}(x_1) 
{f}(x_2) d\sigma(x_1)d\sigma(x_2),
\label{piopo}
\end{equation}
where the point $z(x,\epsilon)$
varies in ${\cal T}^{+}$ according to the following definition  (see Eq. (\ref{delta})): 
$z(x,\epsilon) = \delta(x, \epsilon u_x)$, $u_x$ being a time-like vector in the meridian plane of $X_d$ determined by the conditions $u_x\cdot x = 0 $, 
$u_x \in V^+$, ${u_x}^2 = 1$.
Plugging the expression (\ref{wig1}) of ${\rm W}_\nu$ into (\ref{piopo}) shows that ${\cal I}
(f,\epsilon)\geq 0$ (for all $f$ and $\epsilon$). Condition 
(\ref{positivity}) then follows by taking the limit $\epsilon\to 0$. The hermiticity property, implied by positivity, is also obtained by taking boundary values from ${\cal T}_{12}$ of the identity ${\rm W}_\nu(z_2,z_1)=\overline{{\rm W}_\nu
(\bar{z_1},\bar{z_2})}$, easily checked on Eq. (\ref{wig1}). 

Since all the complex waves $\psi^\xi(z,s)$ satisfy in their domain the complex version of the de{\nobreak}Sitter-Klein-Gordon equation, it 
 follows that Eq. (\ref{kg}) (with 
$\mu^{2}=(\frac{d-1}{2R})^{2}+\left(\frac{\nu}{R}\right)^{2}$) is satisfied in both variables $z_1, z_2$ by the function 
$ {\rm W}^{(\gamma)}_ \nu  $;
correspondingly, by passing to the boundary values, one sees that the kernel 
$ {\cal W}^{(\gamma)}_ \nu  $ is a distribution solution on $X_d\times X_d$ of the equations:
\begin{equation}
(\square_{d,x_1} +{\mu^{2} }) {\cal W}^{(\gamma)}_ \nu(x_1,x_2)
=
(\square_{d,x_2} +{\mu^{2} }) {\cal W}^{(\gamma)}_ \nu (x_1,x_2)= 0.
\label{wigg}
\end{equation}
In order to introduce a bona fide de Sitter field $\phi_\nu(x)$ 
with two-point function 
${\cal W}^{(\gamma)}_ \nu(x_1,x_2)$, it is necessary to show that ${\cal W}^{(\gamma)}_ \nu(x_1,x_2)$, not only
satisfies the positivity   and normal analyticity
conditions a) and d) checked above, but also the locality, covariance
and hermiticity properties 
properties.
Once this is done, one can easily check that, in view of Eqs.
(\ref{wigg}), 
the field $\phi_\nu$ constructed in terms of ${\cal W}_\nu$ by the
Formulae (\ref{crea}) and (\ref{azzo}) satisfies the equation 
$(\square_{d,x} +{\mu^{2} }) {\phi}_\nu(x)=0$
as an operator-valued distribution on the relevant Hilbert space ${\cal
H}$ (in this theory, the subspace ${\cal H}_1$ is characterized as a set
of classes of test functions $h_1$ on $X_d$, modulo the functions 
which are of the form  
$
(\square_{d,x_1} +{\mu^{2} })h(x)$; all the subspaces ${\cal H}_n$ are
characterized similarly).

To prove the covariance property 
 c)  
we first of all recall that in view of Eq. (\ref{tubi1}) 
${\cal T}^+$ and ${\cal T}^-$ 
are invariant under the 
action of the real de Sitter group, like ${\rm T}^+$, ${\rm T}^-$ and
$X^{(c)}_d$.
Therefore, if $z_{g}$ denotes the point obtained by applying a real
de Sitter transformation $g$ to the point $z\in {\cal T}^\pm$, 
the expression
\begin{equation}
{\rm W}_{\nu}({z_1}_g,{z_2}_g) = 
c_{d,\nu}\int_{\gamma}
\left({{z_1}_{g}\cdot \xi_{\gamma}}\right)^{-\frac{d-1}{2} + i\nu}
\left({\xi_{\gamma}\cdot {z_2}_g} \right)^{-\frac{d-1}{2} - i\nu}
d\mu_{\gamma}(\xi_{\gamma}) \label{wig2}
\end{equation}
remains meaningful; let us now show that  
\begin{equation}
{\rm W}_{\nu}(z_1,z_2) = 
{\rm W}_{\nu}({z_1}_g,{z_2}_g ).\label{covariance}
\end{equation}

Let us consider any subgroup $L_e$ of $SO_0(1,d)$ and write Eq. (\ref{wig1}) with an orbital basis $\gamma$ invariant with respect to $L_e$. In view of the invariance of the measure $d\mu_\gamma$ and of $\gamma$ under the $L_e$, it follows that Eq. (\ref{covariance}) holds for $g\in L_e$.

Since any element of $G_d$ may be decomposed as the product of transformations 
belonging to the subgroups $L_e$ we have displayed\footnote{This assertion corresponds to 
the physically intuitive statement that any transformation of the relativity group 
is the composition of a ``space translation'', a ``time translation'' and a 
``Lorentz boost''. A proof can be found in \cite{[Ar]}.},
it follows that Eq. (\ref{covariance}) is true for any $g\in G_d$.
  Taking into account the fact that $ {\rm W}_\nu $ is analytic in 
$ {\cal T}_{12} $
it follows (as in the 
proof of proposition 2.2) that the function $ {\rm W}_\nu $ extends to an invariant 
perikernel on $ X^{(c)}_d $ (analytic in $\Delta$)
which can be identified with a function $ {\rm w}_\nu(
\alpha) $ of the 
single invariant variable $ \alpha  = - {z_1\cdot z_2 \over R^2}. $

In order to prove the locality condition b), let us also introduce the invariant perikernel 
${\rm W}'_\nu({ z_1},{z_2}) ={\rm W}_\nu({ z_2},{z_1})$,  
satisfying 
${\rm W}'_\nu({ z_1},{z_2}) = \overline{{\rm W}_\nu({\overline z_1},{\overline z_2})}$ and such that $bv_{{\cal T}_{21}}{\rm W}'={\cal W}'$. 
 We 
will then show that $ {\rm W}_\nu $ and $ {\rm W}^{\prime}_ \nu $ can both be identified
 (up to a factor) with the generalized Legendre function $
P^{(d+1)}_{-{d-1 \over 2}+i\nu} \left({z_1\cdot z_2 \over R^2} \right). $ 
This is most easily shown by choosing the following points $ \left(z_1(v),z_2
\right)\in{\cal T}_{12}, $ 
such that $ {z_1(v)\cdot z_2 \over R^2} =  {\cosh v} : $
$ z_1(v) = (-iR\  {\cosh v},\ -iR\ {\sinh} \ v,\ 0,...,0)\ ,\ \ \
z_2=(iR,0,...,0)\ ,\ {\rm for} \ v\geq 0. $
At these points, Eq.(\ref{wig1}) is written as follows:
\begin{equation}
{\rm W}_\nu \left(z_1(v),z_2 \right) = \frac{c_{d,\nu} e^{\pi\nu}\omega_{ d-1}}{R^{(d-2)}}
\int^ \pi_ 0( {\cosh}  v+ {\sinh}  v \;{\cos}  \theta)^{ -{d-1
\over 2}+i\nu}( {\sin}  \theta)^{ d-2}d\theta,  
\end{equation}
 where we have chosen to integrate over $\gamma_0$;
by changing the integration variable $ \theta 
\longrightarrow  w $ such that $ e^w = {\cosh}  v+ {\sinh}  v\ {\cos}  \theta , $ 
allows us to recognize the integral representation (\ref{legendre1}) of the function $
P^{(d+1)}_{-{d-1 \over 2}+i\nu}( {\cosh} v) $ 
and thus yields:
${\rm w}_\nu( - {\cosh}  v) = {\rm W}_\nu \left(z_1(v),z_2 \right) = 
C_{d,\nu}
P^{(d+1)}_{-{d-1 \over 2}+i\nu}( {\cosh} v)
$, with $C_{d,\nu} = c_{d,\nu}e^{\pi\nu} \nu^{-1}R^{-(d-2)}\omega_ d$.
This shows that 
${\rm w}_ \nu( - {\cosh}  v) = {\rm W}_\nu (z_1(v), { z_2})$ is real-valued for all $v$ and therefore that one also has:
\begin{equation}
{\rm w}_ \nu( - {\cosh}  v) = \overline{W_ \nu \left({ z_1(v)}, { z_2} \right)} = W'_ {\nu} \left( \overline{ z_1(v)}, \overline{ z_2} \right);\label{opq} 
\end{equation} 
 we note  that since  each real pair $ \left(x_1(v)=(-R\ {\sinh} \ v,0,...,0,-R\  {\cosh v}),x_0 \right) $ 
in the locality region $ {\cal R} $ is on the same orbit of the complex
de\nobreak\ Sitter group 
as the corresponding pairs $ \left(z_1(v),z_2 \right) $ and $ \left(
\overline{ z_1(v)}, \overline{ z_2}
\right), $ Eq.(\ref{opq}) is  
also interpretable (in view of the invariance property c) for ${\rm W}_\nu$ and 
${\rm W}'_\nu$ as the locality relation $ {\cal W}_\nu \left(x_1(v),x_0
\right) = {\cal W}^{\prime}_ \nu \left(x_1(v),x_0 \right) $ at 
any such pair, and  implies  the identity of  
${\rm W}_\nu$ and ${\rm W}'_\nu$. The proof of the theorem is complete.
   \\  

\begin{corollary} The two-point function $ {\rm W}_\nu $ of the de Sitter-Klein-Gordon field is    $ C_{d,\nu} P^{(d+1)}_{-{d-1 \over 2}+i\nu}({z_1\cdot z_2 \over R^2})$.  
 $C_{d,\nu}$ is  fixed by imposing the CCR's or local Hadamard behavior with the relevant coefficient of the dominant term and is given by
\begin{equation}
C_{d,\nu} = 
\frac{
\Gamma\left(\frac{d-1}{2} +i \nu\right)
\Gamma\left(\frac{d-1}{2} -i \nu\right)}{2^d R^{d-2}
\Gamma\left(\frac{d}{2}\right) \pi^{\frac{d}{2}}}. \label{a420}
\end{equation}
\end{corollary}
 {\bf Proof.}
The singular behaviour of $ {\cal W}_\nu $ at coinciding points is given by the behaviour of  $P^{(d+1)}_{-{d-1 \over 2}+i\nu}$ near its singular point $\frac{z_1\cdot z_2}{R^2}= \cos \theta = -1$.
Using the integral representation (\ref{legendre1}) near $\theta = \pi$   for $z_1$ and $z_2$ spacelike separated and $d>2$ one gets:   
\[
{\rm W}_{\nu}(z_1,z_2) \simeq
\frac{C_{d,\nu}  {2^{\frac{d}{2}}
\Gamma(\frac{d-2}{2}})
\Gamma(\frac{d}{2})}{{2\Gamma(\frac{d-1}{2}+i\nu)}{\Gamma({\frac{d-1}{2}}-i\nu)}} 
 \left(\frac{z_1\cdot z_2}{R^2}+1\right)^{\frac{2-d}{2}}
\]
\begin{equation}
=\frac{c_{d,\nu}e^{\pi\nu} \pi^{\frac{d}{2}} {2^{\frac{d}{2}}\Gamma({\frac{d-2}{2}})}}{
R^{d-1}{\Gamma(\frac{d-1}{2}+i\nu)}{\Gamma({\frac{d-1}{2}}-i\nu)}} 
 \left[-\frac{(z_1- z_2)^2}{2R^2}\right]^{\frac{2-d}{2}}.
\label{4.21}\end{equation}
The corresponding dominant term for the two-point function of the Minkowski Klein-Gordon field is given by 
\begin{equation}
-iD^{(-)}({\rm z}_1 -{\rm z}_2)=
\frac{\Gamma(d-2)}{(2i)^{d-2}\pi^{\frac{d-1}{2}}\Gamma(\frac{d-1}{2})}
\left[-\frac{1}{({\rm z}_1-{\rm z}_2)^2}\right]^{\frac{d-2}{2}}.
\label{4.22}\end{equation}
Since at short distances the de Sitter and the Minkowski distances (in a tangent plane)  are asymptotically equal,
we obtain that the two-point function satisfies  the  Hadamard condition   \cite{[H],[KW]}. 
The constant $C_{d,\nu}$   may now be fixed by 
imposing canonical normalisation of the Minkowskian case according to 
Eq. (\ref{4.22}). 
This is equivalent to requiring the CCR's.
By equating coefficients in Eqs. (\ref{4.21}) and (\ref{4.22}) we get
Eq. (\ref{a420}). The constant appearing in Eq. (\ref{wig1}) is then given by
\begin{equation}
 c_{d,\nu}=\,\,\frac{\Gamma(d-1) {{\Gamma(\frac{d-1}{2}+
i\nu)}{\Gamma({\frac{d-1}{2}}-i\nu)}}e^{-\pi\nu}}
{(2\sqrt{\pi})^{2d-1}\Gamma(\frac{d-1}{2})
\Gamma({\frac{d}{2}})}
\label{consta}.
\end{equation} 
This result also covers the case $d=2$, where the corresponding short-distance behaviour of  $ {\cal W}_\nu $ is proportional to $\log \left(\frac{z_1\cdot z_2}{R^2}+1\right)$.
\vskip5pt

\begin{remark} {\em Theorem 4.1 and Corollary 4.1 can be extended to cover also field theories relative to real values of $\kappa$ (see Eq. (4.4)). the various points of the proof of these  statements remain unchanged, except for the proof of the positive-definiteness. We do not give here the details.}
\end{remark}
We can then state:
\begin{theorem} \label{maintheorem2}
For each  $ m, $ with $ 0 < m \leq \frac{d-1}{2R}, $ there exists a 
de\nobreak\ Sitter GFF satisfying all the properties a), b), c), d) 
described in 
section 2 which is a solution of the linear field equation (\ref{kg}) with $ \mu^
2= \left({d-1 \over 2R} \right)^2-m^2. $

The corresponding two-point function $ {\rm W}_{-i\nu} \left(z_1,z_2 \right), $ labelled
by the dimensionless 
parameter $ \nu  = mR, $ is given by the following   integral 
representation:
\begin{equation}
{\rm W}_{-i\nu} \left(z_1,z_2 \right) = c_{ d,i\nu} \int^{ }_ \gamma
\left(z_1\cdot \xi_ \gamma \right)^{-{d-1 \over 2}+\nu} \left(\xi_ \gamma
\cdot z_2 \right)^{-{d-1 \over 2}-\nu} d\mu_ \gamma \left(\xi_ \gamma
\right)\ ,\label{wig1com}
\end{equation}
in which $ \gamma $ denotes {\it any} orbital basis of $ C^+ $   and  $ c_{ d,i\nu} $ is given in Corollary 4.1.
\end{theorem}

\begin{remark} \label{gyu}
{\em The existence of these two classes of Klein-Gordon fields on de Sitter space-time shows that the K\"allen-Lehmann decomposition (\ref{a325}) given in Section 3 is not the most general one for two-point functions on de Sitter space-time.
Indeed theories covered by that decomposition are superpositions involving only the two-point functions of the principal series. We will refer to this fact by calling from now on Eq. (\ref{a325}) the {\em K\"allen-Lehmann principal decomposition}.  
One expects that in the most general case, obtained {\em by  relaxing the growth condition corresponding to the analyticity properties used in Theorem 3.1}, a two-point function satisfying all the properties a), b), c), d) will admit a K\"allen-Lehmann representation of the form
\begin{equation}
\underline{\rm W}(z)= \int^{\infty }_{0 } \rho_p( m) C_{d,mR} 
P^{(d+1)}_{-{d-1 \over 2}-im R} \left({z\cdot z_0 \over R^2}
\right)   {d}m  + \int^{\frac{d-1}{2R} }_{0 } \rho_c( m) C_{d,imR} 
P^{(d+1)}_{-{d-1 \over 2}-m R} \left({z\cdot z_0 \over R^2}
\right)   {d}m .
\label{aiu}\end{equation}}
\end{remark}
\begin{remark}
{\em
One meets decompositions similar to Eq. (\ref{aiu}) in the context of $SO_0(1,d)$ representation theory (see e.g. \cite{[Fa2]}). However in general these decompositions have a supplementary term corresponding to what is called {\em the discrete series of representations}. These representations should be discarded here because of the locality requirement.}
\end{remark}

\begin{remark}{\em      Eq. (\ref{wig1}) and (\ref{wig1com})  gives  a new integral representation of the indicated  class of Gegenbauer functions of the first kind with a clear
geometrical interpretation in terms of two points (i.e. as a kernel) on the $d$-dimensional 
one-sheeted hyperboloid, summarizing several known integral representations 
in one single formula. More precisely this means the following: different integral representations of the Gegenbauer function 
$P^{(d+1)}_{-\frac{d-1}{2} +
i\nu}(w)$ ($\nu$ complex) are obtained by choosing a   integration path on the cone and two suitable points $z_1$ and $z_2$ of the complex hyperboloid in such a way that 
$w = \frac{z\cdot z_2}{R^2}$.   
}\end{remark}

\begin{remark}{\em      The boundary value of 
Eqs. (\ref{wig1}) and  (\ref{wig1com}) gives us the   
following  representation for 
the two-point Wightman function of the   de Sitter-Klein-Gordon field:

\[
{\cal W}_{\nu}(x_1,x_2)=
c_{d,\nu}\int_{\gamma}
(x_1\cdot \xi_\gamma)^{-\frac{d-1}{2}+i\nu}_{+}
+ e^{-i\pi(-\frac{d-1}{2}+i\nu)}(x_1\cdot \xi_\gamma)^{-\frac{d-1}{2}+i\nu}_{-})\cdot\]
\begin{equation}
\cdot(  (x_2\cdot \xi_\gamma)^{-\frac{d-1}{2}-i\nu}_{+}
+ e^{i\pi(-\frac{d-1}{2}-i\nu)}(x_2\cdot \xi_\gamma)^{-\frac{d-1}{2}-i\nu}_{-} )
d\mu_\gamma ,\label{wig4}
\end{equation}
}\end{remark}  
(similarly for the complementary case)
where we adopted the notation introduced by Gel'fand for the distributions 
$(\cdot)_\pm$ \cite{[GEL]}. The important point of this formula is that
it allows a factorization of the two-point function in
terms of global plane waves on the hyperboloid $X_d$ which is completely analogous to the corresponding Fourier representation for two-point function of the Minkowski free field of mass $m$. The latter will be obtained as the limit of Eq. (\ref{wig4}) for $R\to\infty$ (see \S 5.2).
\subsection{Fourier transform of test functions on $X_d$}
An  important consequence of the  representation (\ref{wig1}) is the introduction of  a natural Fourier transform on the
hyperboloid; this transform is crucial for what follows and, in our opinion, it will play a central role in any question connected with 
the harmonic analysis on the hyperboloid. 

Given a function $f \in {\cal D}(X_d)$ we define its Fourier transform  as
the following pair of homogeneuous complex functions on the cone $C^+$:
\[\tilde{f}_{\pm}\left(\xi,-\frac{d-1}{2}-i\nu\right) =
\tilde{f}_{\pm,\nu}(\xi ) = 
\]\begin{equation}
\int_{X_d} 
( (x\cdot \xi)^{-\frac{d-1}{2}-i\nu}_{+}
+ e^{\pm i\pi(-\frac{d-1}{2}-i\nu)}(x\cdot \xi)^{-\frac{d-1}{2}-i\nu}_{-} )
f(x)d\sigma(x)
\label{fourier}
\end{equation}
More generally, we may introduce a Fourier transform depending on a complex parameter $s$:
\begin{equation}
\tilde{f}_{\pm}(\xi,s) = \int_{X_d} 
\left((x\cdot \xi)^{s}_{+} 
+ e^{\pm i\pi s}(x\cdot \xi)^{s}_{-}\right)f(x)d\sigma(x)
\label{fourier2}
\end{equation}
The kernels in these  Fourier transforms are obtained by taking the boundary values of the plane
waves $[(x+iy)\cdot\xi]^s$ from the corresponding tubes ${\cal T}^{\pm}$.
This implies that there is a relationship between the analyticity properties 
of the test functions and the vanishing of the corresponding Fourier transforms.
A detailed presentation of the mathematical  properties of this 
Fourier transform will be given somewhere else. 

We end this section by giving 
an explicit ${\rm L}^2$ representation of the scalar product (\ref{hjk}) for ${\cal W} = {\cal W}_\nu$ belonging to the principal class, which is analogous to the momentum-space representation of one-particle states in Minkowskian QFT. Indeed, with the help of the Fourier transform (\ref{fourier}) we may  rewrite the scalar
product (\ref{hjk})  as follows:
\begin{equation}
\langle \dot{f},\dot{g}\rangle_\nu = \int_{X_d\times X_d} {\cal W}_\nu(x,y)\overline{f}(x)f(y)d\sigma(x)d\sigma(y)=
  c_{d,\nu}\int_{\gamma}
\bar{\tilde{f}}_{+,\nu}(\xi_\gamma)\tilde{g}_{+,\nu}(\xi_\gamma)d\mu_{\gamma},
\end{equation}
which directly shows that 
$\langle \dot{f},\dot{f}\rangle_\nu\geq 0$.
This entails that the ``one-particle'' Hilbert space of the theory  ${\cal H}_1$ may be realized as the space ${\rm L}^{2}(\gamma,d\mu_\gamma)$ of complex-valued  functions 
defined on the orbital basis  $\gamma \subset C^{+}$ and square integrable w.r. to the corresponding measure $d\mu_{\gamma}$ (this assertion will be fully justified in the next subsection; here we can only conclude that the one-particle Hilbert space of the theory is a closed subspace of   ${\rm L}^{2}(\gamma,d\mu_\gamma)$).
Again, the full Hilbert space ${\cal H}$ of the representation can be described as the
Hilbertian sum ${\cal H}_0 \oplus [ \oplus_n S({\cal H}_1)^{\otimes n}]$. 
Finally, we obtain the following Fourier-type representation of the field operator:
\[
\left(\phi(f)\Psi\right)^{(n)}({\xi_\gamma}_1,\ldots,{\xi_\gamma}_n)=\frac{1}{\sqrt{n}}
\sum_{k=1}^{n}\tilde{f}_{+,\nu}({\xi_\gamma}_k)\Psi^{(n-1)}({\xi_\gamma}_1,\ldots, \hat{{\xi_\gamma}}_{k},\ldots,{\xi_\gamma}_n)
\]
\begin{equation}
+{\sqrt{n+1}}
c_{d,\nu}\int_{\gamma} {\overline{{\tilde{\bar f}}_{+,\nu}}}({\xi_\gamma})
{\Psi}^{(n+1)}({\xi_\gamma},{\xi_\gamma}_1,\ldots,{\xi_\gamma}_n)
d\mu_\gamma(\xi)\label{azzolino}
\end{equation}

\subsection{Representations and their contraction}

In this subsection we briefly discuss the representations of the de 
Sitter group which are associated with  the free de Sitter field. 
We will treat    the ``one-particle'' representation (in ${\cal H}_1$), 
the full representation being constructed by the Fock procedure.
Since representation theory is not our principal aim, many mathematical details will be skipped and we will keep the discussion at a formal level; moreover
we will give  explicit formulae  only for the
two dimensional case, which is notationally much simpler than the general one;
however the results are completely general and mathematically rigorous. 
Let us start by showing the following 

\begin{theorem}
The representation of $G_d$ carried by the free de Sitter-Klein-Gordon field
of mass  $\mu^{2}=(\frac{d-1}{2R})^{2}+\left(\frac{\nu}{R}\right)^{2}$ is the unitary irreducible representation of the principal series characterized by the  value $c = -\mu^2 R^2$ of 
the Casimir operator.
\end{theorem}

 {\bf Proof.}
Let us consider the representation $\alpha_g$ of $G_d$ that we defined in \S 2.2, restricted to the one-particle space of the Borchers algebra 
${\cal B}^{(1)}={\cal D}(X_d)$:
$\alpha_g(f)=f(g^{-1}x),\;g \in G_d$.
By taking the Fourier transform (\ref{fourier}) of this equation we obtain a representation of $G_d$ in  the space ${\cal F}^{s}$ of  complex valued functions defined on  $C^{+}$,  homogeneuous of degree $s$
($h(a\xi) = a^{s}h(\xi),  \;\;a>0$), 
which is given by
 \begin{equation}
(T^{s}(g)\tilde{f}_+)(\xi,s) = \tilde{f}_+(g^{-1}\xi,s).
\label{rep1}
\end{equation}
Associated with each $s\in {{\Bbb C}}$ there is a $G_d$-invariant sesquilinear 
form $\langle \cdot, \cdot \rangle$ given by 
\begin{equation}
\langle f, g \rangle = \int_{\gamma}\tilde{\bar{f}}_{+}(\xi,-s-d+1)\tilde{g}_{+}(\xi,s)\,
d\mu_\gamma(\xi).
\label{sesqui}\end{equation}
In this equation $\gamma$ is any orbital basis of $C^{+}$ and $\mu_\gamma$ is the 
(corresponding) measure assuring $G_d$ invariance.
Eq. (\ref{sesqui}) gives actually a pre-Hilbert product only for certain values of the parameter $s$; in particular for $s=-\frac{d-1}{2}-i\nu$ we get by Theorem \ref{maintheorem}  that the sesquilinear form
(\ref{sesqui}) coincides with the pre-Hilbert product associated with the two-point function ${\cal W}_\nu$. 
Since the measure $\mu_\gamma$ lives on the submanifold $\gamma$ of $C^{+}$,  
one can naturally introduce the  
 operation $Q_\gamma$ \cite{[V]} which consists in taking the restriction to $\gamma$
 of a function $\tilde{f}\in {\cal F}^{s}$:
\begin{equation}
\underline{\tilde{f}}(\xi_\gamma) = (Q^s_\gamma \tilde{f}_+)(\xi_\gamma) = \tilde{f}_+(\xi,s)|_{\xi\in \gamma}.
\end{equation}
This induces a representation $\tilde{U}^{s}$ of $G_d$ on a space of (suitably smooth) complex-valued functions defined on the manifold $\gamma$; this representation is obtained as follows:
\begin{equation}
(\tilde{U}^{s}(g)\underline{\tilde{f}})(\xi_\gamma) =
(Q^s_\gamma T^{s}(g) {Q^s_\gamma}^{-1}\underline{\tilde{f}})(\xi_\gamma). \label{repgamma}
\end{equation}
Concrete realizations are then obtained by specifying a choice of $\gamma$.
In particular, the choice $\gamma=\gamma_0=\{\xi \in C^{+},\, \xi^{(0)}=1\}$
allows us to show (by explicit construction) that the representation $\tilde{U}^{-\frac{d-1}{2}+i\nu}$ extends to the unitary irreducible representation ${U}^{-\frac{d-1}{2}+i\nu}$ of the principal series,  corresponding to the value $c=-(\frac{d-1}{2})^{2}-\nu^{2}$ of the Casimir operator and realized in the Hilbert space ${\cal H}^{(1)}={\rm L}^{2}(\gamma_0,\,d\mu_{\gamma_0})$, where $\mu_{\gamma_0}$ is the rotation invariant measure on  $\gamma_0$. 
   \\  
\begin{remark}{\em     
As one can see, the procedure we used for constructing the representation is 
very similar to the one outlined in \cite{[V]}.
The further information we have added consists in the fact that here the whole construction is obtained  naturally as an output of the corresponding 
quantum field theory. Indeed, we simply  started by considering 
(smooth) functions on 
the hyperboloid $X_d$, together with the left regular representation of the corresponding relativity group. All the further steps, including
the introduction of the relevant spaces of homogeneous complex valued functions defined on the cone $C^{+}$ and the restriction of those functions to a suitable submanifold of ${ C}^{+}$, are imposed by the two-point function of the quantum field theory.
}\end{remark}  
 In an analogous way we can state 
\begin{theorem}
The representation of $G_d$ carried by the free de Sitter-Klein-Gordon field
of mass  $\mu^{2}=(\frac{d-1}{2R})^{2}-\left(\frac{\nu}{R}\right)^{2}$ is the unitary irreducible representation of the complemantary series characterized by the  value $c = -\mu^2 R^2$ of 
the Casimir operator.
\end{theorem}

We now specialize the discussion to the case $d=2$ 
and we briefly show how to construct the principal representations 
and take corresponding contraction.
The generators of the  relativity group, which in this case is 
$G_2=SO_0(1,2)$, are given by

\begin{equation}
L_{01}=\left[
\begin{array}{ccc}
0 & 1 & 0 \\
1 & 0 & 0 \\
0 & 0 & 0
\end{array}
\right] \;\;\;\;\;\;\;
L_{02}=\left[
\begin{array}{ccc}
0 & 0 & 1 \\
0 & 0 & 0 \\
1 & 0 & 0
\end{array}
\right]\;\;\;\;\;\;\;
L_{12}=\left[
\begin{array}{ccc}
0 & 0 & 0 \\
0 & 0 & 1 \\
0 & -1 & 0
\end{array}
\right]
 \label{generators}
\end{equation}
A geodesical obsever 
passing through the origin of our de Sitter space-time sees  
$L_{01}$  as the generator of Lorentz transformations,
$L_{02}$ as the generator of time translations
and $L_{12}$  as the generator of space translations. 
A generic element of $SO_{0}(1,2)$ may be  parametrized in the following way:
\begin{equation}
g(\theta,\psi,\phi) = \exp(\theta L_{12})\exp(\psi L_{02})
\exp(\phi L_{01}) \label{factori}.
\end{equation}
It follows from Eqs. (\ref{generators}) and (\ref{factori}) that $SO_0(1,2)$ 
has the following action on $C^{+}$:  $\xi \to g(\theta,\psi,\phi)^{-1}\xi$
\begin{equation} 
\begin{array}{ccl}
\xi^{(0)} & \to & \xi^{(0)} \cosh \psi \cosh \phi + \xi^{(1)}(\sinh\psi \sin\theta -
\cosh\psi \sinh\phi \cos\theta)+\\
 & & + \xi^{(2)}(-\cosh\psi\sinh\phi\sin\theta -\sinh\psi \cos\theta) \\
 & & \\
\xi^{(1)} & \to & -\xi^{(0)}  \sinh\phi + \xi^{(1)} \cosh\phi \cos\theta + 
\xi^{(2)} \cosh\phi\sin\theta  \\
 & & \\
\xi^{(2)} & \to & -\xi^{(0)} \sinh \psi \cosh \phi + \xi^{(1)}(\sinh\psi \sinh\phi\cos\theta  -\cosh\psi \sin\theta) + \\
 & & +\xi^{(2)}(\sinh\psi\sinh\phi\sin\theta +\cosh\psi \cos\theta)\end{array}
\end{equation}
A concrete realization of the representation (\ref{repgamma}) is obtained by choosing a (complete) curve $\gamma$. 
To compare our results with the representations of 
$SO_0(1,2)$ as they are customarily given in literature we choose the curve 
$\gamma_{0}$;   the associate mapping $Q_{\gamma_0}:{\cal F}^{s}\rightarrow {\cal
C}(\gamma_{0},{{\Bbb C}})$ has  the following form :
\begin{equation}
\underline{\tilde{f}}(e^{i\alpha})=(Q_{\gamma_0}\tilde{f})(e^{i\alpha}) = {\tilde{f}}(\xi^{(0)}=1,\,\xi^{(1)}=\cos\alpha,\,\xi^{(2)}=\sin\alpha). 
\end{equation}
The inverse map (defined on the image of $Q_{\gamma_0}$) is easily obtained as
\begin{equation}
(Q_{\gamma_0}^{-1}\underline{\tilde{f}}) (\xi) =
\tilde{f}(\xi)=
({\xi^{(0)}})^{s}\underline{\tilde{f}}\left(\frac{\xi^{(1)}+i\xi^{(2)}}{\xi^{(0)}}\right).
\end{equation}
As in Eq. (\ref{repgamma}) the use of the map $Q_{\gamma_0}$ gives  a representation $\tilde{U}^{s}$
of $SO_{0}(1,2)$ on ${\cal
C}(\gamma_{0},{{\Bbb C}})$. 
Here are explicit  formulae for the three subgroups associated to the generators (\ref{generators}):
 \begin{equation}
(\tilde{U}^{s}(g(\theta,0,0)){\tilde{f}})(e^{i\alpha})=
\tilde{f}(e^{i(\alpha -\theta)}),
\end{equation}

$(\tilde{U}^{s}(g(0,\psi,0))\underline{\tilde{f}})(e^{i\alpha}) =$
\begin{equation}
 =(\cosh\psi -\sinh\psi \sin\alpha)^{s} 
\tilde{f}
\left(\frac{\cos\alpha +i \cosh\psi \sin\alpha-i\sinh\psi}
{\cosh\psi -\sinh\psi \sin\alpha}\right),
\end{equation}

$(\tilde{U}^{s}(g(0,0,\phi))\underline{\tilde{f}})(e^{i\alpha}) =$
\begin{equation}
=(\cosh\phi -\sinh\phi \cos\alpha)^{s} 
\tilde{f}
\left(\frac{ \cosh\phi\cos\alpha -\sinh\phi +i\sin\alpha}
{\cosh\phi -\sinh\phi \cos\alpha}\right).
\end{equation}
One sees from these formulae that the choice of $\gamma_0$ as integration manifold produces   a  representation of $SO_0(1,2)$ on a functional space in which rotations (the subgroup $SO(2)$) are realized in a natural way. This is consequence of the rotation invariance of the manifold $\gamma_0$.

One can  also  show directly  that   when $s = -\frac{1}{2} + i \rho$  the representation  can  be extended to the unitary irreducible representation of the principal series with Casimir $c=-\frac{1}{4} -\rho^{2}$ realized on the Hilbert space
${\cal H} = {\rm L}^{2}(\gamma_0,\,d\alpha)$.

We now turn to the discussion of the contraction of these representations.  To this aim, it is better to realize them by using another manifold, namely the curve $\gamma_2 =\gamma^{+}_{2} \cup \gamma^{-}_{2} = \{\xi\in C^{+},\;\;\xi^{(2)}=m\} \cup
\{\xi\in C^{+},\;\;\xi^{(2)}=-m\} $.
As we have seen in \S 4.2,  
this manifold has a more direct physical interpretation, and furthermore,
this choice  allows a contraction  of the 
principal series of representations of  the de Sitter group  (onto
the representations of the  Poincar\'e group)  which is very direct 
(compare our formalism with the one presented in \cite{[MN]}) 
since it involves only the representations
of the Poincar\'e
group ${\cal P}^{\uparrow}_{+}(1,1)$:
constructed (as usual in QFT) 
{\em \`a la} Wigner and requires no use of  group-theoretical structures. 

To this end we  define the map  $Q_{\gamma_2}:{\cal F}^{s}\rightarrow {\cal
C}(\gamma_{2},{{\Bbb C}})$:
 \begin{equation}
\underline{\tilde{f}}(k) =\left[
\begin{array}{c}
\underline{\tilde{f}} ^{(+)}(k)\\
\underline{\tilde{f}}^{(-)}(k)
\end{array}
\right] 
=(Q_{\gamma_2}\tilde{f})(k) = 
\left[
\begin{array}{c}
\tilde{f}(\xi^{(0)},\,\xi^{(1)}=k,\,\xi^{(2)}=m)\\
\tilde{f}(\xi^{(0)},\,\xi^{(1)}=-k,\,\xi^{(2)}=-m)
\end{array}
\right]
\end{equation}
where obviously $\xi^{(0)}={k^0}$.
The inverse map (defined on the image of $Q_{\gamma_2}$) is  obtained as
\begin{equation}
(Q_{\gamma_2}^{-1}\underline{\tilde{f}}) (\xi) =
\tilde{f}(\xi)=
\left\{
\begin{array}{l}
(\frac{\xi^{(2)}}{m})^{s}\underline{\tilde{f}}^{(+)}
(\frac{m\xi^{(1)}}{\xi^{(2)}})\\
  (-\frac{\xi^{(2)}}{m})^{s}\underline{\tilde{f}}^{(-)} (\frac{m\xi^{(1)}}{\xi^{(2)}})
\end{array}
\right.
\begin{array}{c}
{\rm if}\;\; \xi^{(2)}>0 \\
{\rm if} \;\; \xi^{(2)}<0
\end{array}.
\end{equation}
 The map $Q_{\gamma_2}$ gives an alternative realization of the representation $\tilde{U}^{s}$
of $SO_{0}(1,2)$, which lives  on the space ${\cal
C}(\gamma_{2},{{\Bbb C}})$; as before, when $s = \frac{1}{2}+i\nu$,   $\tilde{U}^{s}$ extends to a unitary irreducible representation, which is now  carried by the Hilbert space ${\cal H} = {\rm L}^{2}(\gamma_2^{+},\frac{dk}{{k^0}})\oplus  {\rm L}^{2}(\gamma_2^{-},\frac{dk}{{k^0}})$. The three previous subgroups of $SO_0(1,2)$ are now represented as follows:

$({U}^{s}(g(\theta,0,0))\underline{\tilde{f}})(k) =(m)^{-s}|k\sin\theta + m\cos\theta|^{s}
$\begin{equation}
\left[
\begin{array}{l}
Y^{(+)}_\theta(k)\underline{\tilde{f}}^{(+)}\left(\frac{k\cos\theta
- m\sin\theta}{m\cos\theta+k\sin\theta}\right)
+
Y^{(-)}_\theta(k)\underline{\tilde{f}}^{(-)}\left(\frac{k\cos\theta
-m \sin\theta}{m\cos\theta+k\sin\theta}\right)\\
Y^{(-)}_\theta(k)\underline{\tilde{f}}^{(+)}\left(\frac{k\cos\theta
- m\sin\theta}{m\cos\theta+k\sin\theta}\right)
+
Y^{(+)}_\theta(k)\underline{\tilde{f}}^{(-)}\left(\frac{k\cos\theta
- m\sin\theta}{m\cos\theta+k\sin\theta}\right)
\end{array}
\right] ,\label{sptr1}
\end{equation}
where $Y^{\pm}_\theta(k)=H(\pm(k\sin\theta + m\cos\theta))$ and $H$ denotes the Heaviside function; 

$({U}^{s}(g(0,\psi,0))\underline{\tilde{f}})(k) =m^{-s}|{k^0}\sinh\psi - m\cosh\psi|^s$
\begin{equation}
\left[\begin{array}{l} J^{(+)}_{\psi}(k)\underline{\tilde{f}}^{(+)}\left(\frac{m k}{m\cosh\psi-{k^0}\sinh\psi}\right)
+
J^{(-)}_{\psi}(k)\underline{\tilde{f}}^{(-)}\left(\frac{m k}{m\cosh\psi-{k^0}\sinh\psi}\right)\\
J^{(-)}_{\psi}(k)\underline{\tilde{f}}^{(+)}\left(\frac{m k}{m\cosh\psi-{k^0}\sinh\psi}\right)
+
J^{(+)}_{\psi}(k)\underline{\tilde{f}}^{(-)}\left(\frac{m k}{m\cosh\psi-{k^0}\sinh\psi}\right)
 \end{array}
\right] \label{ttr1}
\end{equation}
where $J^{\pm}_{\psi}(k) = H(\pm(m\cosh\psi-{k^0}\sinh\psi))$; 
\begin{equation} (\tilde{U}^{s}(g(0,0,\phi))\underline{\tilde{f}})(k) = \left[
\begin{array}{l}
\underline{\tilde{f}}^{(+)}(\cosh \phi\, k-\sinh\phi\, {k^0})\\
\underline{\tilde{f}}^{(-)}(\cosh \,\phi k+\sinh\phi\, {k^0})
\end{array}
\right]\label{Lorentz}\end{equation}
 \begin{remark}{\em     
One  sees that now it is the representation of the Lorentz subgroup $g(0,0,\phi)$ which is realized naturally, and this is due to the Lorentz invariance of the manifold $\gamma_2$.
}\end{remark}

To discuss the contraction we also need the generators of  ${\cal P}^{\uparrow}_{+}(1,1)$:

 \begin{equation}
{\cal L}_{01}=\left[
\begin{array}{ccc}
0 & 0 & 0 \\
0 & 0 & 1 \\
0 & 1 & 0
\end{array}
\right],\;\;\;\ {\cal L}_{02}=\left[
\begin{array}{ccc}
0 & 0 & 0 \\
1 & 0 & 0 \\
0 & 0 & 0
\end{array}
\right],\;\;\;\;  {\cal L}_{12}=\left[
\begin{array}{ccc}
0 & 0 & 0 \\
0 & 0 & 0 \\
1 & 0 & 0
\end{array}
\right].\;\;\;\;\ \end{equation}
A generic element of ${\cal
P}^{\uparrow}_{+}(1,1)$
 may be  parametrized as
\begin{equation}
{\sl g}({a}, a_{0},\phi) =\exp ({a}
 {\cal L}_{12})  \exp (a_{0} {\cal L}_{02}) \exp\phi ({\cal L}_{01})
\end{equation}
The group contraction map is defined in the following way:
\begin{equation}
f_{\epsilon}:SO_{0}(1,2) \rightarrow {\cal P}^{\uparrow}_{+}(1,1),\;\;
f_{\epsilon}( g(\theta,\psi,\phi)) =
{\sl g}(\frac{\theta}{\epsilon},\frac{\psi}{\epsilon},\phi)
\label{function}
\end{equation}
When the point $({a},a_{0},\phi)$ belongs to a suitable ($\epsilon$
depending) neighbourhood of $(0,0,0)$, we can inverte this  map: 
\begin{equation}
f^{-1}_{\epsilon}({\sl g}({a},a_{0},\phi)) =
 g(\epsilon {a},\epsilon a_{0},\phi)
\end{equation}
${\cal P}^{\uparrow}_{+}(1,1)$ is then  a contraction of $SO_{0}(1,2)$
because
\begin{equation}
\lim_{\epsilon \rightarrow 0} f_{\epsilon}(g(\epsilon {a},\epsilon
a_{0},\phi)
g(\epsilon {b},\epsilon b_{0},\eta)) =
{\sl g}({a},a_{0},\phi){\sl g}({b},b_{0},\eta)
\end{equation}
For the reader's convenience we remind the definition of contraction of a family of representations.
Let $G$ be a Lie group and let $G^{c}$ its contraction.
Consider a one-parameter family 
${\cal D}_{\epsilon}(g),\;\;\epsilon\in(0,1]$, of
continuous and unitary representations of $G$ on Hilbert spaces ${\cal
H}_{\epsilon}$. Furthermore consider a family of unitary maps
$U_{\epsilon}:{\cal H}_{\epsilon} \rightarrow {\cal H}
$ such that for each $\Psi \in \cal H$ there exists $\bar{\epsilon}$ for
which if
$\epsilon <\bar{\epsilon}$ then $\Psi \in \makebox{Ran}U_{\epsilon}$.
 If the limit
\begin{equation}
\lim_{\epsilon \rightarrow 0} U_{\epsilon}{\cal
D}_{\epsilon}(f^{-1}_{\epsilon}{\sl g}) U^{-1}_{\epsilon}\Psi \equiv {\cal
D}({\sl g})\Psi \label{limit}
\end{equation}
exists for each $\Psi \in \cal H$ and defines a continuous unitary
representation of $G^{c}$, then the representation $\cal D$ of  $G^{c}$ is
said
a contraction of the representation ${\cal D}_{\epsilon=1}$ of $G$.
In our concrete case we have that
\begin{equation}
{\cal H}_{\epsilon} ={\cal H} =
L^{2}(\gamma_{2},d\mu_{\gamma_2}),
\;\;\;\;\;\;U_{\epsilon}={\bf 1},\;\;\;\;\;{\cal D}_{\epsilon}(g) =
U^{s(\epsilon)}(g) \end{equation}
and $f_{\epsilon}$ as defined in Eq. (\ref{function}).
Our aim is to prove that if $s(\epsilon) = -\frac{1}{2} +
i\frac{m}{ \epsilon} $, $\epsilon = R^{-1}$,
the limit in Eq. (\ref{limit}) exists and that \begin{equation}
{\cal D}({\sl g})\Psi = [{\cal D}^{+}_{m}\oplus {\cal D}^{-}_{m}]({\sl
g})\Psi. \end{equation}

\begin{remark}{\em     
There are two   things in our construction which render the contraction easier:
first, the representations of $SO_0(1,2)$ and  the representation ${\cal D}^{+}_{m}\oplus {\cal D}^{-}_{m}$ of the Poincar\'e group  (which will be obtained by contraction) are realized in the same functional space.  
Secondly, one sees in Eq. (\ref{function}) that the Lorentz subgroup stays
untouched in the contraction. Our construction 
provides a representation of the Lorentz subgroup which   {\em has the same functional form}
for the two groups.
}\end{remark}

It follows from the previous remark 
that for the Lorentz subgroup there is actually no limit to take. Let us examine the space translations subgroup. We prove pointwise convergence assuming all the continuity properties needed.
By setting $\theta={a}/R$ in Eq. (\ref{sptr1}) we then obtain that
\begin{equation}
\lim_{R\rightarrow\infty}(U^{-\frac{1}{2}+i{mR}}g(a/R,0,0))\underline{\tilde{f}})(k)=
\left[
\begin{array}{l}
e^{ika}\underline{\tilde{f}}^{(+)}(k)\\
e^{ika}\underline{\tilde{f}}^{(-)}(k)\\
\end{array}
\right].
\end{equation}
which corresponds exactly to the desired
representation of the Minkowskian space translations.
In exactly the same way we may show that given a  generic ${\sl g} \in
{\cal P}^{\uparrow}_{+}(1,1)$ we obtain the following result:
\[
\lim_{R\rightarrow \infty}
(U^{-\frac{1}{2}+i{mR} }
(f^{-1}_{\frac{1}{mR}}({a},a_{0},\chi))\underline{\tilde{f}})({ k})\;\;=\;\; \]
 \begin{equation}=([{\cal D}^{+}_{m}\oplus {\cal D}^{-}_{m}]\;({
a},a_{0},\chi)\;\underline{\tilde{f}})({k})
\end{equation}
Thus, the unitary representations of the principal series of
the de Sitter group are contracted pointwise to a direct sum of
Wigner
representations of the Poincar\'e group.Furthermore, it can
also be shown that the contraction is obtained  also in the Hilbert space
topology, i.e.
\begin{equation}
\lim_{\epsilon \rightarrow 0}
\|U^{s(1/R)}(
(f^{-1}_{1/R}({a},a_{0},\chi))\underline{\tilde{f}}\;-\;
[{\cal D}^{+}_{m}\oplus {\cal D}^{-}_{m}]\;({ a},a_{0},\chi)\;)\underline{\tilde{f}}
\|_{{\cal H}} = 0 \end{equation}
but the complete proof will not be reproduced here (see e.g. \cite{[MN]}).

\section{Complements to the theory of two-point functions and of generalized free fields. The Reeh-Schlieder property}

\subsection{End of the proof of the K\"allen-Lehmann principal representation}
We begin this section  by completing  the proof of the  K\"allen-Lehmann principal representation we initiated in Section 3, by showing the positivity of the weight $ \rho( \nu) $ 
for general two-point functions. For the sake of simplicity 
we carry out detailed computations in the two-dimensional case but the general case can be easily recovered. 
We use here the parametrization of $ X_d $ given in Eq. (\ref{para1}):
$x=x(\psi,\theta) $, 
($ x^{(0)}= \, \sinh  \psi  $,
$ x^{(1)}=\, \cosh  \psi \,  \sin \theta  $, 
$ x^{(2)} = \, \cosh \psi \, \cos  \theta  $) where we set $R=1$ without loss of generality.
In the following, a special role will be played by the class of rotation invariant functions.
If we put  $\f(\psi,\theta) \equiv f(x(\psi,\theta))$,  rotation invariance of a function $f$ is expressed by $\f=\f(\psi)$.
Of course the Fourier transform of ${f}$ will also be rotation invariant.
Indeed, by parameterizing 
the vector $\xi$  in the following way: 
$\xi$=($\xi^{(0)}$, 
$\xi^{(1)}= \xi^{(0)}\cos\gamma$, $ \xi^{(2)}= \xi^{(0)}\sin\gamma$) one sees that   
\begin{equation}
\tilde{f}_{+}(\xi,s) =\left(\xi^{(0)}\right)^{s} \int_{-\infty}^{\infty}
\int_{0}^{2\pi}[\sinh \psi -\cosh \psi \sin(\theta + \gamma)]^{s}\f(\psi)d\theta \cosh \psi d\psi
\label{trasfoy}
\end{equation}
where the angle $\psi$ has a small positive imaginary part, and one sees that the  integral does not depend on $\gamma$ (analogously for $\tilde{f}_{-}(\xi,s))$. 
To  compute explicitly the Fourier transform (\ref{trasfoy}) we change the variables as follows:
\begin{equation}
\left\{
\begin{array}{ccl}
 x^{(0)} & = & \sinh \psi\\
x^{(1)} & = &   e^{t/2}\sqrt{2(\sinh\psi-\sinh t)}\\
x^{(2)} & = &  e^{t}-\sinh \psi\\
\end{array}
\right.\;\;\;\;t\in {{\Bbb R}},\;\;\psi>t.
\end{equation}
This   parametrization covers only one quarter of $X_2$. The remaining parts
are obtained similarly. It follows that  
\begin{equation}
\tilde{f}_{\pm,\nu}(\xi )= 2\left(\xi^{(0)}\right)^{{-\frac{1}{2} - i \nu}}
\int_{-\infty}^{\infty}e^{ - i \nu t} 
\int_{t}^{\infty} \frac{\f(\psi)  +e^{\pm i\pi(-\frac{1}{2} - i \nu)}\f(-\psi)}{\sqrt{2(\sinh \psi -\sinh t)}} \cosh\psi d\psi;
\end{equation}
thus  the Fourier transform (\ref{trasfoy}) of a rotation invariant function is  written in these variables as a  Weyl transform followed by an ordinary Fourier transform.
\begin{proposition} 
Let  $  f(x(\psi,\theta))= \f(\psi,\theta)  $ be a test-function on $X_2$,
with analyticity properties in $ \psi $, 
namely $\underline{f}(\psi,\theta) = \lim_{\Im \psi \to 0}\underline{f}(\psi ,\theta)$, where 
 $\underline{f}(\psi ,\theta)$ is analytic in the strip\footnote{note that the domain of $ f $ is contained in $ {\cal T}^+ $.} 
$ -\pi R < {\Im {\psi}} < 0, $ 
with the consistency condition:
\begin{equation}
\underline{f}(\psi-i\pi,\theta)  = \underline{f}(-\psi,\theta + \pi)  \label{(2)} 
\end{equation}
Then $\tilde{{f}}_{-,\nu}(\xi) $  { vanishes.}\label{tu1}
\end{proposition}
 {\bf Proof.}  Write the Fourier transform $\tilde{{f}}_{-,\nu}(\xi) $  as an integral in $
\psi $ 
and $ \theta $. At fixed $ \theta  $ the integration cycle in $ \psi $ can be
distorted from $ {\Bbb R} $ to $ {{\Bbb R}} -i\pi , $ 
this integral is then identified (in view of Eq. (\ref{(2)})) with {\em minus} the 
corresponding integral at $ \theta + \pi $, which implies the vanishing of  $\tilde{{f}}_{-,\nu}(\xi) $ and proves the statement. 
   \\

Thus if $ \f(\psi) $ is analytic in the strip $\{-\pi < {\Im}  \psi  < 0\} $ 
and satisfies the condition $ \f(\psi -i\pi)  = \f(-\psi) $ then $ 
\tilde{ {f}}_{-,\nu}(\xi ) =  0 $. This gives that 
\begin{equation}
\tilde{ {f}}_{+,\nu}(\xi ) = 
4e^{\pi\nu}\cosh\pi\nu  \,\xi^{ -{1 \over 2}-i\nu}_0
\int^{ +\infty}_{ -\infty} {e}^{-i\nu t} {d}
t \int^{ +\infty}_{  \sinh \ t}{\f(\psi) {d}   \sinh  \psi \over[ 2(
\sinh \ \psi -  \sinh \ t)]^{1/2}}\ . \label{5.5}
\end{equation}
We shall now make use of the Legendre functions as analytic test-functions.

\noindent Let
$ \underline{h}_{\nu_ 0}(\psi)  = P_{-{1 \over 2}+i\nu_ 0}
\left( {\cosh} \left(\psi +{i\pi \over 2} \right) \right)\ . $
The function 
$ \h_{\nu_ 0} $ is { analytic in the strip }$ \{-\pi  < {\Im} \ \psi  < 0\}
$ and moreover $ \h_{\nu_ 0}(\psi -i\pi) =\h_{\nu_ 0}(-\psi) $ 
(since $ P_s(  {\cosh v}) $ is analytic for $ \Re\cosh v \geq  0
$ and $ P_s \left( {\cosh} \left(\psi +{i\pi \over 2}-i\pi \right)
\right) = P_s \left( {\cosh} \left(-\psi +{i\pi \over 2} \right)
\right). $
$ \h_{\nu_0}(\psi) $ therefore satisfies the above property and therefore $  \widetilde{h_{\nu_0}}_{-}(\xi ,-\frac{1}{2} -i\nu)= 0 $.
Eq. (\ref{5.5}) applies and yields
\[\widetilde{h_{\nu_0}}_{+}\left(\xi ,-\frac{1}{2} -i\nu\right) =\] 
\begin{equation} 4{e}^{\pi \nu} {\cosh} \pi \nu (\xi_0)^{-\frac{1}{2} -i\nu }
\int^{ +\infty}_{ -\infty} {e}^{-i\nu t} {d} t   
\int^{\infty}_{  \sinh  t}   
\frac{ P_{-{1 \over 2}-i\nu_ 0}(i  \sinh  \psi)}{[ 2(  \sinh \psi -  \sinh t)]^{1/2}}d \sinh \psi
\label{tu2}
\end{equation}
To compute the last integral 
we use the following representation of  $ P_{-{1 \over 2}-i\nu_ 0}$ (=$ P_{-{1 \over 2}-i\nu_ 0} $): 
\begin{equation}
P_{-{1 \over 2}-i\nu_ 0}(i  \sinh  \psi)  = {2 \over \pi}  {e}^{-{i\pi
\over 4}} {\coth} \pi \nu_ 0 \int^{ \infty}_ \psi{ {\sin} \nu_ 0
\left(t+{i\pi \over 2} \right) \over[ 2(  \sinh  t-  \sinh  \psi)]^{
1/2}} {d} t
\label{apdamp}
\end{equation}
Eq. (\ref{apdamp})
can thus be identified with the inversion of the  Weyl transform\footnote{Under the appropriate condition, this inversion is given by the following formulae: if the Weyl transform of a function $F(\sinh\psi)$ is written as 
\begin{equation} G(  \sinh t) = \int^{ \infty}_{  \sinh  t}{F(  \sinh
 \psi)  {d\sinh}  \psi \over[ 2(  \sinh  \psi -  \sinh t)]^{1/2}}
\end{equation} 
{then} \begin{equation}F(  \sinh  \psi) = {1 \over \pi}  \int^{ \infty}_ \psi{G(
 \sinh  t) {d} t \over[ 2(  \sinh t-  \sinh  \psi)]^{ 1/2}} \label{a466}
\end{equation}}
occurring at the l.h.s. of Eq. (\ref{tu2}) which allows us to compute the latter; we thus obtain 
\[
\tilde h_{\nu_ 0,+}\left(\xi ,-\frac{1}{2} - i\nu\right)  = 
\]
\begin{equation}
+8\pi \ {e}^{-{3i\pi \over 4}} {
{\cosh}^2 \pi \nu_ 0 \over  \sinh  \pi \nu_ 0} \left[ \left(\xi_ 0
\right)^{-{1 \over 2}-i\nu_ 0} {e}  ^{{\pi \nu_ 0 \over 2}}\delta 
\left(\nu -\nu_ 0 \right)- \left(\xi_ 0 \right)^{-{1 \over 2}+i\nu_ 0} 
{e}^{-{\pi \nu_ 0 \over 2}}\delta \left(\nu +\nu_ 0 \right) \right]
\label{aiuto}
\end{equation}
 
In view of the results of \S 4.2, Formula (\ref{a325}) can be interpreted as a K\"allen-Lehmann representation expressing the two-point functions of a certain class of fields (principal-type) in terms of the two-point 
functions of all the massive Klein-Gordon fields ${\cal W}_\nu$, $\nu \geq 0.$
We now complete the properties of this integral representation by proving

\begin{theorem}
Assume that a de Sitter quantum field has a two-point function of the form 
$ {\cal W} \left(x,x^{\prime} \right) = \int^{\infty }_{0 } \rho( \nu)  {d} \nu
\ {\cal W}_\nu \left(x,x^{\prime} \right) $
and satisfying the positivity condition a), namely
\begin{equation}
\int^{ }_{ X_2\times X_2}{\cal W} \left(x,x^{\prime} \right) \bar f(x)
f (x^{\prime} ) {d}\sigma(x) {d}\sigma( x^{\prime})  \geq  0\ .
\end{equation}
Then $\rho$ is a positive measure.
\end{theorem}
 {\bf Proof.}
Let $ \chi $ be any even
function in $ {\cal D} \left({{\Bbb R}}\right) $, and
let us choose the rotation invariant function 
\begin{equation}
 \underline{f}( \psi)  = \int^{+\infty }_{-\infty }|\nu_0|^{\frac{1}{2}}
 \chi \left(\nu_ 0 \right) P_{-{1
\over 2}+i\nu_ 0}(i \sinh \psi) {d} \nu_ 0
\end{equation}
whose Fourier transform is 
\begin{equation}
\tilde{f}_{+,\nu}(\xi) = \int_{-\infty}^{\infty}
|\nu_0|^{\frac{1}{2}}
\chi \left(\nu_ 0 \right) \underline{\tilde{h}}_{\nu_0 +}(\xi, {-{1
\over 2}+i\nu_ 0})  {d} \nu_ 0.
\end{equation}
Eq. (\ref{aiuto}) then gives, for $\nu>0$: 
\begin{equation} 
{ \tilde f_ +(\xi ,\nu) \vert_{\xi_0=1} 
 = 16\pi e^{-\frac{3}{4}\pi i} \nu^{\frac{1}{2}}
 \;{ {\cosh}^2\pi \nu \over  \sinh\pi \nu}
\chi(\nu) {e}^{\frac{\pi \nu}{2}}}.
\end{equation}
Since
$ \int{\cal W}_\nu(x,x')\bar f(x) f(x')dxdx' =  c_{2,\nu}\int^{ }_{ \gamma_0}  \vert\tilde{f}_{+,\nu}(\xi) \vert^ 2 d\mu_{\gamma_0} \ ,
$
we then  have in view of Eq. (\ref{consta}): 
\begin{equation}
\int_{ X_2\times X_2}{\cal W}_\nu \left(x,x^{\prime} \right) \bar f( x)
f (x^{\prime} ) {d} x\ {d} x^{\prime}   = { 64\pi \nu^2
\vert\Gamma\left(\frac{1}{2} + i\nu\right)\vert^2
{\cosh}^4\pi \nu
 \over  \sinh^2\pi \nu} \chi^ 2(\nu).
\end{equation}
Therefore, 
\[\int_{ X_2\times X_2} {\cal W}(x,x^{\prime}) \bar{ f}(x)
f (x^{\prime} ) {d} x {d} x^{\prime}= \]
\begin{equation}
=64\pi \int_{0}^{\infty} 
\nu^2
\rho( \nu)  \frac{{\cosh}^4\pi \nu}{  \sinh^2\pi \nu}  \chi^ 2(\nu) 
\vert\Gamma\left(\frac{1}{2} + i\nu\right)\vert^2 d\nu
 \geq  0. \end{equation}
Since 
our freedom of choice on $\chi$ allows 
$\chi^2\vert_{\overline{{\Bbb R}^+}}$ to vary in a set which is dense 
in the space of continuous functions with compact support in 
$\overline{{\Bbb R}^+}$,
 it follows that  $ {\rho} $ {is} a {positive} {measure.}
   \\  
\begin{remark}{\em     

{\em The role of locality for general two-point functions on de Sitter space-time.} 

It is known that in Minkowski space the full analytic structure of general two-point functions can be obtained for hermitian scalar fields from Poincar\'e invariance, spectral condition and positivity, {\em without using locality}.
The argument goes through a proof of the K\"allen-Lehmann formula based purely on a spectral decomposition of the Wightman function (in ``intermediate states'') with respect to the energy-momentum variables.
In the present case of de Sitter QFT, there does not exist at the moment a general proof of the K\"allen-Lehmann formula (\ref{a325}) (when it holds) on the basis of an $SO_0(1,d)$-decomposition of the representation carried by the Hilbert space ${\cal H}_1$ of the two-point function considered.
At any rate, the use of locality  in Proposition 2.2 is necessary if one deals with fields whose behaviour at infinity only allows one to write a representation of the form (\ref{a321}) (but not (\ref{a325})) for their two-point function, as it is the case for those considered below in \S 5.4.
}\end{remark}

\subsection{    Minkowski GFF's as a contraction of de Sitter GFF's }
In this paragraph we show how to recover Minkowski GFF's by the knowledge of corresponding de Sitter QFT by taking the limit $R\to \infty$ (contraction).
To this end, only the principal series is relavant since the complementary series tend to disappear when $r$ is very large. Thus it is legitimate to use the K\"allen-Lehmann representation (\ref{a325}).

 As a first step we show that the limit of the Fourier representation given in Eq. (\ref{wig1}) is the corresponding well-known Fourier representation of two-point functions for Klein-Gordon fields.
Indeed  by  taking as integration manifold  
$\gamma_{d}$ we see that when $\nu=mR$ is very large, the dominant term in (\ref{wig4}) is given by
\begin{equation}
{\cal W}_{mR}(x_1,x_2)\simeq
c_{d,mR}\,e^{2\pi mR}\int_{\gamma^{+}_d}
(x_1\cdot \xi_\gamma)^{-\frac{d-1}{2}+imR}_{-}   
(x_2\cdot \xi_\gamma)^{-\frac{d-1}{2}-imR}_{-}
d\mu_{\gamma^{+}_{d}}(\xi), \label{wig5}
\end{equation}
the other terms being multiplied by the factor $\exp(-\pi m R)$ (see Eqs. (\ref{consta}) and (\ref{wig4})).
By inserting in the latter  equation the parametrization  (\ref{para1}),  we then obtain in view of Eq. (\ref{parcondicio}) that
\[
\lim_{R\to \infty} {\cal W}_{{mR}}(x_R({\rm x_1}),x_R({\rm x}_2)) =
-iD^{(-)}_m
({\rm x}_1,{\rm x}_2)=\]
\begin{equation}
=\frac{1}{2(2\pi)^{d-1}}\int e^{-i {\rm k x}_1}
e^{i {\rm k x}_2}\delta({\rm k}^2 - m^2)d{{\rm k}}. \label{conttrac}
\end{equation}
Similarly one can obtain  
each minkowskian GFF as the limit of a family of 
GFF's $ \phi^{(\rho)}_ R(x) $ on the de Sitter universes $ X_d(R) $, 
when the 
radius $ R $ of the latter tends to infinity. We can state:

\begin{proposition} For any K\"allen-Lehmann weight $ \rho( m) $
positive on $ {{\Bbb R}}^+, $ 
there exists a  family of generalized free fields $ \left\{
\left(\phi^{( \rho)}_ R(x);\ {\cal H}_R,\Omega_ R \right);R>0 \right\} , $ 
each field $ \phi^{( \rho)}_ R(x) $ being defined on the corresponding de
Sitter universe $ X_d(R), $ 
which satisfy the following properties:

a) The two-point function $ \underline{\rm W}_R(z) $ of the fields $ \phi^{(
\rho)}_ R $ is given by the following 
integral representation:
\begin{equation}
\underline{\rm W}_R(z)= \int^{\infty }_{0 } \rho( m) C_{d,mR} 
P^{(d+1)}_{-{d-1 \over 2}-im R} \left({z\cdot z_0 \over R^2}
\right)   {d}m 
\label{a328}\end{equation}

b) For each point $ {\rm z}   $ in $ {{\Bbb C}}^d \setminus \left\{
{\rm z}; \;{\rm z}^2  \geq 0 \right\} $ the following limit holds:
\begin{equation}
\lim_{R \longrightarrow +\infty}  \underline{\rm W}_R \left(z_R({\rm z}) \right) =
\underline{\rm W}_M({\rm z})\  \end{equation}
where $ \underline{\rm W}_M({\rm z}) $ is the two-point function of the minkowskian
generalized free 
field $ \left(\phi^{( \rho)}_ M({\rm x}),{\cal H}_M,\Omega_ M \right) $ on $ {\Bbb
R}^d, $ characterized by the K\"allen-Lehmann weight 
$ \rho( m), $ namely:
\begin{equation}
\underline{\rm W}_M({\rm z}) = -i\int^{ \infty}_{0 } \rho(m) 
D^{(-)}_m({\rm z}) {d}m 
\end{equation}
\end{proposition}
  {\bf Proof.}
In view of theorem 4.1 every two-point function of the form (\ref{a328}) with $\rho \geq 0$ satisfies conditions a), b), c), d),  and therefore defines a unique
de Sitter GFF on $X_d(R)$, which proves property a).  
Property b) follows from Eq. (\ref{conttrac}).
   \\  
\begin{remark}{\em     
We notice that the unitary representation of the Poincar\'e group constructed by means of the distribution $-iD^{-}_m({\rm x_1-x_2})$ is actually {\em irreducible}; namely, it is the Wigner representation of mass m and spin zero. 
On the other side, as we have seen in \S 4.4, the 
representations obtained as contraction of the unitary representations of the de Sitter group {\em are not irreducible} \cite{[MN]}. 
{\em Therefore the group-theoretical contraction is not a suitable 
procedure to draw conclusion on the structure of the limiting 
quantum field theory.}
}\end{remark}  

\begin{remark}{\em     

We can complement Proposition 5.2 by applying the analysis of Sections 2
and 3 to the two-point function
${\rm{w}}_R(z^{(d)}/R) =\underline{\rm{W}}_R(z)$ on $X_d(R)$,
to the corresponding propagators 
${r}_R(x^{(d)}/R) =\underline{\cal{R}}_R(x)$, and to their Laplace
transform ${\rm G}_R(\nu)$ (obtained from $r_R(\cosh v)$ via Eq. (\ref{34}).
First of all, it can be checked that the weight function 
$\rho_0(\nu) \equiv \rho_{0,R}(\nu)$ of Theorem 3.1 corresponding to the
field $\phi^{(\rho)}_R$ is given by 
\begin{equation}
\rho_{0,R}(\nu) = \pi\rho(\nu/R)/R^d \label{5.22}
\end{equation} 
since, by putting 
$\nu=mR$ in 
Eq. (\ref{a325}), the latter can be identified with Eq. (\ref{a328}).
One can then see that there exists an $R$-independent  
{\it normalized Laplace transform} $ {\rm G} $ 
which is common to all the fields 
$\phi^{(\rho)}_R$
(including 
the minkowskian limit 
$\phi^{(\rho)}_\infty$,
G being equal to the standard minkowskian Laplace transform of the
corresponding retarded propagator).
To this purpose, it is relevant to associate with the space
time-variables $x$  the complex variable
$\kappa = \nu/R$ which carries a {\it mass dimension} and to define 
the {\it normalized Laplace transform} $ {\rm G}(\kappa) $ 
of ${r}_R(x^{(d)}/R)$,
by
the scaling equation:
\begin{equation}
{\rm G}(\kappa) = {\rm G}_R(\kappa R) R^{d-1}/\pi \label{5.23}
\end{equation}
It can be checked that the function ${\rm G}(\kappa)$, analytic 
in the half-plane $\Im \kappa >0$, 
is uniquely determined from $\rho(m)$ by the fact that its boundary
value
on ${\Bbb R}$
satisfies the following relation
(implied by Eqs. (\ref{a326}),(\ref{5.22}) and (\ref{5.23})):
\begin{equation}
i\rho(m) = m [G(m)-G(-m)]
\end{equation} 
As a matter of fact, for
each mass $m$ the  
corresponding de Sitter-Klein-Gordon field on $X_d(R)$ (solution of Eq.
(\ref{kg}), with $\mu^2=\left(\frac{d-1}{2R}\right)^2 + m^2)$  
admits  the following normalized Laplace transform:
${\rm G}^{(m)}(\kappa) = \frac{1}{2m}\left(\frac{1}{\kappa-m}-
\frac{1}{\kappa+m}\right)$.
The K\"allen-Lehmann formula written for the Laplace transform of the propagators therefore admits the same interpretation as that of the minkowskian case, namely it is equivalent to the Cauchy representation 
${\rm G}(\kappa) = -\frac{1}{\pi} \int_{0}^{\infty} \frac{\rho(m) dm}{\kappa^2-m^2}$.
}\end{remark}  
\subsection{The Reeh-Schlieder property}
One simple but surprising theorem
in local quantum field theory is the Reeh and Schlieder property \cite{[SW]}. 
We will show that an analogous theorem still holds for de Sitter generalized 
free fields. In a sense this could 
be even more surprising, 
due to the presence of horizons in this model of space-time. In the following we give a detailed analysis of the situation for the principal free and generalized free fields. A similar analysis could be performed for the complementary fields.
\begin{proposition}
 Let $\phi$ be the free de Sitter-Klein-Gordon quantum field of parameter $\nu$. The vector valued
distribution
$\phi(x)\Omega$
is the boundary value of a vector-valued function 
$\Phi^{(1)}(z)$
 strongly analytic in the tuboid ${\cal T}^{+}$.
\label{pr9}
\end{proposition}
 {\bf Proof.}
Let us consider  a vector $\Psi^{(1)}\in{\rm D}^{(1)}$ where ${\rm D^{(1)}}$ is the dense
subset of ${\cal H}^{(1)}$ obtained by applying the  fields $\phi(f)$, $f\in 
{\cal D}(X)$  to
the "vacuum" $\Omega$;  we can then put 
$
\Psi^{(1)}_f=\phi(f)\Omega.
$
The analyticity properties of the two-point function then imply that 
 the correlation function between $\Psi^{(1)}_f$ and the "vacuum"
\begin{equation}
{\cal W}_{\Psi^{(1)}_f}(x)=\langle\Psi^{(1)}_f,\phi(x)
\Omega\rangle
\label{npsi}
\end{equation}
is the boundary value of a function ${\rm W}_{\Psi^{(1)}_f}(z)$
which is analytic in the
tuboid ${\cal T}^{+}$, i.e. 
\begin{equation}
\langle \phi(f) \Psi_0, \phi(x) \Psi_0\rangle
= {\cal W}(\bar{f}, x) = b.v.{\rm W}(\bar{f}, z). 
\end{equation}
Using the Fourier representation (\ref{wig1}) for the two-point function we deduce 
the following majorization:
\[
|{\rm W}(\bar{f}, z)| =c_{d,\nu} \left|\int_{\gamma}
\bar{\tilde{f_\nu}}(\xi)(\xi \cdot z)^{-\frac{d-1}{2} -i\nu} d\mu_\gamma \right|\leq
\]
\begin{equation}
\leq c_{d,\nu}
\left(\int_{\gamma} |(\xi_\gamma \cdot z)|^{(1-d)}d\mu_\gamma \right)^{\frac{1}{2}}
\left(\int_{\gamma}|{\tilde{f_\nu}}(\xi_\gamma )|^2  d\mu_\gamma \right)^{\frac{1}{2}}
 \leq 
c(z) \|\phi(f)\Psi_0\|.
\label{maggiorazione}
\end{equation}
Since the set ${\rm D}^{(1)}$ is dense in ${\cal H}^{(1)}$, the
majorization (\ref{maggiorazione}) and the Riesz lemma imply that there 
exists a vector 
$\Phi^{(1)}(z)\in {\cal H}^{(1)}$ such that 
\begin{equation}
{\rm W}_{\Psi^{(1)}_f}(z)={\rm W}(\bar{f}, z) =\langle \Psi^{(1)},
\Phi^{(1)}(z)\rangle\;\;\;\;\forall \Psi^{(1)}_f \in {\rm D}^{(1)}
\end{equation}
Given now  $\Psi^{(1)} \in {\cal H}^{(1)}$, we may choose a sequence 
$\{\Psi^{(1)}_{k},\,k\in {\Bbb N}\}, \Psi^{(1)}_{k} \in {\rm D}^{(1)}$, 
such that 
$\Psi^{(1)}_{k}\rightarrow\Psi^{(1)}$ in ${\cal H}^{(1)}$.
Associated with the sequence $\{\Psi^{(1)}_{k}\}$,  there is a sequence
$\{ {\rm W}_{\Psi^{(1)}_{k}}(z)\} $ of functions  analytic in ${\cal T}^{+}$. 
By applying again the inequality (\ref{maggiorazione}) we obtain that 
the functions $ {\rm W}_{\Psi^{(1)}_{k}}(z) $ satisfy a uniform bound on each compact 
subset of  ${\cal T}^{+}$; moreover, the continuity of the Hilbert product implies that, for each $z\in {\cal T}^{+}$                                                                                                                                                                                                                                                                                                                                                                                                                                                                                                                                                                                                                                                                                                                                                                                                                                                                                                      \begin{equation}
\lim_{k\rightarrow \infty} {\rm W}_{\Psi^{(1)}_{k}}(z)
=\langle \Psi^{(1)},
\Phi^{(1)}(z)\rangle = {\rm W}_{\Psi^{(1)}}(z);
\end{equation}
it then follows (by  Vitali's theorem)  that the function
$\Phi^{(1)}(z)$
is weakly (and therefore   strongly) analytic.
The remaining part of the lemma is then proven as in the flat case,   because the constant $c(z)$ in Eq. (\ref{maggiorazione}) may be bounded as follows
for any $z=x+iy\in {\cal T}^{+}$ :
\begin{equation}
c(z)\leq \frac{const}{\|y\|^{\frac{d-1}{2}}}
\label{uffaulla}
\end{equation}
and therefore 
(in view of the results of appendix A) 
the  analytic function 
$ {\rm W}_{\Psi^{(1)}}(z)$ (with  $\Psi^{(1)}\in {\cal H}^{(1)}$) admits a boundary value    ${\cal W}_{\Psi^{(1)}}(x)$ on $X_d$, in the sense of distributions  such that 
 ${\cal W}_{\Psi^{(1)}}(x)= \lim_{k\to \infty} {\cal W}_{\Psi^{(1)}_k}(x)\equiv \langle\Psi^{(1)},\phi(x)
\Omega\rangle. $    \\  
\begin{remark}{\em      In the $\xi$-space realization of the Fock space that we have given in Eq. (\ref{azzolino})  the vector $\Phi(z)\in {\cal H}$ which has as only nonzero component $\Phi^{(1)}(z)$ may be represented (for $z\in {\cal T}^{+}$) by the vector
\begin{equation}
\Phi^{(1)}(z) = (0,(\xi \cdot z)^{-\frac{d-1}{2} -i\nu},0,\ldots).
\end{equation}
In the same way, if $\Psi=(0, \Psi^{(1)}(\xi_\gamma),0,\ldots)$ the analytic function  ${\rm W}_{\Psi^{(1)}}(z)$ is given in ${\cal T}^{+}$  by its Fourier-type representation
\begin{equation}
{\rm W}_{\Psi^{(1)}}(z) =c_{d,\nu}\int_{\gamma}
\overline{{\Psi^{(1)}(\xi)}}(\xi \cdot z)^{-\frac{d-1}{2} -i\nu}d\mu(\xi).
\end{equation}
We could have used this more direct and simple argument to verify the strong analyticity of $\Phi(z)$ but we went through the previous lemma because it promises to generalise to the interacting case. 
}\end{remark}  

For every bounded open set ${\cal O}\subset X_d$ let ${\cal P}_\nu({\cal O})$
be the polynomial algebra of the free field of parameter $\nu$ affiliated with the region ${\cal O}$. We then have:

\begin{proposition}(Reeh-Schlieder property for the free field)
For any open set ${\cal O}$   of (real) de Sitter space-time  
 the ``vacuum'' is a cyclic vector for ${\cal P}_\nu({\cal O})$.
\end{proposition}
 {\bf Proof.}
 The previous Proposition 5.3 together with Corollary  A.1 
imply that if  the distribution 
${\cal W}_{\Psi^{(1)}}(x)$ (for  $\Psi^{(1)}\in {\cal H}^{(1)}$) vanishes in an 
open set  ${\cal O}$ of $X_d$ then it vanishes on $X_d$; this in turn implies
 that $\Psi^{(1)}=0$ by construction of ${\cal H}_1$ (see \S 2.2). This proves  that the set 
\begin{equation}
{\rm D}^{(1)}_{{\cal O}}= \{ \Psi_f\in {\rm D}^{(1)}:\;\;
\makebox{supp} f \subset {\cal O}\}
\end{equation}
is dense in ${\cal H}^{(1)}$. The proposition is then proved by observing that the 
$n$-particle space ${\cal H}^{(n)}$ is the symmetrized tensor product of $n$ copies of ${\cal H}^{(1)}$.    \\  
 
For every bounded open set ${\cal O}\subset X_d$ let ${\cal P}^{(\rho)}({\cal O})$
be the polynomial algebra of a generalized free field of weight $\rho$ (see Proposition 5.2) affiliated with the region ${\cal O}$.  
Using now the K\"allen-Lehmann representation of the corresponding two-point function, namely ${\cal W}(x,x') = \int_{0}^{\infty} \rho(\nu) {\cal W}_\nu(x,x')d\nu$, one could again justify estimates similar to (\ref{maggiorazione}) and (\ref{uffaulla}), and therefore prove the analyticity and boundary value properties of the vector-valued function $\Phi^{(1)}(z)$ for any such field; the argument used for proving Proposition 5.4 would then yield similarly:

\begin{proposition}(Reeh-Schlieder property for the generalized free fields)
For any open set ${\cal O}$   of (real) de Sitter space-time  
 the ``vacuum'' is a cyclic vector for ${\cal P}^{(\rho)}({\cal O})$.
\end{proposition}

\subsection{ De Sitter QFT's derived from minkowskian QFT's}
 The idea that a natural class of de\nobreak\ Sitter QFT's should be obtained
by taking 
the restriction to $ X_d $ of QFT's defined in the ambient Minkowski space $
{\Bbb R}^{d+1} $ 
has been proposed in  \cite{[WY],[RT],[Fre]}.
The  axiomatic 
approach of Section 6 below, based on analyticity domains of the Wightman functions which are 
the restrictions to $ X^{(c)}_d\times ... \times  X^{(d)}_d $of the standard
axiomatic domains in $ {\Bbb C}^{d+1}\times ...\times {\Bbb C}^{d+1} $ 
of the Wightman functions of a $ (d+1) $-dimensional minkowskian QFT gives a 
priori an additional support to this idea. In this connection, it is then 
useful at first to study the class of de\nobreak\ Sitter fields in $ X_d $
obtained from 
minkowskian free fields (or GFF) in $ {\Bbb R}^{d+1} $ by this procedure. Our comments for this class 
of fields follow.

Let $ \hat W \left(z_1,z_2 \right) $ be the two-point function of a given free
field (or GFF) $ \hat \phi( x) $ in 
$ {\Bbb R}^{d+1}; $ $ \hat W $ is analytic in the cut-domain $ \hat \Delta  =
{\Bbb C}^{d+1}\times {\Bbb C}^{d+1}\setminus \hat \Sigma^{( c)}, $ 
where $ \hat \Sigma^{( c)} = \left\{ \left(z_1,z_2 \right) \in  {\Bbb
C}^{d+1}\times {\Bbb C}^{d+1};\ \left(z_1-z_2 \right)^2 = \rho ,\ \ \rho  \geq  0
\right\} . $ Since the 
section of $ \hat \Delta $ by $ X^{(c)}_d\times X^{(c)}_d $ is $ \Delta , $
the analytic function $ W \left(z_1,z_2 \right) = \hat W_{ \left\vert
X^{(c)}_d\times X^{(c)}_d \right.} $ 
satisfies the normal analyticity condition d); it therefore admits 
distribution boundary values $ {\cal W} \left(x_1,x_2 \right) $ and $ {\cal
W}^{\prime} \left(x_1,x_2 \right) $ from the respective 
tuboids $ {\cal T}_{12} $ and $ {\cal T}_{21} $ which define the restrictions
to $ X_d $ of the corresponding 
Wightman distributions of the field $ \hat \phi( x) $ namely:
\begin{equation}
{\cal W }= \hat{\cal W} \left(x_1,x_2 \right)_{ \left\vert X_d\times X_d
\right.} \ \ {\rm and} \ \ {\cal W}^{\prime} =\hat{\cal W} \left(x_2,x_1
\right)_{ \left\vert X_d\times X_d \right.}
\end{equation}

The positivity condition a) for $ {\cal W} $ then follows from the
corresponding 
condition for $ \hat{\cal W} $ on $ {\Bbb R}^{d+1}\times {\Bbb R}^{d+1} $ by
taking sequences of test-functions 
on $ {\Bbb R}^{d+1} $ which tend to $ f(x)\delta \left(x^2+1 \right), $ with $ f
\in  {\cal D} \left(X_d \right); $ conditions b)  and c)
are trivially satisfied. Therefore in view of the results of section 2, there 
exists a well-defined GFF $ \phi( x) $ on $ X_d $ whose $ n $-point functions
are the 
restrictions to $ X^{(c)}_d\times ...\times X^{(c)}_d $ of those of $ \hat
\Phi $ and which satisfies the thermal 
interpretation of \S 2.4. However
there does not exist a   K\"allen-Lehmann representation of the principal 
type for these fields. In fact, one knows the asymptotic
behaviour of 
the massive free field two-point function $ \hat{\cal W}_M \left(x_1,x_2 \right) $
with mass $ M $ in dimension $ d+1 $ 
 namely
\begin{equation}
\hat{\cal W}_M \left(x_1,x_2 \right) \sim  { {\rm e}^{iM \left[ \left(x_1-x_2
\right)^2 \right]^{1/2}} \over \left[ \left(x_1-x_2 \right)^2 \right]^{d/4}}\
,
\end{equation}
which, by restriction to $ X_d\times X_d $ yields:
\begin{equation}
{\cal W}_{M_{ \left\vert X_d\times X_d \right.}} \left(x_1,x_2 \right)
\sim   { {\rm e}^{iM \left[2 \left( {\rm cosh}  {\delta \over R} -1 \right)
\right]^{1/2}} \over \left[2 \left( {\rm cosh}{\delta \over R}-1 \right)
\right]^{d/4}}\ ,
\end{equation}
For $ d>2, $ this behaviour is strictly larger than the behaviour given 
by the Klein-Gordon de\nobreak\ Sitter two point functions of the principal series  (i.e. $ P^{(d+1)}_{-{d-1 \over 2}+imR} \left( {\rm
cosh}  {\delta \over R} \right)) $ 
which is such that 
\begin{equation} |{\cal W}_m \left(x_1,x_2 \right)| \sim   {\rm e}^{-{d-1 \over 2R}\delta}.
\label{comporta}
\end{equation}
It follows that the Laplace transform $ \hat G_M(\nu) $ of $ \hat{\cal W}_{M_{
\left\vert X_d\times X_d \right.}} $ is analytic (in view 
of proposition 3.1) in the half-plane $ \left\{ \nu ;\ {\rm Im} \ \nu  > - {d
\over 4} + {d-1 \over 2} = {d-2 \over 4} \right\} $ but 
certainly has a singularity on the boundary of the latter. Therefore $
\hat{\cal W}_{M_{ \left\vert X_d\times X_d \right.}} $ 
admits a representation of the form (\ref{a321}) in terms of the Legendre functions 
$ { P}^{(d+1)}_{-{d \over 4}-i\nu} \left({x_1\cdot x_2 \over R^2} \right),
$ which  { cannot} be put in the K\"allen-Lehmann form (\ref{a325}), 
due to the singularities of $ \hat G_M(\nu)  $ (in this respect see also Remark (\ref{gyu})). 
 
On the contrary, the massless field $\hat{\phi}$ in ${\Bbb R}^{d+1}$ (considered in \cite{[WY]}) behaves at infinity as $[(x_1-x_2)^2]^{-\frac{d-1}{2}}$
 and its restriction to $X_d$ actually has a principal-type behaviour. 
In fact one can  produce   \cite{[BG]} a K\"allen-Lehmann decomposition    of these fields in terms of the de Sitter   Klein-Gordon fields of the principal series.

\section{An axiomatic approach to de Sitter Quantum Field Theory}

We shall now propose a general characterization of what could be (in this connection see also
\cite{[I]}) a satisfactory quantum field theory on de Sitter space-time.    
We shall adopt the approach presented in \S 2.2, which is based on the 
GNS-type construction associated with a positive linear functional $ \omega $
on the 
Borchers-Uhlmann algebra $ {\cal B} $ on $ X_d; $ in the general case of
interacting fields, 
the functional $ \omega $ can be characterized by a sequence of Wightman
distributions 
$ {\cal W}_n \left(x_1,...,x_n \right) = \left\langle \Omega ,\phi \left(x_1
\right)...\phi \left(x_n \right)\Omega \right\rangle $ on $ \left[X_d
\right]^n $ satisfying an appropriate set of 
axioms (as in the Garding-Wightman reconstruction theorem for Minkowskian 
theories \cite{[SW]}).

If three of these axioms (listed under 1, 2 and 3 below) are straightforward 
adaptations to the de\nobreak\ Sitter space-time of the corresponding axioms
of 
Minkowskian Q.F.T., the fourth one which plays the role of the Spectral 
Condition must be given a completely new form. In fact, due to the absence of 
a global energy-momentum interpretation on the curved space-time $ X_d, $ our 
fourth axiom will be formulated in terms of analytic continuation properties 
of the distributions $ {\cal W}_n $ in the complexified manifolds $
\left[X^{(c)}_d \right]^n $ corresponding 
to $ \left[X_d \right]^n. $

Of course, the choice of such global analyticity properties may certainly be 
done in a non-unique way, and the corresponding properties of the GNS 
\lq\lq vacuum\rq\rq\ $ \Omega $ of the considered theories will of course
depend on the postulated 
analyticity properties. This is in contrast with the GFF case, which is 
determined unambiguously by the perikernel analytic structure of the 
two-point functions, and equivalently characterized by the thermal 
interpretation of the GNS \lq\lq vacuum\rq\rq . In the general case, the
thermal 
interpretation of the theory might provide a criterion of selection between 
several {\it a priori possible} global analytic structures.

We do not address this kind of question here,  
but we would just like to propose, as an open suggestion, a rather natural 
form for the fourth axiom, which leads to an axiomatic approach of de\nobreak\
Sitter 
QFT very close to the Wightman axiomatic approach by its analyticity 
properties and by the corresponding program which it suggests.

\vskip5pt
{\em The axioms}
\vskip5pt
\noindent
The set of distributions  $ {\cal W}_{n}$ ($n\geq 1$) is assumed to satisfy the following properties (we limit ourselves here to the case of a boson field):
\begin{enumerate}
\item (Covariance). Each ${\cal W}_{n}$ is de Sitter invariant, i.e.
\begin{equation}
{\cal W}_{n}(f_{n\{g\}})
={\cal W}_{n}(f_{n})
\label{cov}
\end{equation}
for all de Sitter transformations $g$, where 
\begin{equation}
f_{n\{g\}} ({x_{1}},\ldots,x_{n})
=f_{n} (g^{-1}{x_{1}},\ldots,g^{-1}x_{n})
\end{equation}
(Eq. (\ref{cov}) implies that the one-point function is a constant, which can be put equal to zero by additive renormalization of the field, as in the Minkowski case).
\item (Locality)
\begin{equation}
{\cal W}_{n}({x_{1}},\ldots,x_{j},x_{j+1},\ldots,x_{n})
={\cal W}_{n}({x_{1}},\ldots,x_{j+1},x_{j},\ldots,x_{n})
\end{equation}
if $(x_{j}-x_{j+1})^{2}<0$.
\item (Positive Definiteness). Given $f_{0} \in {{\Bbb C}}, 
f_{1} \in {\cal D}(X_d),\ldots,
f_{k} \in {\cal D}([X_d]^{k}),$
then
\begin{equation}
\sum_{n,m=o}^{k}
{\cal W}_{n+m}(\bar{f_{n}}\otimes
g_{m})\geq 0
\end{equation}
where
$ ({f_{n}}\otimes g_{m})({x_{1}},\ldots,x_{n+m})=
{f_{n}}({x_{1}},\ldots,x_{n}){g_{m}}(x_{n+1},\ldots,x_{n+m})$
\end{enumerate}
 We shall now give a substitute for the usual spectral condition of axiomatic field theory in Minkowski space, which will be called  ``weak spectral condition''.
We propose under this name analyticity properties of the Wightman functions which reproduce as closely as possible those implied by the usual spectral condition in Minkowskian QFT.
In the latter, it is known that the spectral condition can be
equivalently expressed by the following analyticity properties of the
Wightman functions 
${\cal W}_{n}$ \cite{[SW]}
resulting from the Laplace transform theorem in ${{\Bbb C}}^{d\,n}$: for each
$n\;\;(n\geq2)$ the distribution 
${\cal W}_{n}(\rm{x_{1},\ldots,x_{n}})$
is the boundary value of an analytic function 
${{\rm W}}_{n}({\rm z}_{1},\ldots,{\rm z}_{n})$
defined in the tube 
\begin{equation}
{\rm T}^d_{n-1}=\{{\rm z} = ({\rm z}_{1},\ldots,{\rm z}_{n});\;{\rm z}_{j}=
{\rm x}_{j} + i{\rm y}_{j}\in {{\Bbb C}}^{d},\;{\rm y}_{j+1}-
{\rm y}_{j}\in V^{+},\;1\leq j \leq n-1\}
\label{tubular}
\end{equation}
When the Minkowski space is replaced by  the de Sitter space $X_d$ embedded in ${{\Bbb R}}^{d+1}$, a
natural substitute for this property can be 
proposed by replacing (for each $n$) the tube ${\rm T}^d_{n-1}\subset {\Bbb
C}^{d\,n}$ by the corresponding open subset 
${\cal T}_{n-1}= [X^{(c)}_{d}]^{n}\cap {\rm T}^{d+1}_{n-1}$ of 
$[X^{(c)}_{d}]^{n}$. 
Using the notions introduced in appendix A and an argument similar to the Proof of Proposition 2.1. one can show the following

\begin{proposition}
{i)} The set 
\begin{equation}
{\cal T}_{n-1}=\{{ z} = ({ z}_{1},\ldots,{ z}_{n});\;{ z}_{j}=
{ x}_{j} + i{ y}_{j}\in X^{(c)}_{d},\;{ y}_{j+1}-
{ y}_{j}\in V^{+},\;1\leq j \leq n-1\}
\label{tubular1}
\end{equation}
is a domain of $[X^{(c)}_{d}]^{n}$;\\
{ii)} ${\cal T}_{n-1}$ is a tuboid above  $[X_d]^{n}$, with profile
\begin{equation}
\Lambda^{n} = \bigcup_{\x\in [X_d]^{n}} (\x,\Lambda^{n}_{\x})
\end{equation}
where, for each $\x=(\x_{1},\ldots,\x_{n})\in [X]^{n}$, $\Lambda^{n}_{\x}$ 
is a non-empty open convex cone with apex at the origin in 
${ T}_{\x}([X_d]^{n})$ defined as follows:
\begin{equation}
{\Lambda}^{n}_{\x}=\{{ \y} = ({ \y}_{1},\ldots,{ \y}_{n});\;{ \y}_{j}\in 
{ T}_{\x_{j}}X_d,\;1\leq j \leq n,\;{ \y}_{j+1}-
{ \y}_{j}\in V^{+},\;1\leq j \leq n-1\}
\label{tubular2}
\end{equation}
\label{pr1}
\end{proposition}

Proposition 6.1. allows us to  state the following axiom:
\begin{enumerate}
\setcounter{enumi}{3}
\item (Weak Spectral Condition).
For each $n$ the distribution ${\cal W}_{n}({x_{1}},\ldots,x_{n})$ 
is the boundary value of an analytic 
function ${\rm W}_{n}({z_{1}},\ldots,z_{n})$, defined in the tuboid ${\cal T}_{n-1}$ of  the complex manifold $[X^{(c)}_{d}]^{n}$
\end{enumerate}

By an application of theorem A.2., this axiom implies that for every $\x$ there is some local tube 
$\Omega_{\x}+i\Gamma_{\x}$ around $\x$ in any chosen system of local complex  coordinates on $[X^{(c)}_{d}]^{n}$ whose image in  $[X^{(c)}_{d}]^{n}$ 
is contained in ${\cal T}_{n-1}$ and has a profile very close to the profile of ${\cal T}_{n-1}$ (restricted to a neighbourhood of $\x$), from which the boundary value equation ${\cal W}_{n}=b.v.{\rm W}_{n}$ can be understood in the usual sense.
It implies equivalently that, in a complex neighbourhood of each point $x=(x_{1},\ldots,x_{n})\in [X_d]^{n}$, the analytic function 
${\rm W}_{n}({z_{1}},\ldots,z_{n})$ is of moderate growth (i.e. bounded by a power of $\|y\|$, where $\|y\|$ denotes any local norm of $ y=\Im z = (y_{1},\ldots, y_{n})$) when the point  $  z = (z_{1},\ldots, z_{n})$ tends to the reals inside ${\cal T}_{n-1}$.

\vskip5pt
\noindent{\em Use of locality together with weak spectral condition: the analytic n-point functions}

\vskip5pt
In view of the locality axiom the $n!$ permuted Wightman functions 
${\rm W}^{\pi}_{n}$,
respectively analytic in the permuted tuboids ${\cal T}^{\pi}_{n-1}$ 
have boundary values 
${\cal W}^{\pi}_{n}$
which coincide on the region 
${\cal R}_n$
of space-like configurations
\begin{equation}      
{\cal R}_n = \{ x = (x_1,\ldots, x_n) \in [X_d]^{n};\, (x_i-x_j)^2
<0,\, i\leq i,j\leq n\}.
\end{equation}
By making use of the edge-of-the-wedge theorem for tuboids (Theorem
A.3), one sees that for each  $n$ there is an analytic function 
$\makebox{\Eul W}_{n}(z_1,\ldots,z_n)$ 
which  is the common analytic continuation of all the functions 
${\rm W}^{\pi}_{n}$ in the union of the corresponding tuboids 
${\cal T}^{\pi}_{n-1}$ 
together with a complex neighbourhood of ${\cal R}_n$.

Other consequences of these axioms and the status of the associated
field theories as regards the classical theorems (PCT, Spin-Statistics, 
Bisognano-Wichmann theorem, etc.,) will be presented in a subsequent paper.

\vskip10pt
{\em Acknowledgements}\ \ It is a pleasure to thank D.\,Buchholz, J. Faraut, K. Fredenhagen   J.-P.\,Gazeau 
R.L. Stuller and G.A.\, Viano for useful discussions.
U. M. is a fellow of the EEC -Human capital and mobility program--contract n. ERBCHBICT930675.
\vskip20pt

\catcode `\@=11
\@addtoreset{equation}{section}
\def\theequation{\Alph{section}.\arabic{equation}}
\catcode `\@=12
\catcode `\@=11
\@addtoreset{proposition}{section}
\def\theproposition{\Alph{section}.\arabic{proposition}}
\catcode `\@=12
\catcode `\@=11
\@addtoreset{theorem}{section}
\def\thetheorem{\Alph{section}.\arabic{theorem}}
\catcode `\@=12
\catcode `\@=11
\@addtoreset{remark}{section}
\def\theremark{\Alph{section}.\arabic{remark}}
\catcode `\@=12
\catcode `\@=11
\@addtoreset{definition}{section}
\def\thedefinition{\Alph{section}.\arabic{definition}}
\catcode `\@=12
\catcode `\@=11
\@addtoreset{definition}{section}
\def\thelemma{\Alph{section}.\arabic{lemma}}

\appendix
\section{Tuboids and distributional boundary values}
\subsection{Tuboids}
Let ${\cal M}$ be a real $n$-dimensional analytic manifold, 
$T {\cal M} = \bigcup_{x\in {\cal M}} (x, T_x{\cal M})$ 
the tangent bundle to ${\cal M}$ and ${\cal M}^{(c)}$ a complexification of 
${\cal M}$. If $x_0$ is any point in ${\cal M}$, ${\cal U}_{x_0}$ and 
 ${\cal U}^{(c)}_{x_0}$ will denote open neighbourhoods of $x_0$, respectively in ${\cal M}$ and ${\cal M}^{(c)}$ such that  
${\cal U}_{x_0}$ = ${\cal U}^{(c)}_{x_0}\cap {\cal M}$; a corresponding neighbourhood of $(x_0,0)$ with basis  ${\cal U}_{x_0}$ in $T {\cal M}$
will be denoted $T_{{\rm loc}}{\cal U}_{x_0}$. 
\begin{definition} 
We call admissible local diffeomorphism at a point $x_0$ any diffeomorphism $\delta$ which maps some neighbourhood 
$T_{{\rm loc}}{\cal U}_{x_0}$ of $(x_0,0)$ in  ${ T}{\cal M}$
onto a corresponding neighbourhood ${\cal U}^{(c)}_{x_0}$ of $x_0$ in 
${\cal M}^{(c)}$ (considered as a $2n$-dimensional ${\cal C}^\infty$ manifold) in such a way that the following properties hold:
\begin{description}
\item{a)} $\forall x \in {\cal U}_{x_0}$, $\delta[(x,0)] =x$;
\item{b)} $\forall (x,y) \in {T}_{{\rm loc}}{\cal U}_{x_0}$, with  $y\not= 0,\, (y\in { T}_{x}{\cal M})$, the differentiable function
$t \to z(t) = \delta[(x,ty)]$
 is such that 
\begin{equation}
\frac{1}{i}\frac{dz}{dt}(t)|_{t=0} = \alpha y, \;\;\makebox{with} \;\;\alpha > 0.
\end{equation}
\end{description}
\label{a1}
\end{definition}
The conditions of Definition A.1. are obviously invariant under any real biholomorphic mapping describing a change of chart representation for  ${\cal M}$ and  ${\cal M}^{(c)}$.
The  simplest  example of admissible local diffeomorphism is obtained in a 
chart representation of  ${\cal U}_{x_0}$ (resp.  ${\cal U}^{(c)}_{x_0}$ ) 
by open neighbourhoods of the origin $\Omega$ (resp. $\Omega^{(c)})$ in  
${{\Bbb R}}^{n}$ (resp. ${{\Bbb C}}^{n}$),
such that  $\Omega$ = $\Omega^{(c)}\cap{{\Bbb R}}^{n}$; 
the identity mapping 
$
\underline{\delta}_{0}(\underline{x},\underline{y})= \underline{x}+i\underline{y}
$
from 
$T{{\Bbb R}}^{n} \equiv {{\Bbb R}}^{n}\times{{\Bbb R}}^{n}$
onto
${{\Bbb C}}^{n} = {{\Bbb R}}^{n} + i{{\Bbb R}}^{n}$
defines a local diffeomorphism (on 
$
{T}_{{\rm loc}}{\cal \Omega}_{x_0}= \underline{\delta}^{-1}_{0}(\Omega^{(c)})
$) 
which satisfies conditions a) and b). More general admissible diffeomorphisms
$\underline{\delta}$ (which map some domain $T_{\rm loc}\Omega$ of 
$T{{\Bbb R}}^{n}$ onto the complex domain $\Omega^{(c)}$) 
will allow us to introduce the appropriate definition of {\em local tube}.\\
\vskip5pt
  {\em Local tubes}
\vskip5pt
Let ${\cal C}_r$ be an open truncated cone in ${{\Bbb R}}^{n}$ of the form 
$
{\cal C}_r = \{ \y \in {{\Bbb R}}^{n};\, \y = \lambda \hat{\y},\,0<\lambda<r,\,
\hat{\y}\in \hat{\cal C}\}
$
where
$\hat{\cal C}$
denotes a domain in the sphere  
${{\Bbb S}}^{n-1}$.
Usually a local tube in ${{\Bbb C}}^n$ is defined as a domain of the form 
$\Omega + i {\cal C}_r \equiv \underline{\delta}_{0}(\Omega \times {\cal C}_r)$.
In order to extend this notion to complex manifolds we shall now call 
{\em generalized local tube} in ${{\Bbb C}}^n$ (near the origin) any domain of the form 
$\underline{\delta}(\Omega \times {\cal C}_r)$, where 
$\underline{\delta}$ is any admissible local diffeomorphism at the origin.
If the basis $\hat{\cal C}$
of ${\cal C}_r$ in ${{\Bbb S}}^{n-1}$ contains a given point $\hat{\underline{y}}_0$ we shall say that 
$\underline{\delta}(\Omega \times {\cal C}_r)$ is a {\em neighbouring local tube} of the point $(0,\hat{\underline{y}}_0)$ of the sphere bundle 
$S{{\Bbb R}}^{n} = {{\Bbb R}}^{n} \times {{\Bbb S}}^{n-1}$. \label{loctub}
 
The relevance of this definition  is justified  by the following
\begin{lemma}
If $\Delta = \underline{\delta}(\Omega \times {\cal C}_r)$ is  a 
neighbouring local tube of some point $(0,\hat{\underline{y}}_0)$ of 
$S{{\Bbb R}}^{n}$, then the following properties hold:
\begin{description}
\item{i)} there exists a neighbourhood $(\Omega_0 \times \hat{\cal C}_0)$ 
of $(0,\hat{\underline{y}}_0)$ in $S{{\Bbb R}}^{n}$ and a number 
$r_0>0$ such that the set 
$\Delta_0= \{ x + i \lambda \hat{\underline{y}}, \,  x \in \Omega_0,\,
0<\lambda<r_0,\, \hat{\underline{y}} \in \hat{\cal C}_0 \}$ is contained in 
$\Delta$;
\item{ii)} for every admissible local diffeomorphism 
$\underline{\delta}'$ at the origin there exists a neighbouring local tube of
the form $\Delta'= \underline{\delta'}(\Omega' \times {\cal C}'_r)$ of $(0,\hat{\underline{y}}_0)$ which is contained in 
$\Delta$.
\end{description}
\label{lltt}
\end{lemma}
 {\bf Proof.}

i) Let us consider the action of the diffeomorphism 
$\sigma=\delta\circ\delta^{-1}_0$ (between neighbourhoods of the origin 
in ${{\Bbb C}}^{n} = {{\Bbb R}}^{2n}$) in the polar representation 
$X= {{\Bbb R}}^{n}\times{{\Bbb S}}^{n-1}\times\overline{{\Bbb R}}^{+}$
of ${{\Bbb C}}^{n}$ (namely $z=x+i\rho\hat{y}$, with $\rho \geq 0$ and 
$\hat{y} \in {{\Bbb S}}^{n-1}$): in view of Def. \ref{a1} one can check that 
$\sigma$ defines a homeomorphism 
$\hat{\sigma}$ between neighbourhoods of the subset 
$Y = \{0\}\times{{\Bbb S}}^{n-1}\times\{0\}$ in $X$ (in the topology induced on 
$X$ by ${{\Bbb R}}^{n}\times{{\Bbb S}}^{n-1}\times {{\Bbb R}} $) such that  $\hat{\sigma}[(x,\hat{y},0)]=(x,\hat{y},0)$.
Now, the set $\Delta\cup \Omega$ of ${{\Bbb C}}^{n}$ is represented 
in $X$ by the open set 
$\hat{\Delta} = \hat{\sigma} (\Omega\times\hat{\cal C}\times[0,r[ )$,
which is a neighbourhood of the point $(0,\hat{\y}_0,0)$ in $X$;
since the homeomorphism $\hat{\sigma}$ preserves this point, 
$\hat{\Delta}$ certainly contains a neighbourhood of $(0,\hat{\y}_0,0)$ 
of the form 
$\hat{\Delta}_0 = (\Omega_0\times\hat{\cal C}_0\times[0,r_0[ )$ in X,
which proves that correspondingly 
${\Delta}_0 \subset {\Delta}$.

ii) With every diffeomorphism $\underline{\delta}'$,
the previous procedure associates a homeomorphism 
$\underline{\hat{\sigma}}'$ which transforms the basis of neighbourhoods 
$ (\Omega'\times\hat{\cal C}'\times[0,r'[ )$ of the point $(0,\hat{y},0)$ 
(in $X$) into another basis of neighbourhoods of the same point;
one can therefore find such a set 
$\hat{\Delta}'  = \hat{\sigma}'(\Omega'\times\hat{\cal C}'\times[0,r'[ )$ such that
$\hat{\Delta}'\subset \hat{\Delta}$ which implies correspondingly
${\Delta}'\subset {\Delta}$. 
   \\  

Returning to complex manifolds, one can call {\em local tube} in 
${\cal M}^{(c)}$ any domain which admits a representation by a generalized 
local tube in ${{\Bbb C}}^{n}$ in a certain well-chosen chart. However, 
the more general notion of {\em tuboid} introduced below is more flexible for considering analytic functions with  moderate increase and corresponding 
distributional boundary values on $\cal M$.

A tuboid can be described as a domain in ${\cal M}^{(c)}$ which is bordered 
by the real manifold ${\cal M}$ and whose ``shape'' near each point 
of ${\cal M}$ is (in the space of $\Im z$ and for $\Im z \to 0$) very close to a given cone $\Lambda_x$ of the tangent space $T_x{\cal M}$ to  ${\cal M}$ at the point $x$. The following more precise definitions are needed.

\begin{definition} We call ``profile'' above ${\cal M}$ any open subset $\Lambda$ of ${ T}{\cal M}$ which is of the form 
$\Lambda= \bigcup_{x\in {\cal M}}(x,\Lambda_{x})$,
where each fiber 
$\Lambda_{x}$ is a non-empty cone with apex at the origin  in ${   T}_{x}{\cal M}$ ($\Lambda_{x}$ can be the full tangent space  ${   T}_{x}{\cal M}$).\label{a2}
\end{definition}

It is convenient to introduce the ``projective representation''  ${ \dot{T}}{\cal M}$
of ${ T}{\cal M}$, namely  
${ \dot{T}}{\cal M} = \bigcup_{x\in {\cal M}}(x,{ \dot{T}}_{x}{\cal M})$, 
with ${ \dot{T}}_{x}{\cal M}$ = ${ {T}}_{x}{\cal M}\setminus \{0\}/
{{\Bbb R}}^{+}$. The image of  each point $y \in  { {T}}_{x}{\cal M}$ in 
${ \dot{T}}_{x}{\cal M}$ is  $\dot{y}=\{\lambda y;\,  \lambda >0\}$.
Each profile $\Lambda$ can then be represented by an open subset 
$\dot{\Lambda}= \bigcup_{x\in {\cal M}}(x,\dot{\Lambda}_{x})$ of  
${ \dot{T}}_{x}{\cal M}$ (each fiber $\dot{\Lambda}_{x} = {\Lambda}_{x}/
{{\Bbb R}}^{+}$ being now a relatively compact set).
We also introduce the complement of the closure of $\dot{\Lambda}$ in 
${ \dot{T}}{\cal M}$,
namely the open set 
$\dot{\Lambda}'$ =
${ \dot{T}}{\cal M}\setminus \overline{\dot{\Lambda}}$=
$\bigcup_{x\in {\cal M}}(x,\dot{\Lambda}'_{x})$ (note that 
$\dot{\Lambda}'_{x}\subset{ \dot{T}}_{x}{\cal M}
\setminus \overline{\dot{\Lambda}}_{x}$).
\begin{definition}
A domain $\Theta$ of ${\cal M}^{c}$ is called a tuboid with profile $\Lambda$ 
above ${\cal M}$ if it satisfies the following property.
For every point $x_0$ in $\cal M$, there exists an admissible local diffeomorphism $\delta$ at $x_0$ such that:
\begin{description}
\item{a)} every point $(x_0,\dot{y}_0)$ in $\dot{\Lambda}$ admits a compact neighbourhood $K(x_0,\dot{y}_0)$ in $\dot{\Lambda}$ such that \\
$\delta \left[ \{(x,y);\, (x,\dot{y})\in K(x_0,\dot{y}_0),\,(x,y) \in 
{   T}_{{\rm loc}}{\cal U}_{x_0}\}\right]\subset \Theta$.
\item{b)} every point $(x_0,\dot{y}_0')$ in  $\dot{\Lambda}'$ admits a compact neighbourhood $ K'(x_0,\dot{y}_0')$ in $\dot{\Lambda}'$ such that \\
$\delta\left[\{(x,y);\, (x,\dot{y})\in K'(x_0,\dot{y}_0),\,(x,y) \in 
{   T}_{{\rm loc}}{\cal U}'_{x_0}\}\right]\cap \Theta=\emptyset$
\end{description}
In a) and b) ${   T}_{{\rm loc}}{\cal U}_{x_0}$ and ${   T}_{{\rm loc}}{\cal U}'_{x_0}$ denote sufficiently small neighbourhoods of 
$(x_0,0)$ in ${ T}{\cal M}$ which may depend respectively on $\dot{y}_0$ and $\dot{y}_0'$.\label{a3}
\end{definition}
Since it makes use only of the notion of local admissible diffeomorphism , this definition is invariant under the choice of the 
 analytic chart representation  of ${\cal M}$ and ${\cal M}^{(c)}$.;
moreover, it is very flexible in view of the following 
\begin{proposition}
If $\Theta$ is a tuboid with profile $\Lambda$ above $\cal M$, it satisfies the properties a) and b) of Definition \ref{a3} for {\em all} admissible local diffeomorphisms $\delta$ at all points of ${\cal M}$.
\label{pra1}
\end{proposition}
 {\bf Proof.}
The set described in a) (or similarly in b)) actually contains a neighbouring local tube of the given point $(x_0,\dot{y}_0)$ whose representation in 
${{\Bbb C}}^{n}$ (in a given chart around $x_0$) could be described as in 
Lemma \ref{lltt} by a set of the form $\Delta = \underline{\delta}
(\Omega\times {\cal C}_r)$, $\underline{\delta}$ being the representation of 
$\delta$ in the considered chart. It then follows from part ii) of 
Lemma \ref{lltt} that for any other admissible diffeomorphism 
$\delta'$ (at $x_0$) there exists another neighbouring local tube of $(x_0,\dot{y}_0)$ represented in the chart by 
$\Delta '= \underline{\delta}'
(\Omega'\times {\cal C}'_{r'})$
and such that 
$\Delta '\subset \Delta $. The set inclusion (resp. identity) expressed by 
property a) (resp. b)) will  then be a fortiori satisfied by this new 
neighbouring local tube defined with the help of $\delta'$.
   \\  
\vskip5pt
{\em Local representations of tuboids on manifolds by tuboids in 
${{\Bbb C}}^{n}$}
\vskip5pt
A useful application of Proposition \ref{pra1} is obtained by considering local diffeomorphisms $\delta$ which admit a privileged linear representation of the form $\underline{\delta}=\underline{\delta}_0$ in a certain system of coordinates. In such variables, a tuboid $\Theta$ is represented {\em locally} (i.e. near a point $x_0$) as a tuboid $\underline{\Theta}$ in ${{\Bbb C}}^{n}$ whose profile $\underline{\Lambda}$ (in $T{{\Bbb R}}^{n}$) satisfies the following properties.
\begin{description}

\item{i)} Every point $(\x_0,\hat{\y}_0)$ in $\underline{\Lambda}$ 
(with $\hat{\y}_0 \in {{\Bbb S}}^{n-1}$) admits a 
neighbouring local tube of the form $\Omega + i {\cal C}_r$ = 
$\underline{\delta_0}(\Omega \times{\cal C}_r)$ which is contained in 
$\underline{\Theta}$. 
\item{ii)} Every point $(\x_0,\hat{\y}_0)$ in $\underline{\Lambda}' = T{{\Bbb R}}^{n}\setminus \overline{\underline{\Lambda}}$  admits a neighbouring local tube 
 $\Omega' + i{\cal C}'_{r'}$ = 
$\underline{\delta_0}(\Omega'\times {\cal C}'_{r'})$ such that 
 $(\Omega' + i {\cal C}'_{r'})\cap\underline{\Theta}=\emptyset$. 
\end{description}
In \cite{[BI]} a slightly more restrictive definition of tuboids in ${{\Bbb C}}^{n}$ was given, in which condition ii) was replaced by the following one
(non-invariant under real biholomorphic mappings):
{ii)$'$}$\underline{\Theta}\subset 
\underline{\delta_0}(\underline{\Lambda})$ (the condition {i)} being valid as before).
We can however say that the domains introduced in \cite{[BI]} form a ``basis of tuboids'' of ${{\Bbb C}}^{n}$, since every tuboid $\underline{\Theta}$ in 
${{\Bbb C}}^{n}$ (defined here)  contains a tuboid 
$\underline{\Theta}_b$ of that special family having the same profile as 
$\underline{\Theta}$ (namely  $\underline{\Theta}_b=  \underline{\Theta}\cap
\underline{\delta_0}(\underline{\Lambda})$).

Properties of pseudoconvexity of tuboids which generalize in some sense the standard ``tube theorem'' (see e.g. \cite{[W]}) have been derived in 
\cite{[BI]}; in view of the previous remarks, these results can be readily applied to the general class of tuboids in ${{\Bbb C}}^{n}$ considered here; they can be applied as well (due to their coordinate-independent character)
 to the more general class of tuboids on manifolds introduced above. These properties can be summarized in the following
\begin{theorem}
1) A necessary condition for a tuboid $\Theta$ with profile $\Lambda$ above $\cal M$ to be a pseudoconvex domain (or holomorphy domain) in ${\cal M}^{(c)}$ is that each fiber $\Lambda_x$ of $\Lambda$ is an open {\em convex} cone 
in $T_x{\cal M}$ (possibly equal to the whole space $T_x{\cal M}$).

2) If a tuboid $\Theta$ has a profile $\Lambda$ whose all fibers $\Lambda_x$
are convex, then, for each domain ${\cal M}'$ of ${\cal M}$ 
described by a chart in ${{\Bbb R}}^n$, there always exists a tuboid  $\Theta' $ above ${\cal M}'$  contained in
 $\Theta$ with  profile $\Lambda\cap T{\cal M}'$ which is a holomorphy
 domain in ${\cal M}^{(c)}$
\end{theorem}
The following statement is a direct consequence of Definition A.3.
\begin{proposition}  Let $ \Theta_ i $ $ (i=1,2) $ be tuboids
above $ {\cal M}_i $ in $ {\cal M}^{(c)}_i $ with 
respective profiles $ \Lambda_ i. $ Then $ \Theta_ 1\times \Theta_ 2 $ is a
tuboid above $ {\cal M}_1\times{\cal M}_2 $ in $ {\cal M}^{(c)}_1\times{\cal
M}^{(c)}_2, $ with 
profile $ \Lambda_ 1\times \Lambda_ 2 $ in $ T{\cal M}_1\times T{\cal M}_2. $
\end{proposition}
\subsection{Distribution boundary values of holomorphic functions in tuboids}
We wish to give a generalization for tuboids on manifolds of the notion of 
distribution boundary value of holomorphic functions with moderate growth 
near the reals. We first recall the following standard result for local tubes 
in $ {{\Bbb C}}^n $ $\lbrack$SW$\rbrack$ $\lbrack$W$\rbrack$.
There is an equivalence between the following two properties of a function $
f(z) $ 
analytic in a given local tube $ \Theta =\Omega +i{\cal C}_r $ (with e.g. $
{\cal C}_r $ convex).

\vskip 12pt

i) $ f $ is {\it of moderate growth near the real} in the following sense: in
each 
local tube $ \Theta^{ \prime} =\Omega^{ \prime} +i{\cal C}^{\prime}_ r $ such
that $ \Omega^{ \prime} \subset \subset  \Omega $ and $ \hat{\cal C}^{\prime} 
\subset \subset  \hat{\cal C}, $ $ (\hat{\cal C}, $ $ \hat{\cal C}^{\prime} $
denoting the bases 
of $ {\cal C}_r $ and $ {\cal C}^{\prime}_ r $ on the sphere $ {{\Bbb S}}^{n-1}),
$ there exists an integer $ p\geq 0 $ and a 
constant $ C $ (depending on $ \Theta^{ \prime} ) $ such that:
\begin{equation}
\forall z = x+iy \in  \Theta^{ \prime} \ \ ,\ \ \ \vert f(x+iy)\vert  \leq  C
\vert y\vert^{-p}
\end{equation}

ii) The sequence of distributions $ \left\{ f_y;y \in  {\cal C}_r \right\} $
defined in $ {\cal D}^{\prime}( \Omega) $ by the 
equation:
\begin{equation}
\left\langle f_y,\varphi \right\rangle  = \int^{ }_{ } f(x+iy)\varphi( x) {\rm
d} x
\end{equation}
(for all $ \varphi $ in $ {\cal D}(\Omega) ) $ is weakly convergent for $ y $
tending to 0 in $ {\cal C}_r; $ this 
defines a distribution $ T= bv\ f=\lim_{\y\to 0; {y\in{\cal C}_r}} f_y $ in $ {\cal
D}^{\prime}( \Omega) , $ called the boundary value of 
$ f(z) $ on $ \Omega $ from the local tube $ \Theta . $

Moreover, if the growth order $ p $ of $ f(z) $ (in i)) does not depend on $
\Theta^{ \prime} , $ the 
boundary value $ T $ of $ f $ is a distribution of finite order (linked to $
p) $ in $ \Omega , $ 
and an alternative characterization of $ f(z) $ (which in fact makes the 
connection between i) and ii)) is the following one:

iii) There exists a holomorphic function $ g(z) $ which is defined in $ \Theta
$ and 
uniformly bounded in the closure of each subdomain $ \Theta^{ \prime} $
(defined in i)), and an 
integer $ q $ such that $ f(z) = \left(a_i {\partial \over \partial z_i}
\right)^qg(z), $ where $ a_i{\partial \over \partial z_i} $ denotes a first order 
derivative operator with real constant coefficients (note that a relevant 
property of the vector $ a = \left\{ a_i;1\leq i\leq n \right\} $ is that it
belongs to the basis $ {\cal C}_r $ of $ \Theta ). $ 
Moreover, the corresponding relation $ T = \left(a_i {\partial \over \partial
x_i} \right)^qG $ holds in the sense of 
distributions on $ \Omega , $ with $ T = bv\ f $ and $ G =bv\ g, $ the latter
being identified with 
the usual limit of $ g $ on $ \Omega , $ as a continuous function.

\vskip 12pt

We now consider the general case of tuboids $ \Theta $ with profile $ \Lambda
; $ in view of 
Theorem A.1, $ \Lambda $ can always be assumed to have {\it exclusively convex
fibers.} 
Our generalization of the previous notions requires two steps: we consider successively tuboids in $ {{\Bbb C}}^n $ and 
tuboids in manifolds.

\vskip 12pt

a) In $ {{\Bbb C}}^n, $ let $ \Theta $ be a tuboid above a domain $ D $ of $ {\Bbb R}^n, $ with profile $ \Lambda  = \bigcup^{ }_{ x\in D} \left(x,\Lambda_ x
\right) \subset  TD \subset  T{{\Bbb R}}^n. $

In view of Definition A.3 and Proposition A.1 applied to tuboids in $ {{\Bbb C}}^n
$ 
(with the choice $ \delta =\underline{\delta_ 0}), $ one can say that $ \Theta \supset
\bigcup^{ }_{ \left(x_0,\hat y_0 \right)\in \Lambda} \Theta \left(x_0,\hat y_0
\right), $ where each 
domain $ \Theta \left(x_0,\hat y_0 \right) = \Omega \left(x_0,\hat y_0
\right)+i{\cal C}_r \left(x_0,\hat y_0 \right) $ is a neighbouring
local tube of {\it 
the point} $ \left(x_0,\hat y_0 \right)$ of $\Lambda$  (with $ \left\vert\hat y_0
\right\vert  = 1)$. 
For a holomorphic function $ f $ in $ \Theta , $ it is therefore meaningful to
keep the 
previous definition of the {\it property of moderate growth near the real}
(see i)), the uniform majorizations of the form (A.2) being now required to 
hold in all the local tubes $ \Theta \left(x_0,\hat y_0 \right) $ (thus
implementing all the directions of 
the profile $ \Lambda $ of $ \Theta ). $

Similarly, we shall say that a holomorphic function in $ \Theta $ {\it admits
a 
distribution boundary value on $ D $ from $ \Theta $} if the restriction of $
f $ to each 
local tube $ \Theta \left(x_0,\hat y_0 \right) $ satisfies the previous
property ii) for functions $ \varphi $ in the 
corresponding space $ {\cal D} \left(\Omega \left(x_0,\hat y_0 \right)
\right); $ if this holds for all $ \left(x_0,\hat y_0 \right)\in \Lambda , $
it is clear 
that the resulting local distributions $ T_{ \left(x_0,\hat y_0 \right)} \in 
{\cal D}^{\prime} \left(\Omega \left(x_0,\hat y_0 \right) \right) $ satisfy 
consistency conditions in all the two-by-two intersections of the family of
open 
sets $ \Omega \left(x_0,\hat y_0 \right); $ since this family forms a covering
of $ D, $ the set of 
distributions $ \left\{ T_{ \left(x_0,\hat y_0 \right)}; \left(x_0,\hat y_0
\right)\in \Lambda \right\} $ therefore defines a unique distribution $ T $ 
on $ D $ which coincides with each of them on the corresponding open sets $
\Omega \left(x_0,\hat y_0 \right). $

The equivalence between properties i) and ii) then reduces in a trivial way 
to the corresponding equivalence for holomorphic functions in local tubes 
which has been stated above.

b) Since a tuboid in a manifold $ {\cal M}^{(c)} $ can be represented as an
atlas of 
tuboids in $ {{\Bbb C}}^n $ satisfying consistency conditions in the two-by-two 
intersections of charts of $ {\cal M}^{(c)}, $ the theory of distribution
boundary values 
of holomorphic functions from a tuboid in $ {\cal M}^{(c)} $ should reduce to
the theory in 
$ {{\Bbb C}}^n $ (given above in a)) provided the conservation of the previous
notions 
under (real) biholomorphic mappings be established. We shall now examine this 
point.

It is sufficient to consider in $ {{\Bbb C}}^n $ a biholomorphic mapping $ z
\longrightarrow  Z = \Phi( z) $ 
which transforms a tuboid $ \Theta $ above $ D, $ with profile $ \Lambda $
onto a tuboid $ \tilde \Theta $ above $ \tilde D, $ 
with profile $ \tilde \Lambda  $;  $ D $ and $ \tilde D $ are {\it bounded}
domains of $ {{\Bbb R}}^n $ such that $ \tilde D = \Phi( D) $ 
and (in the corresponding tangent spaces) $ \tilde \Lambda  = (\partial \Phi)(
\Lambda) . $ Let $ f(z) $ be a 
holomorphic function in $ \Theta , $ its image $ \tilde f = \Phi f $ defined by $ \tilde f(Z)= \left(f\circ \Phi^{ -1} \right)(Z) $ 
being holomorphic  in $ \tilde \Theta . $

At first, it is easy to check the invariance of property i) under the mapping 
$ \Phi ; $ namely, $ f $ is of moderate growth near the real in $ \Theta $ iff
the same property 
holds for $ \tilde f $ in $ \tilde \Theta ; $ moreover, if there is a uniform
growth order $ p, $ it is 
preserved by $ \Phi $ and $ \Phi^{ -1}. $ This follows from two simple facts:
the first one is 
that near any real pair $ (x_0,X=\Phi( x_0)) $ in $ D\times\tilde D, $ any
complex pair $ (z,Z=\Phi( z)) $ 
is such that $ {\ Im} \ Z = \left({\partial \Phi \over \partial x}(x_0)\right) {\rm
Im} \ z + 0 \left(\vert {\ Im} \ z\vert^ 2 \right), $ with $ 0 < C_1 \leq 
\left\vert{ D(X) \over D(x)} \right\vert  \leq  C_2 $ on the 
bounded set $ D; $ therefore inequalities of the form (A.2) are preserved
under 
the mappings $ \Phi $ and $ \Phi^{ -1}. $ The second fact is that (in view of
proposition A.1) 
any local tube $ \Theta \left(X_0,\hat Y_0 \right) $ in $ \tilde \Theta $
(neighbouring a profile element $ \left(X_0,\hat Y_0 \right)\in\tilde \Lambda
) $ has 
an inverse image $ \Phi^{ -1} \left[\Theta \left(X_0,\hat Y_0 \right) \right] $
in $ \Theta $ which contains a neighbouring local tube 
$ \Theta \left(x_0,\hat y_0 \right) $ of the profile element $ (x_0 = \Phi^{
-1} \left(X_0 \right), $ $ \hat y_0=y_0/ \left\vert y_0 \right\vert , $ $
y_0={\partial \Phi^{ -1} \over \partial X} \left(X_0 \right)\hat Y_0) $ $
\left( \left(x_0,\hat y_0 \right)\in \Lambda \right). $

Therefore, the exact conditions of property i) are satisfied by the function 
$ f $ in $ \Theta , $ if they are satisfied by $ \tilde f $ in $ \tilde \Theta
$ (and conversely).

In view of the equivalence of properties i) and ii) for tuboids in $ {{\Bbb C}}^n
$ 
(see a)), we can then say that the moderate growth property of $ f $ and $
\tilde f $ in 
their respective tuboids $ \Theta $ and $ \tilde \Theta $ is equivalent to the
fact that both $ f $ and $ \tilde f $ 
admit distribution boundary values $ T =bv\ f $ and $ \tilde T = bv\ \tilde f
$ respectively in $ {\cal D}^{\prime}( D) $, 
$ {\cal D}^{\prime} \left(\tilde D \right) $ from the tuboids $ \Theta $ and $
\tilde \Theta . $ 

The remaining point to establish is that $ \tilde T $ is the relevant
transform $ \Phi T $ of 
the distribution $ T $ under the (real $ C^{\infty} ) $-mapping $ \Phi , $
according to what is needed 
in the theory of distributions on $ C^{\infty} $-manifolds (see e.g.
\cite{[CDD]}); 
this transformation law is:
$$ \left\langle \Phi T,\tilde \varphi \right\rangle  = \langle
T,\varphi\rangle \ \ ,\ \ \ {\rm where} \ \varphi  = \left\vert{ D(X) \over
D(x)} \right\vert  \tilde \varphi \circ \Phi $$
(the transformation $ \tilde \varphi \to \varphi $ being a continuous mapping
from $ {\cal D} \left(\tilde D \right) $ to $ {\cal D}(D)); $ this 
law coincides with the usual one when $ T $ is a continuous function $ (\Phi T
= T\circ \Phi^{ -1}). $ 

In order to show that $ bv{\tilde f} = \Phi \ bv\ f, $ we shall make use of the
property iii) of 
the function $ \tilde f $ which we apply in any given local tube $ \Theta
\left(X_0,\tilde Y_0 \right) $ in $ \tilde \Theta ; $ let us 
assume for simplicity that $ q=1, $ so that
\begin{equation}
\tilde f_0 = \tilde f_{ \left\vert \Theta \left(X_0,\hat Y_0 \right) \right.}
= \left(a_j {\partial \over \partial Z_j} \right) \tilde g\ ,
\end{equation}
where $ \tilde g(Z) $ is holomorphic in $ \Theta \left(X_0,\hat Y_0 \right) $
and continuous in the closure of this 
set; correspondingly, there holds the following relation for the boundary 
values in $ {\cal D}^{\prime} \left(\Omega \left(X_0,\hat Y_0 \right) \right):
$
\begin{equation}
bv\ \tilde f_0 = \tilde T_{ \left\vert \Omega \left(X_0,\hat Y_0 \right)
\right.} = \left(a_j {\partial \over \partial X_j} \right) \tilde G\ ,
\end{equation}
where $ \tilde G = bv\ \tilde g $ is continuous on $ \Omega \left(X_0,\hat Y_0
\right). $

Let us now define the holomorphic function $ f_0(z) = \hat f_0(\Phi( z)) $
which, by 
convenience, we consider in a local tube $ \Theta \left(x_0,\hat y_0 \right) $
as introduced above (with $ x_0=\Phi^{ -1} \left(X_0 \right), $ 
$ y_0 = {\partial \Phi^{ -1} \over \partial X} \left(X_0 \right) \hat Y_0). $

Let also $ g(z) = \tilde g(\Phi( z)); $ in the domain $ \Theta \left(x_0,\hat
y_0 \right), $ there holds the following 
functional relation, equivalent to (A.4):
\begin{equation}
f_0 = f_{ \left\vert \Theta \left(x_0,\hat y_0 \right) \right.} = \sum^{ }_{
i,j}a_j {\partial z_i \over \partial Z_j} (\Phi( z)) {\partial g \over
\partial z_i}
\end{equation}

By taking the boundary values of both sides of Eq.(A.6) on $ \Omega
\left(x_0,\hat y_0 \right), $ one 
obtains the corresponding relation in $ {\cal D}^{\prime} \left(\Omega
\left(x_0,\hat y_0 \right) \right): $
\begin{equation}
bv\ f_0 = T_{ \left\vert \Omega \left(x_0,\hat y_0 \right) \right.} = \sum^{
}_{ i,j}a_j  {\partial x_i \over \partial X_j} (\Phi( x)) {\partial G \over
\partial x_i}
\end{equation}
where $ G = bv\ g $ is continuous on $ \Omega \left(x_0,\hat y_0 \right) $ and
such that $ \tilde G = \Phi G. $ Now, since the 
partial derivatives of distributions are transformed by the usual rule under 
the mapping $ \Phi $ (see e.g. \cite{[CDD]}), the image by $
\Phi $ of the r.h.s. of 
Eq.(A.7) is equal to the r.h.s. of Eq.(A.5) and therefore $ \Phi \ bv\ f_0 =
bv\ \tilde f_0. $ 
Since this holds for every pair of elementary local tubes $ (\Theta
\left(x_0,\hat y_0 \right), $ $ \Theta \left(X_0,\hat Y_0 \right)) $ 
in $ \left(\Theta ,\tilde \Theta \right), $ this proves the desired identity $
\tilde T\equiv bv\ \tilde f = \Phi \cdot bv\ f $ in $ {\cal D}^{\prime}
\left(\tilde D \right). $ In the 
general case, the use of property iii) for the function $ \tilde f $ will
involve a 
derivation order $ q $ $ (q\geq 1); $ the same argument as above can then be
extended in 
an obvious way by recursion over $ q. $

To summarize the previous study, we can state:

\vglue 0.5truecm

\begin{theorem} Let $ f $ be a holomorphic function in a
tuboid $ \Theta $ of 
a complex $ n $-dimensional manifold $ {\cal M}^{(c)}; $ the following
properties are equivalent 
and independent of the local systems of complex variables, chosen to 
represent any given local part of $ \Theta $ as a tuboid in $ {{\Bbb C}}^n: $

i) $ f $ is of moderate growth near the real submanifold $ {\cal M}, $ in the
sense of (A.2) 
for each choice of local charts of $ {\cal M}^{(c)}. $

ii) $ f $ admits a distributional boundary value $ T $ on $ {\cal M}, $
obtained in all the 
local charts of $ {\cal M}^{(c)} $ by the procedure described in (A.3), in a
consistent 
way.

iii) in each local chart (of a sufficiently refined atlas), $ f $ admits a 
(non-unique) representation of the form $ f= \left(a_i {\partial \over
\partial z_i} \right)^qg, $ with $ g $ holomorphic in the restriction of $\Theta$ to this local chart and 
continuous on the real.
\end{theorem}

Close relations exist between the local order $ p $ of moderate growth of $ f
$ in 
i), the local order of the distribution $ T $ in ii) and the derivation order
$ q $ 
in the local representation of $ f $ in iii).

The previous notions lead to the following form of the edge-of-the-wedge 
property for tuboids on complex manifolds and more generally to the theory of 
\lq\lq singular spectrum\rq\rq\ (or \lq\lq essential support\rq\rq ) of
distributions on analytic 
manifolds.

\begin{theorem}(Edge-of-the-wedge for tuboids).  Let $
\Theta_ 1, $ $ \Theta_ 2 $ be 
two tuboids above $ {\cal M} $ in $ {\cal M}^{(c)}, $ with respective profiles
$ \Lambda_ 1, $ $ \Lambda_ 2 $ in $ T{\cal M}. $ Then 
there exists a tuboid $ \Theta $ containing $ \Theta_ 1\cup \Theta_ 2 $ whose
profile $ \Lambda  = \bigcup^{ }_{ x\in{\cal M}} \left(x,\Lambda_ x \right) $
is such 
that each fiber $ \Lambda_ x $ is the convex hull of $ \left(\Lambda_ 1
\right)_x\cup \left(\Lambda_ 2 \right)_x $ and which enjoys the 
following property.

For every  pair of functions $ f_1, $ $ f_2, $ respectively holomorphic in $
\Theta_ 1, $ $ \Theta_ 2 $ and 
admitting distribution boundary values on $ {\cal M} $ such that $ bv\ f_1 =
bv\ f_2 = T, $ 
there exists a unique holomorphic function $ f $ such that $ f_{ \left\vert
\Theta_ i \right.} = f_i, $ $ i=1,2 $ which 
is defined in $ \Theta $ and such that $ bv\ f = T $ (from $ \Theta  $).
\end{theorem}
 
The proof of the latter theorem is a direct application  of the edge-of-the-wedge 
theorem in the general situation of oblique local tubes ($\lbrack$E$\rbrack$,
$\lbrack$B.I$\rbrack$) which 
one applies to any  pair of neighbouring local tubes $ \left\{ \Theta_ i
\left(x_0,\hat y^i_0 \right) \subset  \Theta_ i;\ i=1,2 \right\} $ 
of arbitrary profile elements $ \left(x_0,\hat y^i_0 \right)\in \left(\Lambda_
i \right)_x $ (represented appropriately in 
arbitrarily chosen local charts of $ {\cal M}). $

We notice that if $\Theta_2 = \overline{\Theta_1}$, then ${\cal M}$ belongs to the analyticity domain $\Theta$; this implies 
\begin{corollary}
If a function $f$ is holomorphic in a tuboid $\Theta$ above $\cal M$ in ${\cal M}^{(c)}$ and if $ bv\; f=0$ on $\cal M$ then $f=0$.
\end{corollary}
 {\bf Proof.}
Applying Theorem A.3 to the functions $f(z), \; \overline{f}(\overline{z})$ in 
$\Theta\cup \overline{\Theta}$ reduces the property to the usual principle of analytic continuation.
   \\

\end{document}